\documentclass[fleqn,usenatbib,useAMS]{mnras}

\usepackage{graphicx}	
\usepackage{amsmath}	
\usepackage{amssymb}
\usepackage{subfigure}

\usepackage{bm}		
\usepackage[dvipsnames]{xcolor}

\newcommand{\pow}[1]{\ifmmode{}^{#1}\else ${}^{#1}$\fi}

\newcommand{\cm}{\,\ifmmode{{\mathrm{cm}}}\else cm\fi}

\newcommand{\ergps}{\,{\rm erg}\,{\rm s}\ifmmode{}^{-1}\else${}^{-1}$\fi}
\newcommand{\Mpch}{\,{\rm Mpc}\,\ifmmode h^{-1}\else $h^{-1}$\fi}
\newcommand{\snru}{\,\ifmmode{\mathrm{Myr}^{-1}}\else Myr${}^{-1}$\fi}
\newcommand{\kms}{\,\ifmmode{\mathrm{km}\,\mathrm{s}^{-1}}\else km\,s${}^{-1}$\fi\xspace}

\newcommand{\red}[1]{{\bf{\textcolor{red} {#1}}}}
\renewcommand{\red}[1]{#1}

\newcommand{\tcc}{\,\ifmmode{t_{\mathrm{cc}}}\else $t_{\mathrm{cc}}$\fi}
\newcommand{\lshatter}{\ifmmode{\ell_{\mathrm{shatter}}}\else $\ell_{\mathrm{shatter}}$\fi}
\def\gsim{\;\rlap{\lower 2.5pt
 \hbox{$\sim$}}\raise 1.5pt\hbox{$>$}\;}
\def\lsim{\;\rlap{\lower 2.5pt
   \hbox{$\sim$}}\raise 1.5pt\hbox{$<$}\;}

\usepackage[T1]{fontenc}
\usepackage{ae,aecompl}

\usepackage{newtxtext,newtxmath} 

\title{Beyond the Diffusion Coefficient: Propagators and Memory in Cosmic Ray Transport}

\author[Liang \& Oh]{
Naixin Liang\thanks{E-mail: naixin@ucsb.edu} and S. Peng Oh
\\
Department of Physics, University of California, Santa Barbara, CA 93106, USA}

\begin{document}
\label{firstpage}
\maketitle

\begin{abstract}
Cosmic ray (CR) transport is usually modeled with a single diffusion coefficient, but this description captures only the growth of the variance. Distinct transport mechanisms can share the same effective diffusion coefficient while producing different particle distributions and approaches to the diffusive limit. 
The distinction matters in multiphase, structured and time-dependent media, and is reflected in the observed variation of CR transport near pulsar wind nebulae, supernova remnants and molecular clouds, and in particle-tracing studies that show both superdiffusion and subdiffusion. 
We therefore develop a propagator-based framework centered on $P(x,t)$, the probability distribution of particle positions, or equivalently its Fourier--Laplace transform $P(k,s)$. 
\red{This object is compact and complete for the observable spatial transport. It naturally reveals if large scale transport has memory:} \red{when the CR flux depends on earlier rather than just the instantaneous pressure gradient. Such memory arises when dynamically relevant transport states, such as trapping or phase changes, are not retained in the macroscopic description.} 
Using the Montroll--Weiss formalism, we measure $P(k,s)$ directly from trajectories, recover the associated memory kernel, and represent broad kernels with a Prony expansion. 
Applied to a multiphase medium, the framework shows that slow regions can regulate escape without dominating the total residence-time budget. We also introduce an accelerated Monte Carlo method for coarse-grained transport, showing that if trapping structures evolve while particles are still sampling them, the static long-time limit need not be reached. This provides a foundation for observational applications, particle-tracing measurements and CR-MHD closures.
\end{abstract}


\section{Introduction}
In recent years, it has been increasingly recognized that cosmic rays (CRs) play a crucial role in galaxies. Their energy density can be comparable to that in gas, radiation, and magnetic fields, and through their coupling to magnetic fluctuations they can heat gas, provide pressure support, drive winds, and reshape the phase structure of the interstellar medium (ISM) and circumgalactic medium (CGM) \citep[e.g.][]{zweibel17,ruszkowski23,hopkins25-review}. Understanding how CRs propagate is essential for any predictive theory of CR feedback.

In the standard picture, the complicated microphysics of CR transport is compressed into a spatially and energetically dependent diffusion coefficient $\kappa(E,\mathbf{x})$, so that the CR flux is given by: 
\[
\mathbf{F}=-\kappa(E,\mathbf{x})\,\nabla P_c,
\]
with $\kappa(E)$ inferred phenomenologically from observables. Empirically, one often writes
\[
\kappa(E)\simeq 3\times 10^{28}\left(E/{\rm GeV}\right)^{\delta}\ {\rm cm^2\,s^{-1}},
\]
with $\delta \simeq 0.3{-}0.6$ \citep[e.g.][]{evoli2023}. This description has had major phenomenological success. But it also hides much of the transport physics.

The basic limitation is simple: a diffusion coefficient only controls the growth of the second moment. It does not uniquely determine the full spatial distribution of particles, and it is incomplete. Distinct stochastic processes can therefore share the same mean-squared displacement while differing in their full spatial profiles, higher moments, and dependence on past history \citep{metzler00,shin+10,cressoni+12,chechkin+17}. A linear mean-squared displacement is often taken as the signature of standard diffusion, but that criterion is not sufficient: transport can look diffusive in variance while remaining non-Gaussian, nonlocal, or history-dependent. Throughout this paper, we will refer to such dependence on past transport history simply as \emph{memory}.

There are now several reasons to suspect that the reduction to a single homogeneous $\kappa$ is incomplete in realistic astrophysical environments. First, the microphysics is uncertain: standard scattering theories struggle to reproduce the observationally inferred amplitude and energy dependence of $\kappa(E)$ without fine-tuning, whether scattering is dominated by self-generated waves or by extrinsic turbulence \citep{kempski22,hopkins22}. Second, the medium is highly structured. The ISM and CGM are multiphase, magnetically intermittent environments in which CRs can be mirrored, trapped, or scattered by shocks, current sheets, field-line wandering, magnetic bottles, and cloud interfaces, not just by a sea of weak waves \citep[e.g.][]{lazarian99,yan08,xu13,sampson22,lemoine23,kempski23,kempski25,butsky24-CRscatter,lubke25_mag,lubke25b,yan+26}. Because density, ionization, and magnetic topology vary strongly across phases, one naturally expects large fluctuations in scattering rate and effective opacity. Observationally, diffusion coefficients inferred near pulsar wind nebulae and supernova-remnant/molecular-cloud interfaces can be $10^2$--$10^3$ times smaller than the Galactic average \citep[e.g.][]{abeysekara17,aharonian+19,yang23}. Third, particle-tracing studies already show non-Gaussian transport, transient subdiffusion and superdiffusion, and broad distributions in specific transport diagnostics such as the time between pitch-angle reversals \citep[e.g.][]{lazarian21,zhang+23,effenberger+25,kempski25,lubke25b}. What remains less developed in most of this literature is the systematic definition of coarse-grained transport events and the direct measurement of the corresponding jump and waiting-time distributions. That is one of the motivations for the present paper. Fourth, several observations remain awkward for the simplest homogeneous-diffusion picture, notably the energy-dependent amplitude and phase of the CR dipole and the broad radial extent of Galactic $\gamma$-ray emissivity \citep{gabici19,ackermann11}. 

The natural response to this problem is not to search for a slightly better diffusion coefficient. It is to keep the full spatio-temporal spreading law. The relevant object is the propagator $P(x,t)$: the probability density that a particle which starts at the origin is found at position $x$ after time $t$. Its Fourier--Laplace transform, $P(k,s)$, which we shall henceforth refer to as the propagator, will often be more useful, because it organizes the transport simultaneously by spatial scale ($k^{-1}$) and timescale ($s^{-1}$). 
Unlike $\kappa$, the propagator retains the full statistical content of the transport: how broad the particle distribution is, how fast it spreads, whether the profile is Gaussian, and whether the response depends on what happened earlier. Thus, even when transport ultimately converges to ordinary diffusion at late times when jump-length variances and mean trapping times are finite \citep{liang25}, the diffusion coefficient $\kappa$ captures only that final asymptotic limit, whereas the propagator $P(k,s)$ retains the physically important crossover toward it. Time dependence of the medium can also matter: if the structures that trap CRs evolve while the particles are still sampling them, the large-scale transport need not approach the static asymptotic limit (\S\ref{subsec:renewal}).

This is why $P(k,s)$ is a much better organizing object for CR transport. It is practical, because it can be measured directly from particle trajectories. It is compact, because it summarizes the full distribution in one object rather than in an infinite hierarchy of moments. And it is interpretable, because different physical effects leave different signatures in different parts of $(k,s)$ space. If quoting a diffusion coefficient is analogous to quoting only the rms amplitude of $\delta T/T$ in the CMB, then the propagator $P(k,s|E)$ is more like the angular power spectrum $C_\ell$: still compact, but statistically far more informative.

A particularly useful framework for constructing and interpreting $P(k,s)$ is the continuous-time random walk (CTRW) formalism of Montroll \& Weiss \citep{montroll65}, which has a broad statistical-physics foundation \citep{bouchaud90,metzler00}. At an appropriate coarse-grained level, transport is described as a sequence of jumps separated by waiting times, and the Montroll--Weiss (MW) formalism links the statistics of those jumps and waits directly to the propagator. Related anomalous-transport ideas have been pursued in cosmic-ray and energetic-particle transport; for a recent review see \citet{effenberger+25}. Early CR applications of CTRW/MW ideas to describe anomalous transport include \citet{ragot+97,zimbardo+13}. 

Our goal here is different: not merely to reinterpret CR transport in Montroll–Weiss language, nor simply to replace one transport coefficient with another. Rather, we develop a practical framework linking microscopic trajectory statistics, full propagators, memory kernels, and macroscopic transport laws in a form that can be applied directly to simulations and, ultimately, to observations. We show how to measure $P(k,s)$ directly from trajectories without first reconstructing jump and waiting-time distributions; how the large-scale limit yields a frequency-dependent diffusivity $\tilde K(s)$ and a time-domain memory kernel $K(t)$; and how transport through a multiphase medium naturally generates bottlenecks, space--time coupling, and delayed convergence to ordinary diffusion. In particular, we will show that slow regions can control the overall resistance to escape without necessarily dominating the total residence-time budget, and that time-dependent resetting can keep the effective transport above the usual static harmonic-mean limit.

This paper is the first in a broader program. Here we establish the language, diagnostics, and analytic machinery: propagators, memory kernels, trajectory-based estimators, coarse-grained multiphase models, and efficient representations of memory. Companion papers will use these tools in three directions: to confront propagator-based transport models with Galactic and heliospheric observables; to measure transport kernels directly from particle tracing in realistic turbulent fields; and to embed these kernels in two-fluid CR-MHD through efficient closures. The hope is to move from fitting an effective $\kappa(E)$ toward a transport framework that can predict several thorny observables at once and can be deployed in dynamical simulations.

The outline of this paper is as follows. In \S\ref{sec:MW}, we introduce the Montroll--Weiss framework for CR transport, including scalar, coupled, and multistate generalizations, and show how the full propagator can be measured directly from particle trajectories. In \S\ref{sec:memory}, we 
show how history-dependent CR transport, or memory, arises, and how it can be quantitatively characterized. In \S\ref{sec:multiphase}, we apply this formalism to transport in a multiphase medium, where slow regions act as bottlenecks that regulate escape and memory. We introduce an accelerated Monte Carlo method that coarse-grains unresolved intra-cell scattering into event-driven transitions. We discuss the implications of this framework in \S\ref{sec:discussion} and summarize our conclusions in \S\ref{sec:conclusion}.

\section{The Montroll-Weiss Propagator}
\label{sec:MW}
To see why the propagator is the central object in our framework, it is useful to recall the limitations of standard transport diagnostics. Traditionally, random walks are often characterized by the growth of the mean-squared displacement. Two related diagnostics are the instantaneous and running diffusion coefficients,
\begin{equation}
\kappa_{\rm inst}(t) \equiv \frac{1}{2}\frac{d}{dt}\langle x^2(t)\rangle,
\qquad
\kappa_{\rm eff}(t) \equiv \frac{\langle x^2(t)\rangle}{2t}.
\end{equation}
These coincide only when $\langle x^2(t)\rangle \propto t$. The diffusion coefficient only tells us how fast the variance grows, not how particles are distributed. That information lives in the full Green's function $P(x,t)$, or equivalently in its Fourier--Laplace transform $P(k,s)$, \red{both of which
encode the full particle distribution and all moments that exist.} As we shall show, $P(k,s)$ also makes memory and space--time coupling explicit. However, the biggest payoff for working in Fourier--Laplace space is that Montroll--Weiss theory allows us to express $P(k,s)$ in terms of jump and waiting-time statistics. This transforms $P(k,s)$ into a compact and physically interpretable object. Crucially, the diffusion coefficient is not a unique fingerprint of transport: distinct processes can have the same $\kappa$ while producing very different propagators. Only the full $P(k,s)$ retains that information. We therefore begin with the Montroll--Weiss formalism (\S\ref{subsec:MWbasics}), then show how to measure $P(k,s)$ directly from particle trajectories (\S\ref{subsec:measureMW}), and finally illustrate why the diffusion coefficient by itself is insufficient (\S\ref{subsec:2_examples}).

\subsection{Basics of Montroll-Weiss Formalism}
\label{subsec:MWbasics}

\subsubsection{Single-State Montroll-Weiss Theory}
At a suitable coarse-grained level, particle transport can be viewed as a sequence of jumps separated by intervals during which the particle remains in its current state. Montroll--Weiss theory assumes that these coarse-grained steps form a renewal process: after each jump, the statistics of the next step are independent of the earlier history. In the simplest case, each step is described by a jump length $r$, drawn from a distribution $\lambda(r)$, and by the time $\tau$ until the next jump, drawn from a distribution $\psi(\tau)$. The goal is then to express the Green's function $P(x,t)$---the probability that a particle starting at the origin is at position $x$ at time $t$---in terms of these underlying step statistics.

At time $t$, there are two possibilities: (i) the particle has not jumped yet, so it remains at $x=0$. This has probability density $S(t) \delta(x)$, where 
\begin{equation} 
S(t) = 1 - \int_0^t \psi(\tau) d\tau
\label{eq:survival}
\end{equation}
is the survival probability of no jumps by time $t$. By direct substitution, we see that the survival function has Laplace transform: 
\begin{equation}
\tilde{S}(s) = \int_0^{\infty} dt \ e^{-st} S(t) = \frac{1 - \tilde{\psi}(s)}{s}.
\label{eq:Stilde}
\end{equation}
(ii) The particle has made {\it at least} one jump (it could have made many). Suppose the {\it last} jump occurred at time $\tau < t$, and at position $x^{\prime}$. Integrating over all possible jump positions $x^{\prime}$ and jump times $\tau$, we obtain: 
\begin{equation}
P(x,t) = S(t) \delta(x) + \int_0^t d\tau \ \psi(t-\tau) \int_{-\infty}^{\infty} dx^{\prime} \lambda(x-x^{\prime}) P(x^{\prime},\tau)
\label{eq:convolution}
\end{equation}
If we Fourier transform equation \ref{eq:convolution} in space and Laplace transform in time, then from the convolution theorem, the convolutions in $(x^{\prime}, \tau)$ become multiplications in $(k,s)$: 
\begin{equation}
P(k,s) = \tilde{S}(s) + \tilde{\psi}(s)\hat{\lambda}(k) P(k,s).
\end{equation} 
Solving for $P(k,s)$ gives us the celebrated Montroll-Weiss formula \citep{montroll65}: 
\begin{equation}
P(k,s) =  \frac{\tilde{S}(s)}{1-\hat{\lambda}(k) \tilde{\psi}(s)} = \frac{1 - \tilde{\psi}(s)}{s} \frac{1}{1-\hat{\lambda}(k) \tilde{\psi}(s)},
\label{eq:MW-equation}
\end{equation}
where we use equation \ref{eq:Stilde} in the second step. 

The Montroll-Weiss formula provides a fundamental link between microscopic jump statistics and macroscopic transport behavior. Assuming we have a waiting-time distribution $\psi(\tau)$ and a jump-length distribution $\lambda(\mathbf{r})$, we can directly obtain their Laplace and Fourier transforms and construct Eq.\ref{eq:MW-equation}. Once ${P}(k,s)$ is known, the real-space Green's function $P(\mathbf{r},t)$ can be obtained by an inverse Fourier-Laplace transform. 
\red{Equation \ref{eq:MW-equation} assumes a single state renewal process, and that the jump and waiting time distributions are independent. We discuss multi-state and coupled generalizations in \S\ref{subsec:generalization} and \S\ref{subsec:matrix_MW}.} 

A few points are worth noting. Equation \ref{eq:MW-equation} can be viewed as the infinite series expansion: 
\begin{equation}
P(k,s) = \tilde{S}(s) \left( 1 + \hat{\lambda}(k) \tilde{\psi}(s) + [\hat{\lambda}(k) \tilde{\psi}(s)]^2 +  ... \right) 
\end{equation}
where the first term corresponds to no jump, the second to one jump, the third to two jumps, and so forth. In real space, the Green's function is constructed from repeated convolutions of the joint jump--waiting kernel with itself\footnote{This underlies the emergence of stable distributions, which remain self-similar under repeated convolution.}; in Fourier--Laplace space, these convolutions reduce to simple multiplications. As we further emphasize in \S\ref{subsec:understand}, this is the reason for operating in Fourier-Laplace space: it diagonalizes the diffusion operator and turns intractable convolutions into straightforward multiplications. One might also wonder why space and time are treated differently: why Laplace transform in time, rather than Fourier transform? The reason is that diffusion is an initial value problem with a clear arrow of time, whose natural temporal modes are decaying exponentials $e^{-st}$, not oscillatory plane waves $e^{-i \omega t}$. Accordingly, the one-sided Laplace transform $\tilde{f}(s) = \int_0^{\infty} e^{-st} f(t) dt$ is the appropriate representation, whereas the two-sided transform $\tilde{f}(\omega) = \int_{-\infty}^{\infty} e^{-i \omega t} f(t) dt$ is more appropriate for stationary, time-translation invariant problems.

\subsubsection{Coupled and Multi-State Generalizations}
\label{subsec:generalization}

Equation \ref{eq:MW-equation}, the standard form of the Montroll-Weiss formula, contains two major simplifications. Firstly, it assumes that the jump-size and waiting-time distributions are independent, which need not be true-- for instance, longer jumps could have longer waiting times, as in the L\'evy walk model. In this case, we must consider the Fourier-Laplace transform of the joint jump size/waiting time distribution $w(\Delta x,t)$: 
\begin{equation}
\Phi(k,s) = \int_0^{\infty} dt \ e^{-st} \int_{-\infty}^{\infty} d(\Delta x) e^{i k \Delta x} w(\Delta x,t).
\end{equation}
Note that: 
\begin{equation}
\Phi(0,s) = \int_0^{\infty} dt \ e^{-st} \left[ \int_{-\infty}^{\infty} d(\Delta x) w(\Delta x,t) \right]= \tilde{\psi}(s), 
\end{equation}
since the term in brackets, which marginalizes $w(\Delta x,t)$ over $\Delta x$, is simply the waiting time distribution $\psi(t)$. Thus, for coupled jump size and waiting time distributions, the Montroll-Weiss equation becomes: 
\begin{equation}
P(k,s) = \frac{1-\Phi(0,s)}{s} \frac{1} {1-\Phi(k,s)}
\label{eqn:mw}
\end{equation}

Secondly, the scalar Montroll-Weiss formula, equation \ref{eq:MW-equation}, assumes single-state renewal, where jump length and waiting time distributions are the same in each event. In reality, renewal can be multi-state, where either or both of these distributions depend on the state the particle is in. An example might be traps (e.g., magnetic mirrors) of different depths and hence waiting times. Later, we calculate a specific example of diffusion in a two state system where the particle diffuses in either a fast or slow medium (\S\ref{sec:multiphase}). 

Once there are multiple states, matrix Montroll-Weiss is necessary. 
\red{Let $w_{ij}(\Delta x,t)$ denote the joint density of the process starting in state $i$, waiting a time $t$, taking a jump $\Delta x$, and transitioning to state $j$. Its Fourier--Laplace transform is $\Phi_{ij}(k,s)$. The vector component $g_i(k,s)$ is then the contribution to the propagator from a walk that starts in state $i$.}
\begin{equation}
\mathbf g(k,s)
=
\begin{pmatrix}
g_1(k,s)\\
\vdots\\
g_N(k,s)
\end{pmatrix}
\end{equation}
The vector $\mathbf{g}$ satisfies the matrix Montroll-Weiss equation: 
\begin{equation}
\mathbf g(k,s)
=
\left[I-\Phi(k,s)\right]^{-1}
\mathbf S(s).
\end{equation}
where the state resolved survival transform is:

\begin{equation}
\tilde{\mathbf S}(s)=\frac{\mathbf 1-\tilde{\bm\psi}(s)}{s},\qquad
\tilde{\psi}_i(s)=\sum_j \Phi_{ij}(0,s).
\end{equation}

where $\mathbf{1}$ is a vector of ones. Thus, for an initial state distribution $\mathbf g^{(0)}$, the Fourier-Laplace propagator is: 
\begin{equation}
P(k,s)=
\mathbf g^{(0)\,T}
\left[I-\Phi(k,s)\right]^{-1}
\mathbf{\tilde{S}}(s).
\label{eq:Pks_matrix}
\end{equation}
\red{Here $g^{(0)}$ is the initial distribution over states.} Note that even though the internal dynamics lives in a multi-state space, the final observable $P(k,s)$ is scalar, since it is just the Fourier-Laplace transform of the scalar spatial probability distribution $P(x,t)$. Equation \ref{eq:Pks_matrix} is the dot product between the initial conditions $\mathbf g^{(0)}$ and the evolved state vector $\mathbf g(k,s) = \left[I-\Phi(k,s)\right]^{-1} \mathbf S(s)$, and therefore sums over all internal states. 

\subsubsection{Consistency Checks: Benchmark Propagators}

Before turning to more complicated transport models, it is useful to verify that the Montroll--Weiss construction reproduces a set of benchmark continuous-time random walks (CTRWs) with known propagators. This serves as a consistency check of both the formalism and our numerical inverse transforms: starting from prescribed coarse-grained jump and waiting-time statistics, we construct $P(k,s)$ from the Montroll--Weiss formula, invert it to obtain $P(x,t)$, and compare the result against both the known Green's function and direct Monte Carlo realizations of the same process. The point here is to show that the Montroll-Weiss framework can reproduce the well-known Green's functions of the standard and fractional diffusion equations. 

We consider three one-dimensional examples. For standard diffusion, we take a Gaussian jump-length distribution and an exponential waiting-time distribution. For subdiffusion, we keep Gaussian jumps but adopt a heavy-tailed waiting-time distribution, $\psi(\tau)\propto \tau^{-1-\beta}$ with $0<\beta<1$, so that long trapping times produce memory. For superdiffusion, we instead take exponential waiting times and a symmetric heavy-tailed jump-length distribution, $\lambda(r)\propto |r|^{-1-\alpha}$ with $0<\alpha<2$, so that rare long jumps drive non-local transport. Applying the Montroll--Weiss formula (Equation~\ref{eq:MW-equation}) to these choices of jump and waiting-time statistics yields the standard propagators in the long-time, large-scale limit (small $s$ and small $k$; derived more systematically in \S\ref{sec:memory}): 
\begin{align}
P(k,s) &= \frac{1}{s+\kappa k^2}, && \text{standard diffusion}, \label{eq:pks-standard} \\
P(k,s) &= \frac{s^{\beta-1}}{s^\beta + K_\beta k^2}, && \text{subdiffusion}, \label{eq:pks-sub} \\
P(k,s) &= \frac{1}{s+K_\alpha |k|^\alpha}, && \text{superdiffusion}. \label{eq:pks-super} 
\end{align}
Here $\kappa$ is the ordinary diffusion coefficient, while $K_\beta$ and $K_\alpha$ are generalized transport coefficients that set the width of the subdiffusive and superdiffusive propagators, respectively.

\begin{figure}
    \centering
    \includegraphics[width=0.999\linewidth]{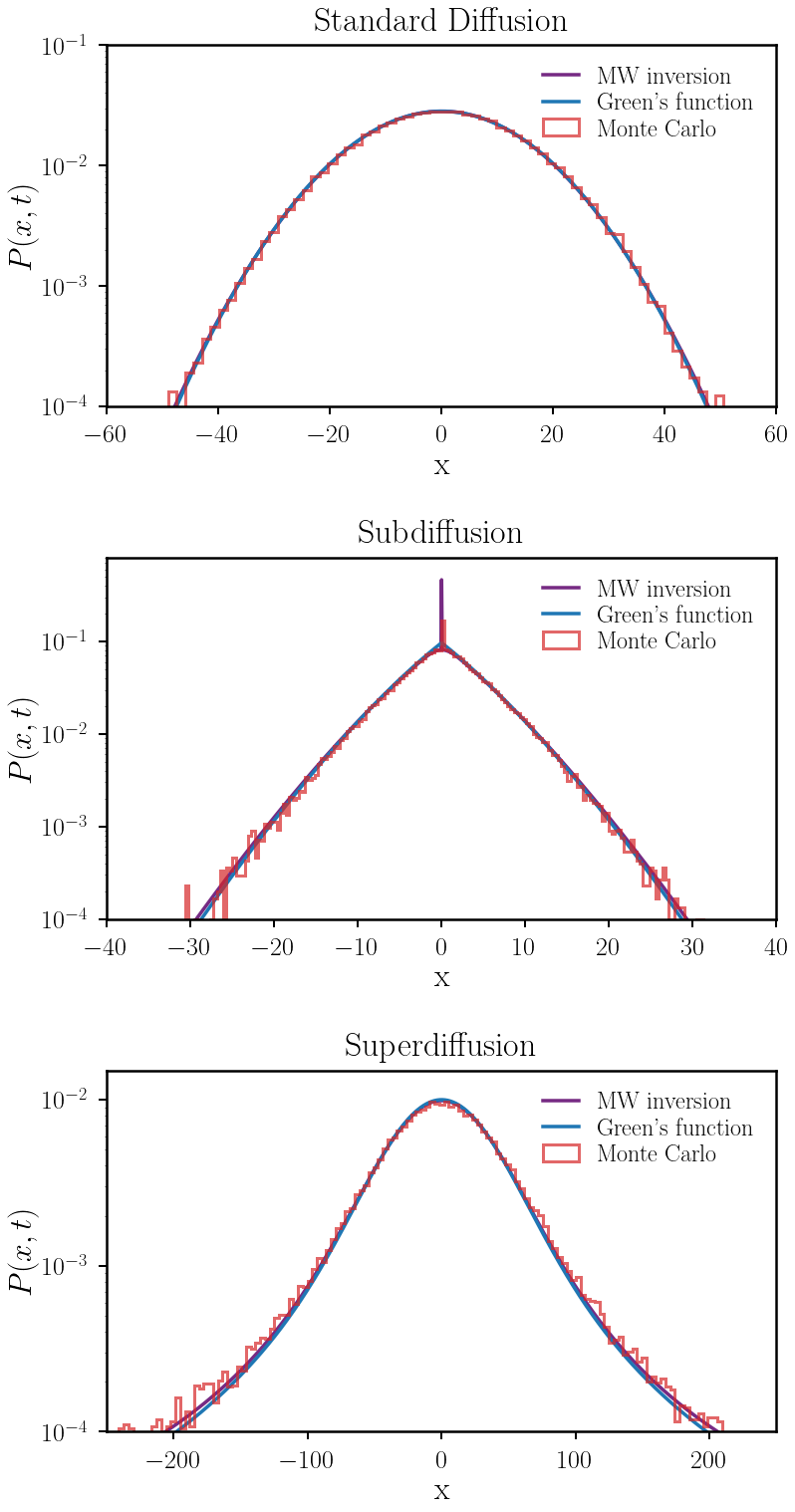}
    \caption{Green's functions $P(x,t)$ for three benchmark CTRWs: standard diffusion (top), subdiffusion (middle), and superdiffusion (bottom). Solid curves show the inverse Fourier--Laplace transform of the Montroll--Weiss propagator constructed from the prescribed jump and waiting-time statistics; dashed curves show the known analytic Green's functions; points show direct Monte Carlo realizations of the same CTRWs. The agreement demonstrates that the Montroll--Weiss formula reproduces the correct propagator across standard, subdiffusive, and superdiffusive limits.}
    \label{fig:green}
\end{figure}

For each case, we also generate direct Monte Carlo realizations of the underlying CTRW by repeatedly drawing jump lengths and waiting times from the chosen $\lambda(r)$ and $\psi(\tau)$. Figure~\ref{fig:green} compares the inverse Fourier--Laplace transform of the Montroll--Weiss propagator against these Monte Carlo results and against the standard analytic Green's functions at a fixed time. The agreement across all three regimes shows that the Montroll--Weiss formula faithfully reproduces standard diffusive transport, as well as canonical subdiffusive and superdiffusive limits, when supplied with the appropriate coarse-grained step statistics.

\subsection{Measurement from Particle Trajectories}
\label{subsec:measureMW}

The discussion above is a forward problem: given coarse-grained jump and waiting-time statistics, the Montroll--Weiss formula yields $P(k,s)$. In realistic systems, however, those statistics are rarely known \textit{a priori}. What we often have instead is a collection of particle trajectories. The practical question is therefore the inverse one: how do we estimate $P(k,s)$ directly from trajectory data? Once $P(k,s)$ is known, it can be used to diagnose memory, space--time coupling, and departures from ordinary diffusion without first specifying the underlying jump and waiting-time statistics.

Given $N$ particle trajectories $X_j(t)$, $j\in\{1,\dots,N\}$, sampled on a uniform time grid, we compute the empirical characteristic function of particle displacements: 
\begin{equation}
    \phi(k,t)=\frac{1}{N}\sum_{j=1}^N e^{i k [X_j(t)-X_j(0)]}.
    \label{eq:phi}
\end{equation}
The Montroll-Weiss (MW) propagator in Fourier-Laplace space is then
\begin{equation}
    P(k,s)=\int_0^{\infty} e^{-s t}\,\phi(k,t)\,dt,
\end{equation}
which in practice is approximated by a Riemann sum:
\begin{equation}
    \label{eqn:measure_mw}
    P(k,s)\approx \sum_n e^{-s t_n}\phi(k,t_n)\Delta t.
\end{equation}
Thus, the full propagator can be measured directly from trajectory data, without first reconstructing the underlying jump and waiting-time distributions.

\begin{figure}
    \centering
    \includegraphics[width=0.99\linewidth]{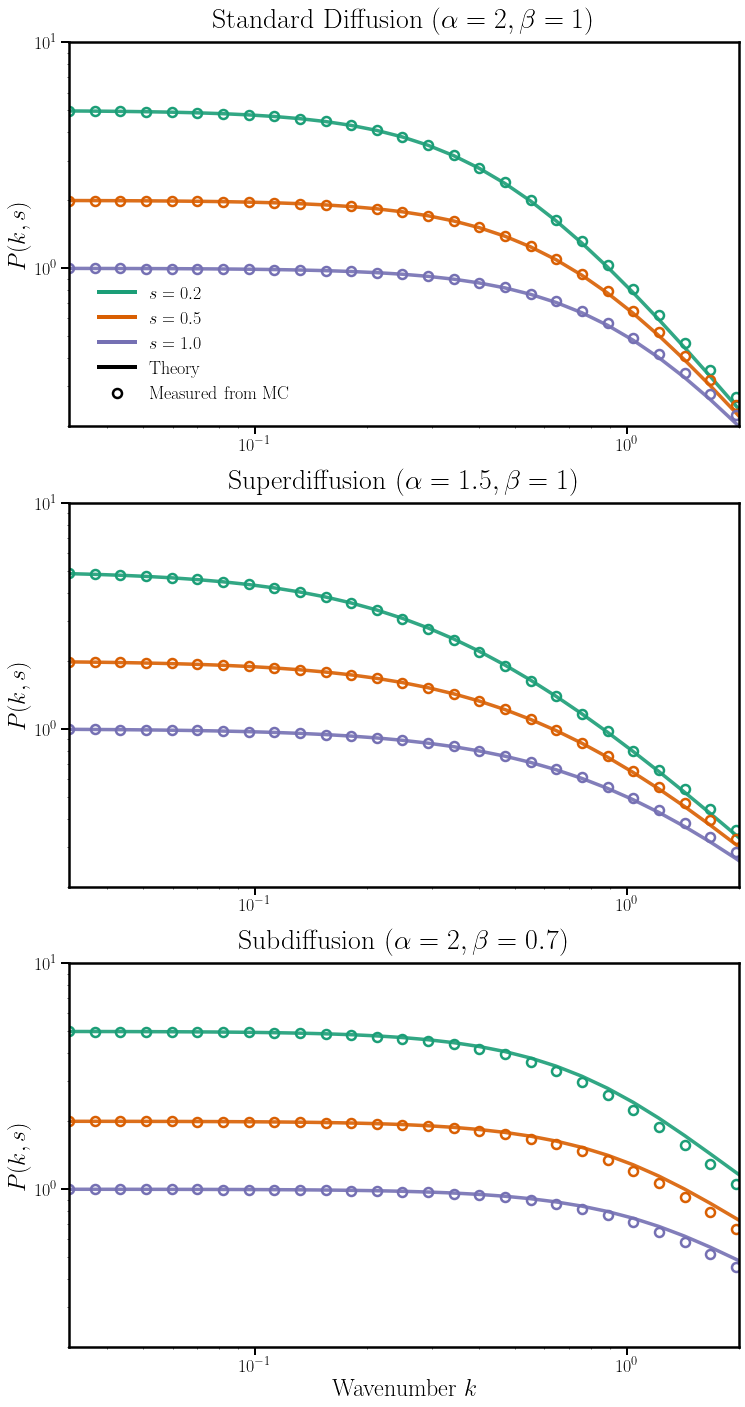}
    \caption{The MW propagator against wavenumber k, measured from Monte Carlo simulations (dot) and constructed by analytics (solid line, equations \ref{eq:pks-standard}--\ref{eq:pks-super}). \emph{Top:} standard diffusion ($\alpha=2,\beta=1$); \emph{Middle:} superdiffusion ($\alpha=1.5,\beta=1$); \emph{Bottom:} subdiffusion ($\alpha=2,\beta=0.7$). All panels show agreement at three different Laplace frequencies $s$.}
    \label{fig:mc_trajectory}
\end{figure}

To validate this estimator, we apply it to the same three benchmark CTRWs considered in Figure~\ref{fig:green}: standard diffusion, superdiffusion, and subdiffusion. For each case, we generate Monte Carlo trajectories from the prescribed step statistics and then measure $P(k,s)$ directly from those trajectories using Eq.~(\ref{eqn:measure_mw}). Figure~\ref{fig:mc_trajectory} shows that the measured propagators agree closely with the theoretical Montroll--Weiss predictions given by equation \ref{eq:pks-standard}--\ref{eq:pks-super} over a range of wavenumbers $k$ and Laplace frequencies $s$. This demonstrates that the trajectory-based estimator accurately recovers the full propagator in standard, subdiffusive, and superdiffusive regimes, and is therefore suitable for later application to multi-phase transport where the coarse-grained transport statistics are not known in advance.

Up to this point, our validation has focused on diffusion. In realistic astrophysical environments, CRs frequently undergo bulk directional motion, either by streaming along magnetic field lines or being advected by background plasma flows \citep[e.g.][]{bloemen93,zweibel17}.
If the particle ensemble has a nonzero mean drift velocity, whether from advection or streaming, the propagator acquires an advective $O(k)$ contribution that dominates the diffusive $O(k^2)$ term at sufficiently small $k$. For a constant drift velocity $v_{\rm d}$, we have:
\begin{equation}
    P(k,s) \sim \frac{1}{s - i k v_{\rm d} + k^2 \tilde K(s)},
\end{equation}
which reduces to $P(k,s) = [s - i k v_{\rm d} + k^2 \kappa]^{-1}$ for ordinary diffusion. To isolate the transport kernel and remove the advective term, we therefore construct the propagator in the co-moving frame,
\begin{equation}
    x'(t) = x(t) - \langle x(t)\rangle,
\end{equation}
or equivalently replace the MSD by the true variance $\langle (x-\langle x\rangle)^2\rangle$. All propagators and kernels below are understood to be in this frame whenever a nonzero mean drift is present.

\subsection{Why the Diffusion Coefficient is Not Enough}
\label{subsec:2_examples}

As we previously emphasized, the diffusion coefficient determines only the growth of the second moment. It does not uniquely specify the full propagator, nor does it reveal whether the transport retains memory of previously sampled environments. To make this concrete, we now consider two simple toy models in which the increase in the mean-squared displacement (MSD) is identical to standard diffusion, \(\langle x^2(t)\rangle \propto t\), yet the underlying transport is not fully characterized by a single effective diffusion coefficient. This is concerning: a linear MSD is the de facto criterion for diagnosing standard diffusion in the literature.

\begin{figure}
    \centering
    \includegraphics[width=0.99\linewidth]{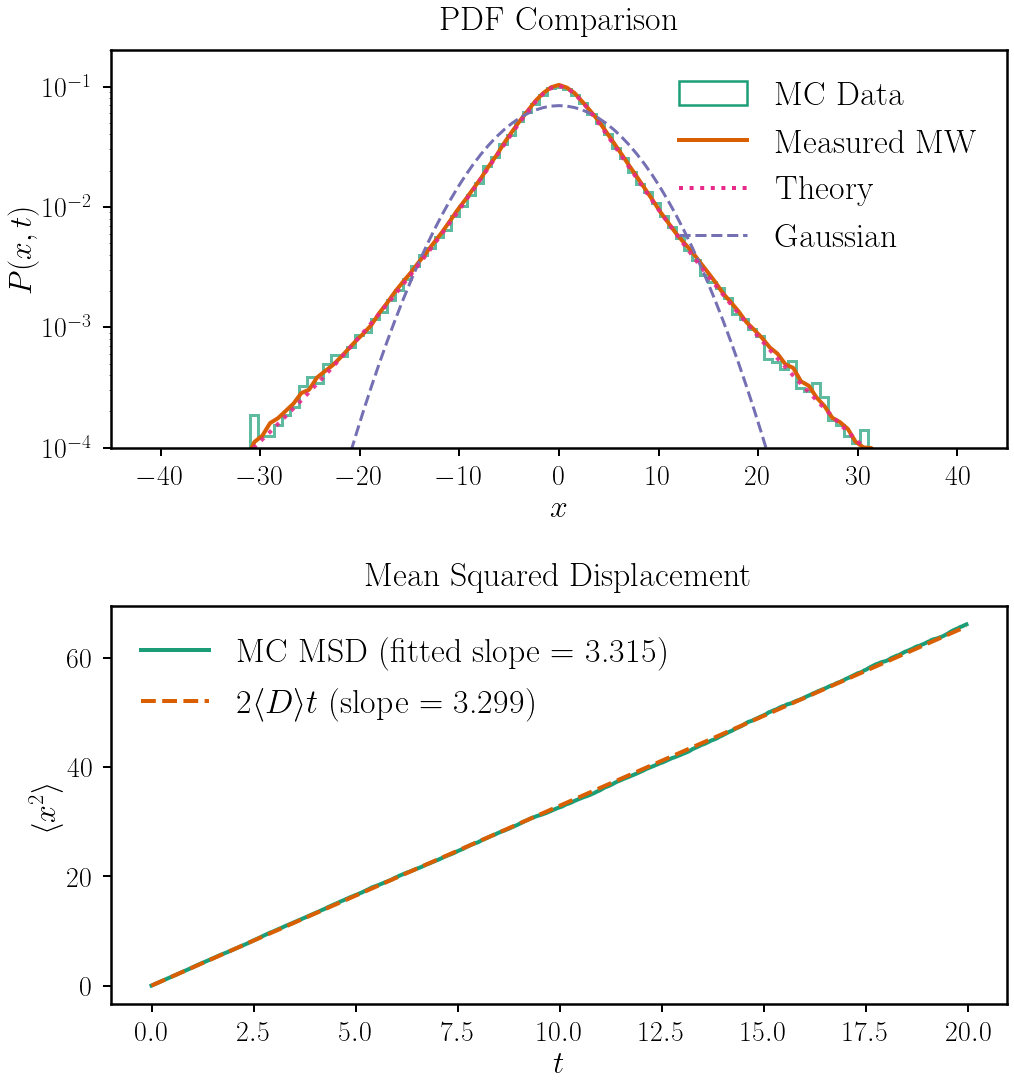}
    \caption{Statistics from the Monte Carlo (MC) data and the inverse Laplace transform of the MW propagator measured by Eq.\ref{eqn:measure_mw}. \emph{Top:} Probability density function $P(x,t)$ for standard diffusion at various diffusivities. The MC data (stepped histogram) show agreement with the measured MW inversion (solid) and theory (dotted). The Gaussian approximation (dashed) is provided for comparison, highlighting the difference in propagators. \emph{Bottom:} Mean squared displacement (MSD) as a function of time $t$. The MC result (solid line) shows a linear growth characteristic of normal diffusion with slope aligning closely with the theoretical prediction $2\langle D \rangle t$ (dashed).}
    \label{fig:logn_D}
\end{figure}

As a first example, consider a static heterogeneous medium. Each particle undergoes ordinary Brownian motion, but with a diffusion coefficient \(D_i\) drawn once from a lognormal distribution and then held fixed for that particle. The ensemble-averaged MSD is
\begin{equation}
    \langle x^2(t) \rangle = \frac{1}{N} \sum_{i=1}^N 2 D_i t = 2\langle D \rangle t .
\end{equation}
An observer measuring only \(\langle x^2(t)\rangle\) would therefore infer perfectly ordinary diffusion with effective coefficient \(\kappa_{\rm eff}=\langle D\rangle\). The full Green's function, however, is
\begin{equation}
    P(x,t)=\int dD\,p(D)\,\frac{1}{\sqrt{4\pi D t}}
    \exp\!\left(-\frac{x^2}{4Dt}\right),
\end{equation}
namely a superposition of Gaussians with different widths rather than a single Gaussian. This mixture model has a narrow core from particles in low-\(D\) environments and broad tails from particles in high-\(D\) environments and is manifestly non-Gaussian. This is an example of what has been termed `Brownian yet non-Gaussian diffusion' \citep{chechkin+17}. As shown in Figure~\ref{fig:logn_D}, the measured MW inversion accurately recovers the non-Gaussian propagator, while a Gaussian with \(\kappa_{\rm eff}\) does not.

To better mimic CRs crossing a multiphase ISM/CGM, where particles spend finite times in different environments, we next allow the local diffusivity \(D\) to be redrawn after a residence time \(t_D\). We draw \(D\) from the same lognormal distribution as above, and draw \(t_D\) from an exponential distribution with mean correlation time \(t_{\rm corr}\). Once \(t_D\) has elapsed, the particle enters a new region and a new pair \((D,t_D)\) is drawn. 
The previous model is recovered in the quenched limit $t_{\rm corr}\to\infty$, in which each particle draws a diffusivity once and retains it indefinitely. By contrast, for finite $t_{\rm corr}$ the diffusivity is repeatedly re-sampled (an annealed medium).

\begin{figure}
    \centering
    \includegraphics[width=0.99\linewidth]{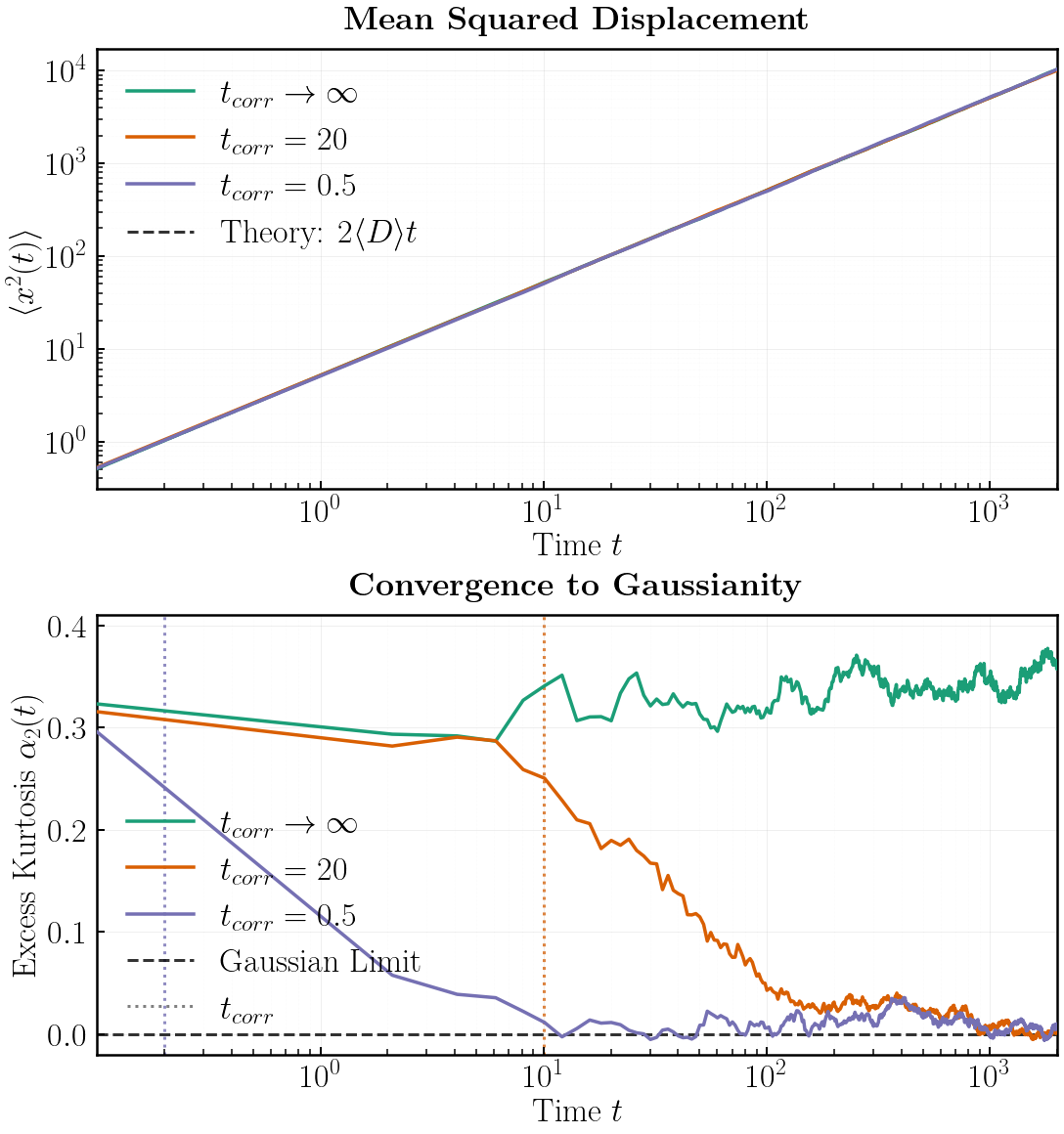}
    \caption{The Monte-Carlo results across different correlation times $t_{\rm{corr}}$. \emph{Top:} MSD $\langle x^2(t) \rangle$ as a function of time $t$. The simulated data for varying $t_{\rm{corr}}$ values (solid) overlap entirely, exhibiting the linear growth characteristic of standard diffusion, in perfect agreement with the theoretical prediction $2\langle D \rangle t$ (dashed). \emph{Bottom:} Convergence to Gaussianity measured by the excess kurtosis $\alpha_2(t)$. For finite correlation times, the excess kurtosis (solid) decays toward the Gaussian limit of zero (dashed), with the transition occurring at timescales corresponding to the $t_{\rm corr}$ values (dotted). In contrast, the static case ($t_{\rm{corr}} \to \infty$) never converges to a Gaussian.}
    \label{fig:tf_ts}
\end{figure}

This second model again has a linear MSD with slope \(2\langle D\rangle\), so the diffusion coefficient alone still fails to distinguish it from ordinary diffusion. However, the higher-order structure evolves differently. As shown in Figure~\ref{fig:tf_ts}, the static case (\(t_{\rm corr}\to\infty\)) remains permanently non-Gaussian, while for finite \(t_{\rm corr}\) the excess kurtosis \(\alpha_2(t)\) decays toward zero on a timescale set by \(t_{\rm corr}\) as particles progressively sample many environments. Thus, systems with the same effective diffusion coefficient can have very different propagators \red{and finite scale structure}. This is precisely the information retained in \(P(k,s)\) and discarded by \(\kappa_{\rm eff}\).


\begin{figure}
    \centering
    \includegraphics[width=0.99\linewidth]{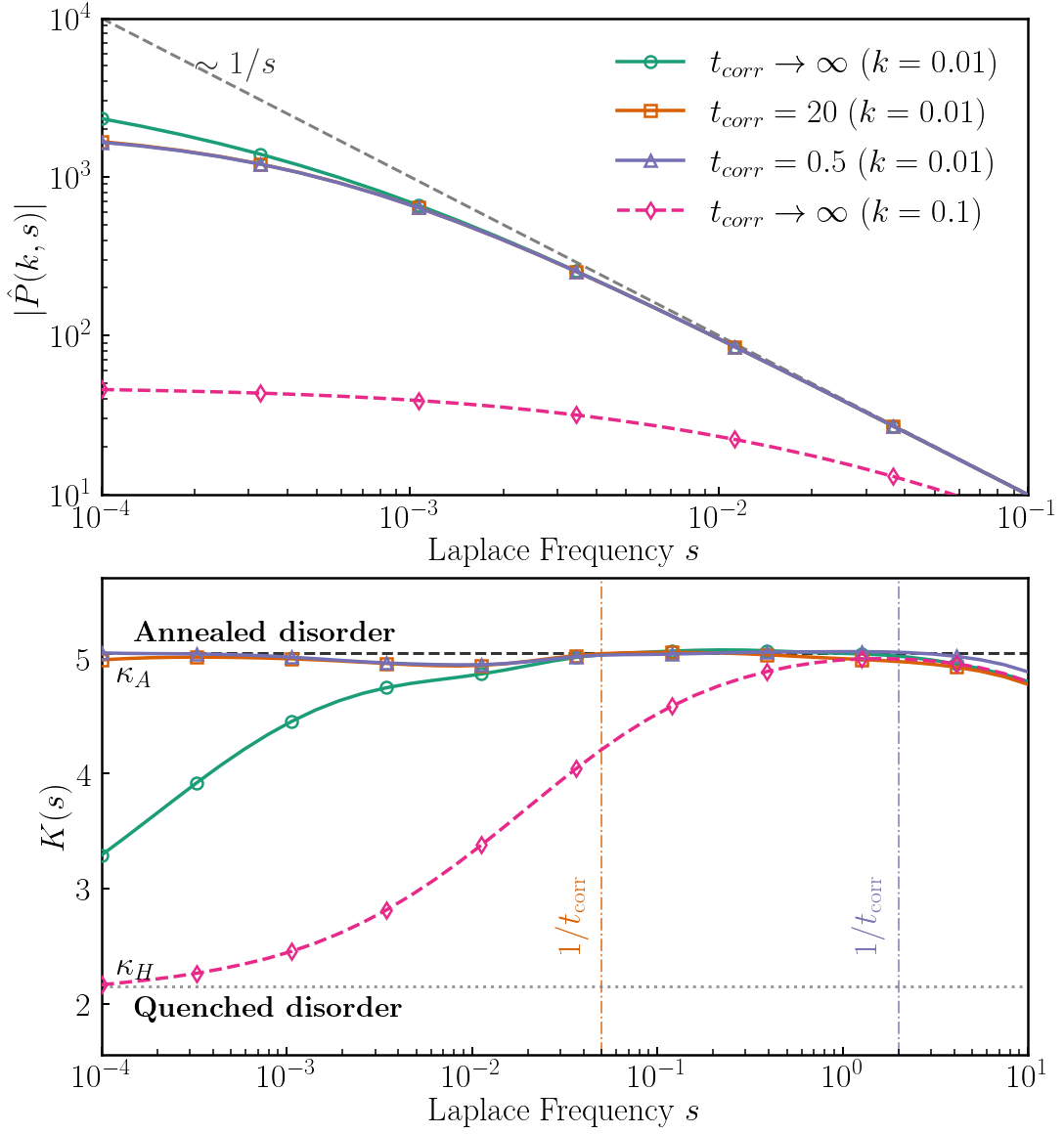}
    \caption{\red{Finite scale structure} in the finite correlation time simulation (\S\ref{subsec:2_examples}). \emph{Top:} The propagator in Fourier-Laplace space, $P(k,s)$, at wavenumber $k=0.1,0.01$. \red{\emph{Bottom:} Memory kernel $\tilde{K}(s)$, which crosses over from the arithmetic mean $\kappa_A=\langle D\rangle$ at high $s$ toward the harmonic mean $\kappa_H$ at low $s$, showing a continued $k$ dependence at small $k$, which is a finite-scale, non-Gaussian signature. The curves do not collapse: the crossover from $\kappa_A=\langle D\rangle$ at high $s$ to $\kappa_H$ at low $s$ shifts to smaller $s$ as $k$ decreases, and in the strict $k\to0$ limit the kernel is the constant $\kappa_A$ at all $s$, consistent with the linear MSD.}}
    \label{fig:tf_ts_memory}
\end{figure}

We now discuss how the Laplace-space kernel $\tilde K(s)$ can be measured directly from particle trajectories. In principle, the most direct route is to first measure the full propagator $P(k,s)$, and then extract $\tilde K(s)$ from its large-scale limit. In practice, this requires working at sufficiently small wavenumber $k$, where the transport has become scale-independent and the asymptotic form
\begin{equation}
    P(k,s) \approx \frac{1}{s + k^2 \tilde K(s)}
    \label{eq:Pks_large_scale}
\end{equation}
is valid. If there is any mean drift or streaming, one must first shift to the co-moving frame, $x'(t)=x(t)-\langle x(t)\rangle$, so that the extracted kernel isolates the diffusive $O(k^2)$ contribution rather than the advective $O(k)$ term.

\red{Given trajectories, one may therefore estimate the finite-scale quantity
\begin{equation}
   \tilde K_{\rm eff}(k,s)
   =
   \frac{1}{k^2}
   \left[
   \frac{1}{P(k,s)} - s
   \right]
   \label{eq:Ks_from_P}
\end{equation}
for several of the smallest resolved wavenumbers, and verify whether the
results collapse onto a common curve. This collapse is the practical
signal that one has reached the low-$k$ regime where the large-scale
kernel
\begin{equation}
\tilde K(s)
=
\lim_{k\rightarrow0}\tilde K_{\rm eff}(k,s)
\label{eq:ks-limit}
\end{equation}
is well defined.}

\red{Writing the quenched propagator for this model as
$P(k,s)=\langle 1/(s+k^2D)\rangle_D$, equation~\ref{eq:Ks_from_P}
gives
\begin{equation}
\tilde K_{\rm eff}(k,s)
=
\frac{1}{k^2}
\left[
\frac{1}{\langle 1/(s+k^2D)\rangle_D}-s
\right]
=
\langle D\rangle
-k^2\frac{{\rm Var}(D)}{s}
+O(k^4).
\label{eq:Ktilde_smallk}
\end{equation}
However, this is a small-$k$ expansion at fixed $s$. As $s$ decreases,
the correction term $k^2{\rm Var}(D)/s$ grows, and the truncated expansion
eventually breaks down. The limit $s\rightarrow0$ at fixed non-zero $k$
must therefore be obtained from the full propagator. In that limit,
$P(k,s)\rightarrow\langle D^{-1}\rangle/k^2$, so that
\begin{equation}
\tilde K_{\rm eff}(k,s)
\rightarrow
\langle D^{-1}\rangle^{-1}
\equiv \kappa_H, 
\end{equation}
the harmonic mean. Therefore the two limits do not commute. Taking $k\rightarrow0$ at fixed
$s$ gives $\langle D\rangle$, while taking $s\rightarrow0$ at fixed
non-zero $k$ gives $\kappa_H$. The crossover between them occurs at
$s\sim k^2D$ and moves to smaller $s$ as $k$ decreases. This is why the
quenched curves in Figure~\ref{fig:tf_ts_memory} do not collapse onto a
common frequency-dependent kernel as $k$ is reduced. Their frequency
dependence is therefore finite-scale structure rather than large-scale
memory. Note that collapse of $P(k,s)$ itself is not a useful diagnostic,
because $P(k,s)\rightarrow1/s$ for any model as $k\rightarrow0$. The
transport information lies in the $O(k^2)$ correction, which
equation~\ref{eq:Ks_from_P} isolates by removing the $1/s$ term and
dividing by $k^2$.}

\red{We discuss the reason for these asymptotic limits in
\S\ref{subsec:interpret}. By contrast, when the mean-squared displacement
is genuinely nonlinear, such as for the subdiffusive CTRWs
($\tilde K(s)\propto s^{1-\alpha}$, Fig.~\ref{fig:Ks}) and the
multiphase medium of \S\ref{sec:multiphase}, the frequency dependence
survives the $k\rightarrow0$ limit and represents true large-scale
memory.}

\section{Memory in CR Transport}
\label{sec:memory}
At large scales, we are often not interested in the full propagator $P(k,s)$, but in how the CR flux responds to a large-scale density gradient. In standard diffusion, that response is instantaneous:
\begin{equation}
    F(x,t) = -\kappa \nabla n(x,t),
\end{equation}
so the present flux is determined entirely by the present gradient. Memory arises when this is no longer true. After coarse-graining over trapping, multi-phase structure, or other unresolved transport states, the flux can depend on the \emph{past history} of the gradient, not just its instantaneous value: 
\begin{equation}
    F(x,t) = -\int_0^t K(t-t')\,\nabla n(x,t')\,dt',
\end{equation}
where $K(t)$ is the time-domain memory kernel. Standard diffusion corresponds to the special case $K(t)=\kappa \delta(t)$, while a broader kernel corresponds to delayed, history-dependent transport.

This distinction matters because two transport processes can have the same effective diffusion coefficient, yet produce very different large-scale behavior if one has memory and the other does not. In particular, trapping and multi-phase structure can generate long-lived correlations that are missed by any treatment based only on $\kappa$. In Laplace space, this same physics appears as a frequency-dependent diffusivity $\tilde K(s)$, the Laplace transform of $K(t)$, which controls the small-$k$ limit of the propagator.

In this section, we first derive the corresponding Laplace-space kernel $\tilde K(s)$ from the large-scale limit of the Montroll--Weiss propagator (\S\ref{subsec:memory-kernal}), and show how it can be measured in simple Monte Carlo examples (\S\ref{subsec:measure_K}). Next, we discuss the physical origin of memory (\S\ref{subsec:understand}). Finally, we show a compact representation of the time-domain kernel $K(t)$ which is well suited for macroscopic transport models (\S\ref{subsec:approx_K}).

\subsection{Deriving the Memory Kernel} 
\label{subsec:memory-kernal}

We now derive the large-scale transport law implied by the scalar, separable Montroll--Weiss propagator. For simplicity, we consider an unbiased walk, with any mean drift removed, and assume that the jump-length distribution has a finite second moment, $\langle \ell^2 \rangle < \infty$. This is the regime relevant for ordinary diffusion and subdiffusion. Superdiffusive L\'evy-flight cases, for which $\langle \ell^2 \rangle$ diverges, require a different large-scale expansion.

In the long-wavelength limit $k \to 0$, the Fourier transform of the jump-length distribution can be expanded as
\begin{equation}
\hat{\lambda}(k)
= \int d\ell\, e^{ik\ell}\lambda(\ell)
= 1 - \frac{k^2}{2}\langle \ell^2\rangle + O(k^4),
\label{eq:lambda-smallk}
\end{equation}
where the linear term vanishes because the walk is unbiased. Substituting this into the Montroll--Weiss formula (Eq.~\ref{eqn:mw}) gives
\begin{equation}
P(k,s)
\approx
\frac{1-\tilde{\psi}(s)}
{s\left[1-\tilde{\psi}(s)+\frac{k^2}{2}\langle \ell^2\rangle \tilde{\psi}(s)\right]}
=
\frac{1}{s+k^2 \tilde K(s)},
\label{eq:MW-approx}
\end{equation}
where
\begin{equation}
\tilde K(s)
=
\frac{s\,\tilde\psi(s)}{1-\tilde\psi(s)}\,\frac{\langle \ell^2\rangle}{2}.
\label{eq:Ks}
\end{equation}
plays the role of a frequency-dependent diffusion coefficient. It is the Laplace-space object that encodes temporal memory in the large-scale limit. Note from equation \ref{eq:MW-approx} and \ref{eq:const_K} that $[\tilde{K}(s)]=L^2 T^{-1}$ has units of diffusivity, and indeed as $k, s \rightarrow 0$ transitions to the standard constant diffusion coefficient. By contrast, the memory kernel has units $[K(t)] = L^2 T^{-2}$.

If the mean waiting time is finite, then as $s\to0$,
\begin{equation}
\tilde\psi(s)=1-s\langle \tau\rangle + O(s^2),
\label{eq:psi-small-s}
\end{equation}
so that
\begin{equation}
\lim_{s\to0}\tilde K(s)=\frac{\langle \ell^2\rangle}{2\langle \tau\rangle}.
\label{eq:const_K}
\end{equation}
Thus, when both $\langle \ell^2\rangle$ and $\langle \tau\rangle$ are finite, the long-time limit gives the standard diffusion coefficient, as expected from Gaussianization and the central limit theorem \citep{liang25}.

The significance of a non-constant $\tilde K(s)$ is that the coarse-grained transport law becomes non-local in time. For an initial density field $n(x,0)$, linear evolution implies
\begin{equation}
n(k,s)=P(k,s)\,n(k,0),
\end{equation}
and therefore
\begin{equation}
\left[s+k^2\tilde K(s)\right]n(k,s)=n(k,0).
\label{eq:MW-evolution}
\end{equation}
Using
\begin{equation}
\mathcal{L}\!\left[\partial_t n_k(t)\right]
=
s\,n(k,s)-n(k,0),
\end{equation}
we obtain
\begin{equation}
\mathcal{L}\!\left[\partial_t n_k(t)\right]
=
-k^2\tilde K(s)\,n(k,s).
\end{equation}
Taking the inverse Laplace transform yields
\begin{equation}
\partial_t n_k(t)
=
-k^2\int_0^t K(t-t')\,n_k(t')\,dt',
\label{eq:nk-memory}
\end{equation}
where
\begin{equation}
K(t)=\mathcal{L}^{-1}\!\left\{\tilde K(s)\right\}
\end{equation}
is the time-domain memory kernel. Inverse Fourier transforming then gives the conservation law
\begin{equation}
\partial_t n(x,t)+\nabla\cdot \mathbf{F}(x,t)=0,
\label{eq:n-conservation}
\end{equation}
where the flux is given by: 
\begin{equation}
\mathbf{F}(x,t)
=
-\int_0^t K(t-t')\,\nabla n(x,t')\,dt'.
\label{eq:flux-memory}
\end{equation}
Thus, the flux is no longer determined only by the instantaneous gradient, as in Fick's law, but by the entire history of past gradients. Standard diffusion is recovered as the special case $\tilde K(s)=\kappa$, or equivalently $K(t)=\kappa \delta(t)$.

For a heavy-tailed waiting-time distribution
\begin{equation}
\psi(\tau)\propto \tau^{-1-\alpha}, \qquad 0<\alpha<1,
\end{equation}
the Laplace transform has the small-$s$ asymptotic form: 
\begin{equation}
1-\tilde\psi(s) \propto s^\alpha
\qquad (s\to0). 
\end{equation}
Equation \ref{eq:Ks} then gives  
\begin{equation}
\tilde K(s)\propto s^{1-\alpha},
\label{eq:subdiff_Ks}
\end{equation}
with inverse Laplace transform
\red{
\begin{equation}
K(t)=\mathcal{L}^{-1}\{\tilde{K}(s)\}\ \propto\ -\,t^{-(2-\alpha)}\qquad(t>0).
\label{eq:subdiff_Kt}
\end{equation}
}
\red{The negative sign has a simple interpretation (see \S\ref{subsec:approx_K}). For a subdiffusive
process, the effective rate of transport decreases as the ensemble
ages, because an increasing fraction of particles is caught in long
waiting intervals. The negative tail of $K(t)$ encodes this declining
response to a previously imposed gradient. Its slow power-law decay
means that the flux retains memory of gradients over a broad range
of past times.} By contrast, for an exponential waiting-time distribution,
\begin{equation}
\psi(\tau)=\frac{1}{\langle \tau\rangle}e^{-\tau/\langle \tau\rangle},
\end{equation}
one finds $\tilde K(s)=\langle \ell^2\rangle/(2\langle \tau\rangle)=\kappa$ and hence $K(t)=\kappa\delta(t)$: the transport is memoryless and Markovian.

\subsection{Measuring the Memory Kernel}
\label{subsec:measure_K}
\red{The test of large-scale memory is whether the frequency dependence of $\tilde K(s)$ survives as $k\to0$. In this section we follow the relaxation of a single Fourier mode.} The advantage is that one prepares the system so that only one long-wavelength mode is present initially, avoiding contamination by mode mixing and reducing noise. We therefore consider a periodic one-dimensional box of size $L$ initialized with a small sinusoidal perturbation,
\begin{equation}
    n(x,0)=n_0\left[1+\epsilon\cos(kx)\right],
    \qquad
    k=\frac{2\pi m}{L},
    \label{eq:single_mode_init}
\end{equation}
where $m$ is an integer and $\epsilon \ll 1$ ensures linear evolution. In Monte Carlo simulations, this corresponds to sampling initial particle positions from a probability density proportional to $1+\epsilon \cos(kx)$.

The amplitude of the mode is then
\begin{equation}
    n_k(t)=\frac{1}{N}\sum_{j=1}^N e^{ikx_j(t)},
    \label{eq:nk_measure}
\end{equation}
which is simply the empirical characteristic function evaluated at a single wavenumber. \red{This should be distinguished from the displacement propagator (equation \ref{eq:phi}), which operates on particle displacements $X_j(t) - X_j(0)$.} Taking its Laplace transform gives $\hat n_k(s)$, and from the large-scale evolution equation one obtains
\begin{equation}
    \tilde K(s)
    =
    \frac{1}{k^2}
    \left[
    \frac{n_k(0)}{\hat n_k(s)} - s
    \right].
    \label{eq:Ks_from_nk}
\end{equation}
This is equivalent to Eq.~(\ref{eq:Ks_from_P}), but is often more robust in practice because the initial condition isolates a single mode from the outset.

\begin{figure}
    \centering
    \includegraphics[width=0.99\linewidth]{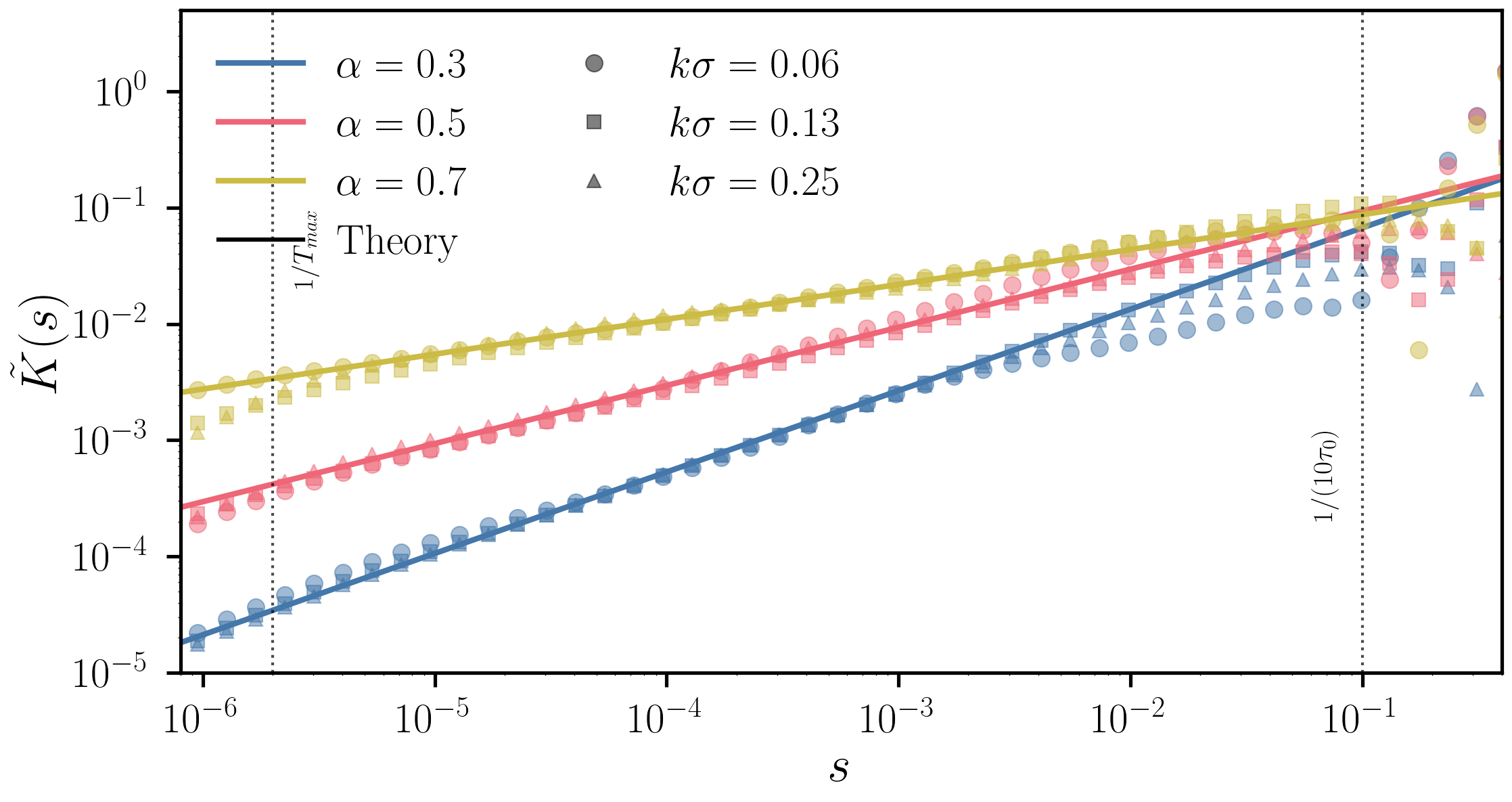}
    \caption{\red{The memory kernel $\tilde{K}(s)$ in the Laplace domain for subdiffusion exponents $\alpha = 0.3$, $0.5$, and $0.7$. Circles, squares, and triangles show $k = 2\pi/L$, $4\pi/L$, and $8\pi/L$, which corresponds to $k\sigma=0.06$, $0.13$, and $0.25$, where $\sigma$ is the jump width. The three coincide over the reliable frequency range, showing that the extracted kernel is $k$-independent. Solid lines are the power laws $\tilde K(s)\propto s^{1-\alpha}$ of Eq.~\ref{eq:subdiff_Ks}. The upper limit in $s$ is roughly $1/(10\tau_0)$, above which particles have made too few jumps; the lower limit is roughly $1/T_{\rm max}$, set by the run length.}}
    \label{fig:Ks}
\end{figure}

We illustrate this method using subdiffusive CTRWs with waiting time exponent $\alpha=0.3,\,0.5,$ and $0.7$. Figure~\ref{fig:Ks} shows the measured $\tilde K(s)$ \red{at the three smallest wavenumbers $k=2\pi m/L$, $m=1,2,4$, or $k\sigma=0.06$, $0.13$, and $0.25$, where $\sigma$ is the jump width,} together with the theoretical scaling
\begin{equation}
    \tilde K(s) \propto s^{1-\alpha}.
\end{equation}
The agreement confirms that the estimator recovers the expected Laplace-space kernel over the accessible frequency range. \red{The curves at the three wavenumbers lie on top of one another. The frequency dependence of $\tilde K(s)$ survives the large-scale limit, which is the signature of large-scale memory, whereas the quenched model of Figure~\ref{fig:tf_ts_memory} fails this test.}

The two estimators above measure the kernel itself. It is also useful to
visualize directly what memory means in real time. To do so, we consider a one-dimensional system with a source at one boundary and an absorbing
boundary at the other, and then reverse the boundaries at time
$T_{\rm flip}=T/2$. We compare (i) standard diffusion with
$\kappa=1$, (ii) subdiffusion with $\alpha=0.7$, and
(iii) two-phase transport with alternating fast and slow regions
($\kappa_h/\kappa_l=100$, cell sizes drawn from
$p(\ell)\propto \ell^{-\alpha_\ell}$ with $\alpha_\ell=2$,
$\ell\in[0.3,\,3.0]$). The last is the same setup we will use in
\S\ref{sec:multiphase}. 
\red{In each case, we measure the particle flux $J(t)$ from the net number of particles crossing the midpoint of the domain per unit time. For the standard and subdiffusive cases, the density profile is smoothed with a Gaussian filter with $\sigma=2$ grid bins, and the gradient is evaluated at the midpoint. For the multiphase case, we instead measure the macroscopic gradient from a linear fit over the central half of the domain. The resulting flux and gradient time series are smoothed with a Gaussian filter before plotting.}

\begin{figure}
    \centering
    \includegraphics[width=0.99\linewidth]{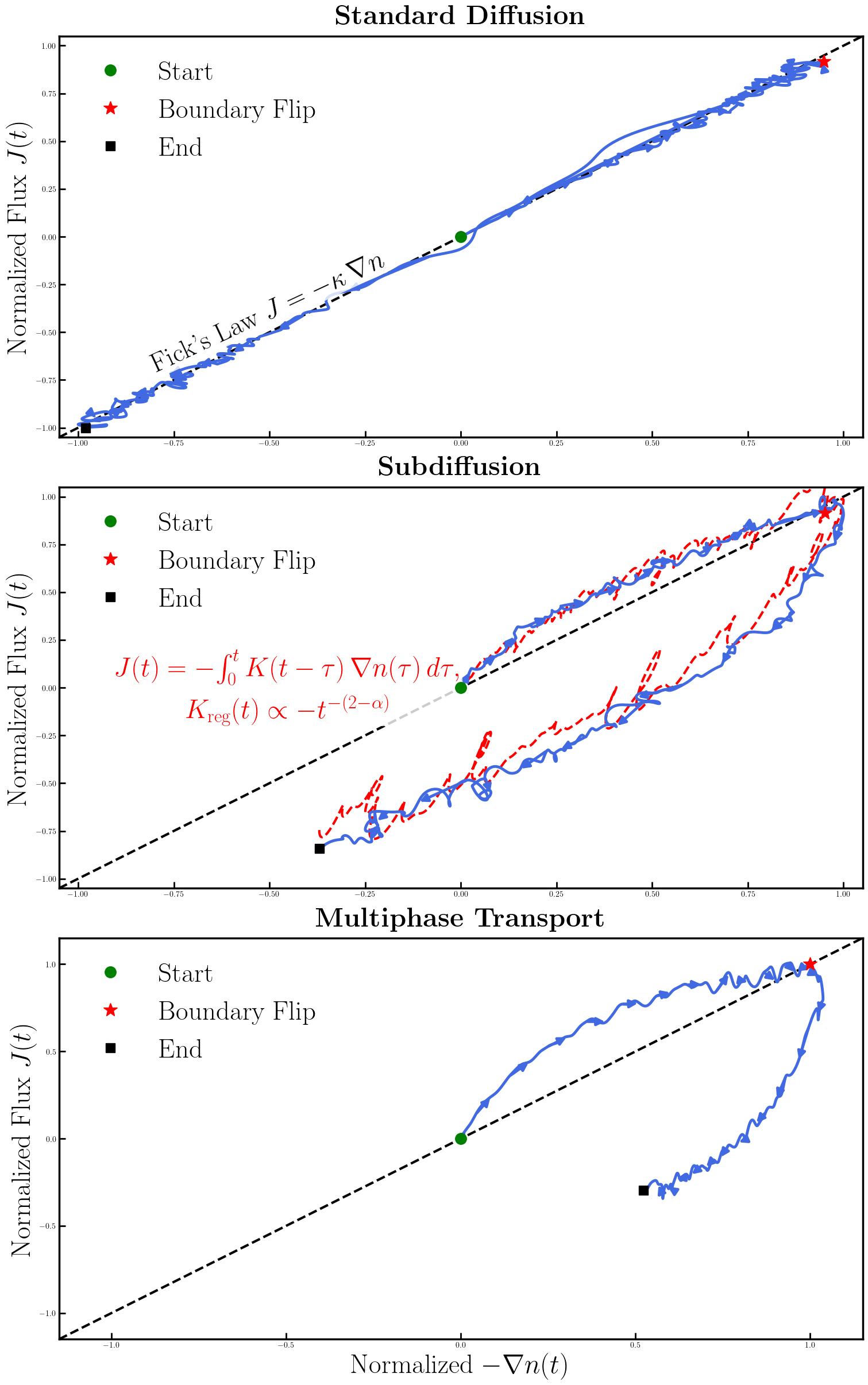}
    \caption{Flux $J(t)$ against the negative density gradient $-\nabla n(t)$. \emph{Top:} standard diffusion ($\alpha=2$); \emph{Middle:} subdiffusion ($\alpha=0.7$); \emph{Bottom:} multiphase transport. 
    \red{In the standard diffusion and subdiffusion panels, the flux and gradient are normalized separately by their maximum absolute values over the run. In the multiphase panel, they are normalized by their absolute values at the boundary flip.} 
    Time evolution is shown by arrows, with a boundary flip happening at half of the simulation time. Standard diffusion adheres to Fick's Law (dashed line), tracing a linear path. In contrast, subdiffusion shows a loop due to memory effects, deviating from the instantaneous Fickian relation but agreeing with the fractional Fick's law prediction (dashed curve). Multiphase transport also follows a similar loop.}
    \label{fig:bdry_flip}
\end{figure}

As shown in Figure~\ref{fig:bdry_flip}, standard diffusion adjusts rapidly after the boundary flip, and the relation between $J$ and $-\nabla n$ remains approximately linear, consistent with Fick's law. By contrast, both subdiffusion and multiphase transport exhibit a delayed response: after the gradient reverses, the flux continues to reflect the earlier state of the system. This produces a hysteresis loop in the $(F,-\nabla n)$ plane, demonstrating directly that the flux depends on the past history of the gradient rather than its instantaneous value alone. In this sense, Figure~\ref{fig:bdry_flip} provides the most direct real-time picture of memory, while Figures~\ref{fig:Ks} quantify it through the Laplace-space kernel $\tilde K(s)$.

\subsection{Understanding the Memory Kernel}
\label{subsec:understand}

Now that we have seen clearly that memory exists, we should step back and reflect on its physical origin. Where does memory come from? Memory in coarse-grained transport is not an ad hoc modification of diffusion, but a structural consequence of projection. At the microscopic level, particle motion in prescribed electromagnetic fields is Markovian: given the full state of the particle and field, its future evolution is independent of the past. However, macroscopic transport models do not track the full microscopic state. They retain only a reduced description, such as the particle density $n(x,t)$ or flux $F(x,t)$, after averaging over unresolved velocities, phases, trapping regions, or magnetic substructure. Once the hidden degrees of freedom are eliminated, the reduced dynamics is generically non-local in time: the future depends on the past through a memory kernel. This is codified in the Mori--Zwanzig formalism \citep{Mori1965}.

A simple example is ordinary Brownian motion. Consider
\begin{equation}
    \dot x = v, \qquad
    \dot v = -\frac{v}{\tau_c} + \eta(t),
\end{equation}
where $\eta(t)$ is white noise and $\tau_c$ is the velocity correlation time. The joint process $(x,v)$ is Markovian, but the position $x(t)$ alone is not:
\begin{equation}
x(t)=x_0+\tau_c v_0\!\left(1-e^{-t/\tau_c}\right)
+\tau_c \int_0^t \left[1-e^{-(t-t^{\prime})/\tau_c}\right]\eta(t^{\prime})\,dt^{\prime}, 
\end{equation}
\red{so that $x(t)$ depends on the history of the noise through the response function $1-e^{-(t-t')/\tau_c}$. This response function illustrates how eliminating an unresolved variable, here the velocity, produces history dependence in the reduced dynamics.}
In the long-time limit $t \gg \tau_c$, this memory becomes unimportant and one recovers ordinary diffusion with an effective diffusivity $D \sim \langle v^2\rangle \tau_c$. At intermediate times, however, the transport is history-dependent even though the propagator remains Gaussian. Thus Gaussianity does not imply the absence of memory; memory simply reflects unresolved dynamics with finite correlation time.

The same logic applies, more generally, to CR transport in a heterogeneous medium. A CR may spend long periods trapped within a magnetic structure, diffuse rapidly through one phase and slowly through another, or encounter regions with different scattering properties. If these internal states are not tracked explicitly, then the coarse-grained dynamics of the large-scale density cannot be Markovian in general. What appears macroscopically as a delayed flux response is simply the imprint of unresolved transitions among many microscopic transport states. A useful recent example is CR pitch angle transport in compressible MHD turbulence, where magnetic mirroring can generate transient trapping and non-Markovian behavior at large pitch angle \citep{yan+26}. The key lesson is not limited to that specific mechanism: once the unresolved spatial configuration of magnetic wells is projected out, the reduced dynamics in a variable such as pitch angle need not remain Markovian. Thus, memory reflects hidden transport states.

The Montroll--Weiss framework provides a transparent way to represent this reduced dynamics when the coarse-grained evolution can be organized around discrete renewal events. In this picture, the complicated motion within a trapping region or phase is compressed into a waiting time, while the transition to the next region defines a jump. Coarse-graining therefore does not eliminate memory; it reorganizes it. The memory appears either as a broad waiting-time distribution in the event-based description, or equivalently as a non-trivial kernel $K(t)$ in the expression for the flux (equation \ref{eq:flux-memory}). Standard diffusion is recovered only in the special limit where the waiting times are effectively exponential, so that the kernel collapses to $K(t)=\kappa\delta(t)$ and the flux responds instantaneously to the gradient.

This renewal picture also makes clear in what sense a non-exponential waiting-time distribution generates memory. After a jump, the process restarts, but between jumps the residual waiting time generally depends on the elapsed age. If a particle has already waited for a time $t$ since its last jump, the probability that it must wait at least an additional time $\Delta t$ is
\begin{equation}
    P(\tau>t+\Delta t \mid \tau>t)=\frac{S(t+\Delta t)}{S(t)},
\end{equation}
where $S(t)=P(\tau>t)$ is the survival function. Only for an exponential waiting law, $S(t)=e^{-\gamma t}$, is this conditional probability independent of $t$. For any non-exponential waiting law, however, the residual waiting time depends on how long the particle has already been trapped. In particular, waiting time distributions with power law tails have $S(t) \sim t^{-\alpha}$, so that: 
\begin{equation}
\frac{S(t+\Delta t)}{S(t)}
=
\left(\frac{t+\Delta t}{t}\right)^{-\alpha}
\end{equation}
depends on $t$. If $t$ is already large, the ratio changes slowly: the longer you have already waited, the longer you expect to continue waiting. Similarly, systems with a hierarchy of relaxation times which can be represented by a Prony series (see \S\ref{subsec:approx_K} below) have $S(t)=\sum a_n e^{-\lambda_n t}$, so that: 
\begin{equation}
\frac{S(t+\Delta t)}{S(t)}
=
\frac{\sum a_n e^{-\lambda_n(t+\Delta t)}}
{\sum a_n e^{-\lambda_n t}},
\end{equation}
which also depends on $t$, and changes more slowly with $\Delta t$ if $t$ is large. The longer you have been waiting, the more likely it is that you are trapped in a mode with a slow relaxation rate $\lambda_n$, and the longer you are likely to continue to wait. Broad waiting-time distributions therefore generate memory in a precise sense: particles that have already waited a long time are statistically biased toward continuing to wait longer.

\subsection{Representing the Memory Kernel} 
\label{subsec:approx_K}

Measuring the Laplace-space kernel $\tilde K(s)$ from trajectories provides a complete large-scale description of the transport, but it is not yet a convenient form for macroscopic simulations. In principle one could recover $K(t)$ from $\tilde K(s)$ by a numerical inverse Laplace transform, but in practice this is often ill-conditioned in the presence of finite particle noise. Furthermore, the flux gradient relation,
\begin{equation}
    J(x,t) = -\int_0^t K(t-t')\,\nabla n(x,t')\,dt',
\end{equation}
is non-local in time, so a direct implementation requires continuous convolution over the entire history of the system. This is both computationally expensive and numerically awkward when the kernel is broad.  We therefore seek a compact representation of the kernel that is both numerically stable and easy to evolve in time.

\red{Before doing so,
it is useful to first examine the quantity $\tilde{G}(s) \equiv \tilde K(s)/s$.
For a single-state, separable renewal process, equation \ref{eq:Ks} gives: 
\begin{equation}
\tilde{G}(s) \equiv \frac{\tilde K(s)}{s}
=
\frac{\langle \ell^2\rangle}{2}
\frac{\tilde\psi(s)}
     {1-\tilde\psi(s)} = \frac{\langle \ell^2\rangle}{2}
\sum_{n=1}^{\infty}\tilde\psi(s)^n.
\label{eq:G_renewal_laplace}
\end{equation}
Multiplication in Laplace space corresponds to convolution in
time, so $\tilde\psi(s)^n$ is the Laplace transform of
$\psi^{*n}(t)$, the $n$-fold convolution of the waiting-time
distribution with itself. Since $\psi^{*n}(t)\,dt$ is the probability that the $n$th jump
occurs during the interval $[t,t+dt]$, summing over $n$ gives the
expected number of jumps occurring during that interval,
\begin{equation}
\left\langle N(t+dt)-N(t)\right\rangle
=
\left[\sum_{n=1}^{\infty}\psi^{*n}(t)\right]dt.
\label{eq:jumps}
\end{equation}
The inverse Laplace transform of $\tilde{G}(s)$ (equation \ref{eq:G_renewal_laplace}) and equation \ref{eq:jumps} gives:
\begin{equation}
G(t)=\frac{\langle \ell^2\rangle}{2}\, \sum_{n=1}^{\infty}\psi^{*n}(t)=  \frac{\langle \ell^2\rangle}{2}\, \frac{d}{dt}\langle N(t)\rangle
\label{eq:G_renewal}
\end{equation}
Thus, for this single state renewal process, $G(t)$ is proportional to the jump rate. For exponential waiting
times this rate is constant. For a subdiffusive waiting-time
distribution it decreases as the ensemble ages, because an
increasing fraction of particles is caught in long waits. Hence
$G(t)$ is positive but decreasing.} 

\red{We now seek a compact parametric representation of $G(t)$. Why $G(t)$, and not $K(t)$? Recall that division by $s$ in Laplace
space corresponds to integration in time, while multiplication by $s$ corresponds to differentiating in time. \(G(t)\) is therefore the time-integrated memory response, while \(K(t)=dG/dt\) is its rate of change. $G(t)$ is positive and,
for subdiffusive transport, decreases as the jump rate
declines.  It therefore has the form of a relaxation function and can
naturally be approximated by a positive superposition of decaying
modes.  By contrast, as seen in equation~(\ref{eq:subdiff_Kt}), $K(t)=dG/dt$ is negative at late times (since $G(t)$ declines with time), so it cannot be represented by the non-negative sum of decaying exponentials we introduce below.}

\red{We represent $G(t)$ as a superposition of
relaxation modes, each of which decays exponentially.
More generally,
if the transport approaches ordinary diffusion at late times,
$G(t)\rightarrow\kappa_\infty$. We therefore write
\begin{equation}
G(t) \simeq H(t)\left[
\kappa_\infty+
\int d\ln\tau\, A(\tau)e^{-t/\tau}
\right],
\end{equation}
where the Heaviside function $H(t)$ enforces causality and $A(\tau)$ specifies the weight
carried by each relaxation timescale. In practice, we approximate the continuous spectrum by a finite number
of modes,
\begin{equation}
G_N(t)=
H(t)\left[
\kappa_\infty+
\sum_{i=1}^{N}a_i e^{-t/\tau_i}
\right],
\qquad
a_i\geq0,\quad \tau_i>0.
\end{equation}
This finite sum of exponentials is a Prony representation. Its Laplace
transform is analytic,
\begin{equation}
\tilde G_N(s)
=
\frac{\kappa_\infty}{s}
+
\sum_{i=1}^{N}
\frac{a_i}{s+1/\tau_i}.
\label{eq:G_prony_s}
\end{equation}}

\red{Since $\tilde G(s)\equiv\tilde K(s)/s$, the corresponding
Laplace-space memory kernel is
\begin{equation}
\tilde K_N(s)
=
s\tilde G_N(s)
=
\kappa_\infty+
s\sum_{i=1}^{N}
\frac{a_i}{s+1/\tau_i}.
\end{equation}
Equivalently,
\begin{equation}
\tilde K_N(s)
=
\kappa_\infty+\sum_{i=1}^{N}a_i
-
\sum_{i=1}^{N}
\frac{a_i/\tau_i}{s+1/\tau_i}.
\end{equation}
Taking the inverse Laplace transform gives
\begin{equation}
K_N(t)
=
\left(
\kappa_\infty+\sum_{i=1}^{N}a_i
\right)\delta(t)
-
H(t)\sum_{i=1}^{N}
\frac{a_i}{\tau_i}e^{-t/\tau_i}.
\label{eq:K_N_t}
\end{equation}
The instantaneous $\delta(t)$ term is the consequence of the finite
jump of the causal Prony approximation at the origin,
$G_N(0^+)=\kappa_\infty+\sum_i a_i$. A microscopic $G(t)$ need not
have this finite jump; if its short-time behavior is important, the
basis can be modified to reproduce the appropriate $t\rightarrow0$
asymptotic form.}



\begin{figure}
    \centering
    \includegraphics[width=0.99\linewidth]{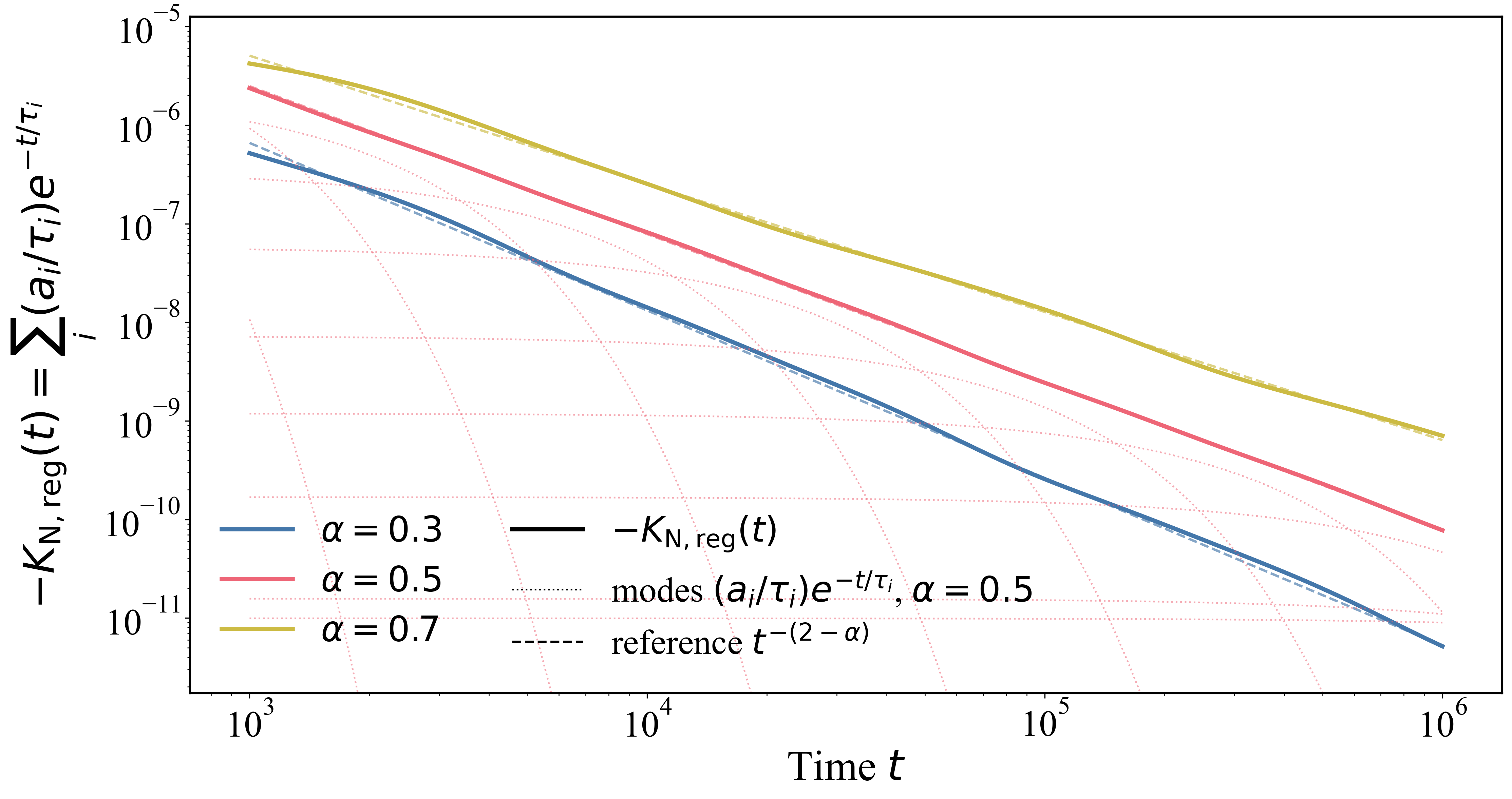}
    \caption{
    \red{Prony representation of the subdiffusive memory kernel, obtained from the measured Laplace-space kernel. The positive transform $\tilde{K}(s)/s$ is fit with a non-negative sum. Thick curves show the regular part; thin dotted curves its individual modes for $\alpha=0.5$; dashed lines the reference $t^{-(2-\alpha)}$, for $\alpha=0.3,0.5,0.7$. The regular part of $K(t)$ is negative, so $-K_{N,\mathrm{reg}}$ is plotted. The reconstructed kernels accurately recover the expected power-law scaling $K(t)\propto t^{-(2-\alpha)}$ for $\alpha=0.3$, $0.5$, and $0.7$ over several decades in time with only $N=10$ modes, demonstrating that broad temporal memory can be represented accurately and compactly over a given interval.}
    }
    \label{fig:Kt_subdiff}
\end{figure}

We illustrate the method using the subdiffusive test problem from \S\ref{subsec:measure_K}. \red{In this strictly subdiffusive toy example, $\kappa_\infty=0$.} For a waiting-time distribution with index $\alpha$, the exact time-domain kernel scales as
\red{
\begin{equation}
K_{\mathrm{reg}}(t)\propto -\,t^{-(2-\alpha)}.
\end{equation}
}
Figure~\ref{fig:Kt_subdiff} compares this target behavior with the kernel reconstructed from a Prony fit to the measured Laplace-space kernel $\tilde K(s)/s)$. 
\red{For $\alpha=0.3,0.5,0.7$ the reconstructed regular tail $-K_{N,\mathrm{reg}}(t)=\sum_i (a_i/\tau_i)e^{-t/\tau_i}$ is consistent with the expected $t^{-(2-\alpha)}$ scaling over the recoverable interval with only $N=10$ modes.}
\red{Over the reliable interval $s\in[10^{-6},10^{-3}]$ the non-negative Prony fit ($N=10$ modes) reproduces
the measured $\tilde{K}(s)/s$ with maximum fractional residual $\lesssim2\times10^{-4}$ 
for all three $\alpha$, as $\tilde{K}(s)$ is a Laplace-integrated and smooth quantity. The limiting uncertainty is the finite reliable window. The fitted coefficients $\{a_i,\tau_i\}$ and the instantaneous coefficients $\sum_i a_i=0.017,\,0.036,\,0.076$ (for $\alpha=0.3,0.5,0.7$) are listed in Table~\ref{tab:prony_coefficients}.}
Thus, a finite sum of local, Markovian relaxation modes can accurately capture long-range temporal memory. This approach is especially well suited to the tempered power-law kernels (i.e., power laws with exponential cut-offs) expected in many heterogeneous transport problems, for which Prony expansions provide accurate approximations with relatively few modes \citep{beylkin10}. More broadly, the fitted set $\{a_i,\tau_i\}$ provides a compact summary of how the memory structure changes with environment, geometry, and transport regime.

Another practical advantage of Eq.~(\ref{eq:K_N_t}) is that the temporal convolution can be replaced by a small set of auxiliary variables. Defining
\begin{equation}
    q_i(t)=\int_0^t e^{-(t-t')/\tau_i}\,\nabla n(t')\,dt',
    \label{eq:q_i}
\end{equation}
the flux becomes
\red{
\begin{equation}
J=-\Bigg(\kappa_{\infty} + \sum_{i=1}^{N}a_i\Bigg)\nabla n+\sum_{i=1}^{N}\frac{a_i}{\tau_i}\,q_i.
\end{equation}
}
while each mode obeys the local evolution equation
\begin{equation}
    \dot q_i=-\frac{q_i}{\tau_i}+\nabla n.
    \label{eq:q_i_ode}
\end{equation}
In other words, the full memory integral is replaced by a small set of local ODEs, with no need to store the entire history of $\nabla n(t)$. We will exploit this representation in future work to build practical macroscopic transport models with memory.

In the next section, we apply this expansion to multiphase transport, constructing macroscopic models from waiting times across cells of differing diffusivity. For numerical convenience, we represent the exact first-passage kernels in the time domain with a Prony-series expansion. We apply it for the waiting time distributions $U(s)$ for exit to the left and right, which are known analytically in Laplace space but do not have simple closed-form inverses in the time domain. Tests against the analytic Laplace-space expressions for $U_l(s)$ and $U_h(s)$ show that a modest number of exponentials accurately reproduces the low-$s$ behavior that controls the macroscopic transport, while the residual mismatch is confined to the high-$s$ limit.

\section{Multiphase CR Transport}
\label{sec:multiphase}

We now apply the propagator-based framework developed above to a controlled model of CR transport through a multiphase medium. This section uses that model as a test bed for the main ideas of the paper: coarse-graining into renewal events, reconstruction of the propagator $P(k,s)$ from particle trajectories, the emergence of memory in the large-scale transport law, 
\red{and the matrix Montroll--Weiss description that is required once the step statistics depend on the interface at which a step starts.} 
We begin by defining the multiphase setup (\S\ref{subsec:setup}) and its coarse-grained transport unit (\S\ref{subsec:two_cells}), then analyze constant- (\S\ref{subsec:const}) and variable-cell (\S\ref{subsec:variable}) media before turning to the physical interpretation of the results (\S\ref{subsec:interpret}), and understanding the impact of a time-dependent transport environment (\S\ref{subsec:renewal}).  

\subsection{Multiphase Medium: Setup}
\label{subsec:setup}

We consider CR transport through an idealized one-dimensional multiphase medium composed of alternating regions with different diffusivities \citep{ewart25}, meant to approximate transport along a magnetic field line. In this limit, slow regions act as bottlenecks in strict series; in higher dimensions (i.e., with cross-field diffusion) particles can in general bypass them, so the harmonic-mean limit becomes a transport-network problem rather than a simple series sum. In the simplest case, the medium consists of a slow phase with diffusivity $\kappa_s$ and a fast phase with diffusivity $\kappa_f$, arranged in an alternating sequence of patches. We begin with the simplest constant-cell model, in which all patches have the same size $L$, and later generalize to variable cell sizes $l$ drawn from a distribution $P(l)$. Although idealized, this setup captures the essential bottleneck physics of transport through a heterogeneous ISM/CGM while remaining simple enough to analyze directly and compare against Monte Carlo simulations. 

The key step is to coarse-grain the transport at cell interfaces rather than follow every microscopic scattering event within a patch. Starting from an interface, a particle diffuses through the two neighboring cells until it first exits through one of the two outer interfaces. This defines a single coarse-grained event, characterized by a displacement $\Delta x=-L_l$ or $+L_h$ and a waiting time $\tau$. The waiting time includes all unresolved motion within the neighboring cells, including multiple recrossings of the internal interface before final escape. In this way, the microscopic dynamics inside a heterogeneous medium is reduced to a sequence of interface-to-interface transitions. 
\red{The step statistics depend on which side of the interface the slow cell lies. For equal-sized cells the step kernel pooled over the two states factorizes into a spatial and a temporal part. For variable-sized cells the geometry couples step lengths and waiting times, and this factorization is lost.}

This model therefore provides a controlled setting in which to derive the coarse-grained step statistics analytically from a two-sided first-passage problem and to test them directly against particle simulations. It also serves as the foundation for the accelerated Monte Carlo scheme developed below, in which particles evolve from interface to interface rather than by small timesteps.

\subsection{Coarse Graining: Two-Cell Transport as a Building Block}
\label{subsec:two_cells}

The elementary coarse-grained event in our multiphase medium is a first-passage problem across two neighboring cells. Without loss of generality, we consider a slow cell of size $L_l$ and diffusivity $\kappa_l$ on the left of an interface, and a fast cell of size $L_h$ and diffusivity $\kappa_h$ on the right. Here and below, subscripts $h$ and $l$ denote the high- and low-diffusivity phases. A particle is injected at the interface $x=0$ and diffuses until it first exits through one of the two outer boundaries, at $x=-L_l$ or $x=+L_h$. At that point we record both the displacement, $\Delta x=-L_l$ or $+L_h$, and the total escape time $\tau$. This waiting time includes all unresolved motion within the two cells, including repeated recrossings of the interface before final escape.

This is a two-sided first-passage problem. We derive the full solution in Appendix \ref{sec:two-cell}. The key quantities are the splitting probabilities $\alpha_l$ and $\alpha_h$, and the Laplace-transformed waiting-time distributions $U_l(s)$ and $U_h(s)$ for exit through the slow and fast sides, respectively. 
\red{It is convenient to define the half-patch resistance of each layer as}
\begin{equation}
    R_l \equiv \frac{L_l}{2\kappa_l},
    \qquad
    R_h \equiv \frac{L_h}{2\kappa_h},
\end{equation}
\red{so that a larger $R_i$ corresponds to a more difficult layer to traverse. This half-patch resistance differs by a factor of two from the full-patch resistance introduced in \S\ref{subsec:interpret} and the common factor of two cancels from the splitting probabilities below.}
As shown in Appendix \ref{sec:two-cell}, the splitting probabilities are then
\begin{equation}
    \alpha_h = \frac{R_l}{R_l+R_h},
    \qquad
    \alpha_l = \frac{R_h}{R_l+R_h}.
    \label{eq:alpha_two_cell}
\end{equation}
Thus, the probability of exiting through one side is controlled by the resistance on the \emph{opposite} side: if the slow side is harder to cross, the particle is more likely to escape through the fast side.

The corresponding Laplace-transformed exit-time distributions are
\begin{equation}
    U_{i}(s) =
    \frac{A_i\,\mathrm{csch}(a_i)}
    {A_l \coth(a_l) + A_h \coth(a_h)},
\qquad i\in\{l,h\}
\label{eq:Us_two_cell}
\end{equation}
where
\begin{equation}
    A_i = \kappa_i \lambda_i,
    \qquad
    \lambda_i = \sqrt{\frac{s}{\kappa_i}},
    \qquad
    a_i = \lambda_i L_i,
    \qquad i\in\{l,h\}.
\end{equation}
The total waiting-time transform is therefore
\begin{equation}
    \tilde{\psi}(s)=U_l(s)+U_h(s).
    \label{eq:two-cell-waiting}
\end{equation}

To verify these expressions, we simulate a one-dimensional diffusion process on the interval $[-L_l,L_h]$ with piecewise diffusivity
\begin{equation}
    \kappa(x)=
    \begin{cases}
        \kappa_l, & x<0,\\
        \kappa_h, & x\ge 0.
    \end{cases}
\end{equation}
We initialize particles at the interface and evolve them in Monte Carlo simulations,
regularizing the diffusivity jump with a narrow $\tanh$ smoothing layer for numerical stability. Each trajectory is followed until first passage to either absorbing boundary, at which point we record the exit time $\tau$ and exit side. The empirical Laplace transform of the waiting-time distribution is then
\begin{equation}
    U(s)=\langle e^{-s\tau}\rangle .
\end{equation}

\begin{figure}
    \centering
    \includegraphics[width=0.99\linewidth]{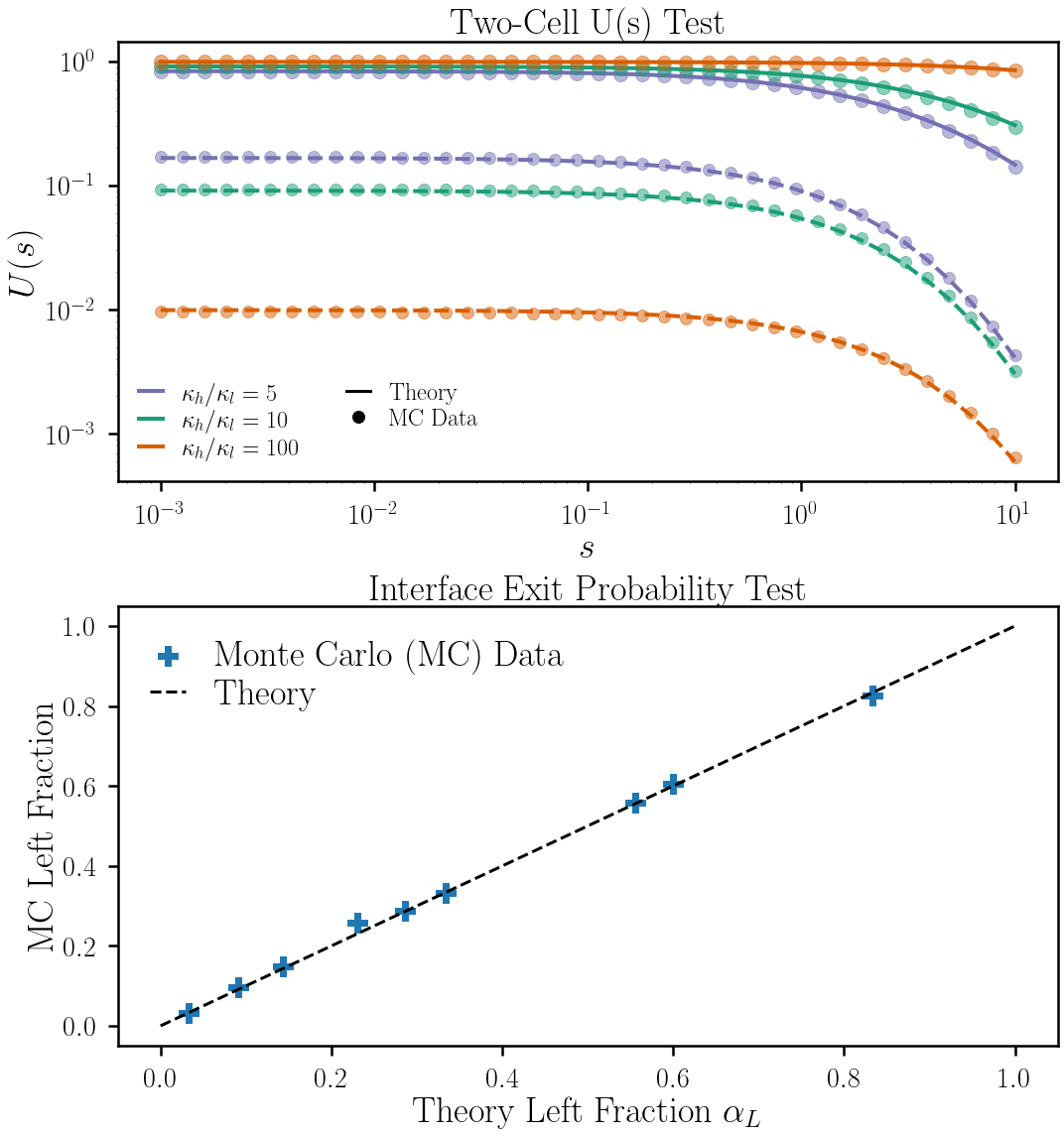}
\caption{
Validation of the coarse-grained two-cell first-passage model. \emph{Top:} Laplace-transformed exit-time distributions $U_l(s)$ and $U_h(s)$ from Eq.~(\ref{eq:Us_two_cell}) compared with Monte Carlo measurements for several diffusivity contrasts $\kappa_h/\kappa_l$ at fixed cell sizes. \emph{Bottom:} Monte Carlo measurements of the slow-side exit probability $\alpha_l$ compared with the analytic prediction in Eq.~(\ref{eq:alpha_two_cell}). Here $R_l=L_l/(2\kappa_l)$ and $R_h=L_h/(2\kappa_h)$ 
\red{are the half-patch resistances of the slow and fast cells.} The agreement shows that the two-cell problem provides the correct coarse-grained building block for the multiphase Montroll--Weiss description.
}
    \label{fig:Us_alpha}
\end{figure}

As shown in Figure \ref{fig:Us_alpha}, the Monte Carlo results agree well with the analytic expressions over a range of diffusivity contrasts. The top panel validates the exit-time transforms $U_l(s)$ and $U_h(s)$, while the bottom panel validates the splitting probabilities in Eq.~(\ref{eq:alpha_two_cell}). These quantities are the fundamental coarse-grained ingredients needed for the accelerated Monte Carlo scheme in the next subsection.

For later use, it is helpful to note that the exact inverse Laplace transform of Eq.~(\ref{eq:two-cell-waiting}) is an infinite sum of spatial eigenmodes. In practice, however, the time-domain waiting-time distributions are well approximated by a Prony expansion (\S\ref{subsec:approx_K}) dominated by the two characteristic crossing timescales, $L_l^2/\kappa_l$ and $L_h^2/\kappa_h$. This provides an efficient route to sampling coarse-grained waiting times in the Accelerated Monte-Carlo scheme below.

This two-cell waiting-time distribution is distinct from the well-known L\'evy--Smirnov form
\begin{equation}
    P(t)\propto t^{-3/2}\exp\!\left(-\frac{L^2}{4 \kappa t}\right),
\end{equation}
the continuous version of Sparre-Andersen scaling, which describes first passage to a \emph{single} absorbing boundary in a semi-infinite domain. 
\red{Short passage times $t \ll L^2/\kappa$ are exponentially suppressed, while there is a $t^{-3/2}$ scaling at late times. This $t^{-3/2}$ scaling is seen in many astrophysical contexts \citep{monter24,liang25,kempski25}, and it reflects the possibility of very long waiting times in the one-sided barrier problem if the particle wanders off into the wild blue yonder. By contrast, in our finite
two-sided interval, there is also an exponential cutoff at late times,
$t\gtrsim L^2/\kappa$, reflecting the fact that the wandering particle will inevitably hit a barrier in a two-sided problem in finite time. Thus there is no extended regime in which the
waiting-time distribution follows a pure $t^{-3/2}$ power law.} The cutoff is what makes the coarse-grained two-cell dynamics effectively finite-variance and well suited to the Prony representation.

\subsection{\red{Constant Cell Size}}
\label{subsec:const}

We now turn to simulating transport in a multi-phase medium and recovering Montroll-Weiss statistics from our Monte-Carlo simulations. We begin with the simplifying case of constant cell sizes, for both fast and slow patches. 
\red{With constant cell sizes, every renewal step is a jump of fixed magnitude $\pm L$. The step statistics are nonetheless different at alternating interfaces. Let $A$ denote an interface with the slow cell on its left and the fast cell on its right, and $B$ the opposite arrangement. From $A$ the particle exits to the right with probability $\alpha_h$, and from $B$ it exits to the left with the same probability.}
\red{
The two alternating interface states $A$ and $B$ make even the constant cell medium a two-state (matrix) Montroll--Weiss process. Constant cells are, however, the \emph{separable} special case. The jump distribution pooled over two states is $\lambda(\Delta x)=\frac12\,\delta(\Delta x-L)+\frac12\,\delta(\Delta x+L)$, so that 
\begin{equation}
\hat\lambda(k)\equiv \int d(\Delta x)\,e^{ik\Delta x}\,\lambda(\Delta x)= \frac{e^{ikL}+e^{-ikL}}{2}=\cos(kL).
\end{equation}
The pooled step kernel factorizes,
\begin{equation}
\tfrac12\big[\Phi_A(k,s)+\Phi_B(k,s)\big]=\cos(kL)\,\big[U_h(s)+U_l(s)\big]=\hat\lambda(k)\,\tilde\psi(s),
\end{equation}
into a spatial factor $\hat\lambda(k)=\cos(kL)$ and a temporal factor $\tilde\psi(s)=U_h(s)+U_l(s)$. 
Here a step to the right from $A$ lands on $B$, from which the next step returns to the left with probability $\alpha_h$, so successive steps are anticorrelated. 
These anti-correlations between successive steps drive the crossover from $\kappa_A$ to $\kappa_H$, and are retained in the matrix description.
Separability of the pooled step kernel and the need for a matrix description are therefore distinct properties. The fact that there are two different interface states A \& B introduces the need for matrix MW, while variable cell size breaks separability as well.
}

\subsubsection{Accelerated Monte-Carlo}
\begin{figure}
    \centering
    \includegraphics[width=0.99\linewidth]{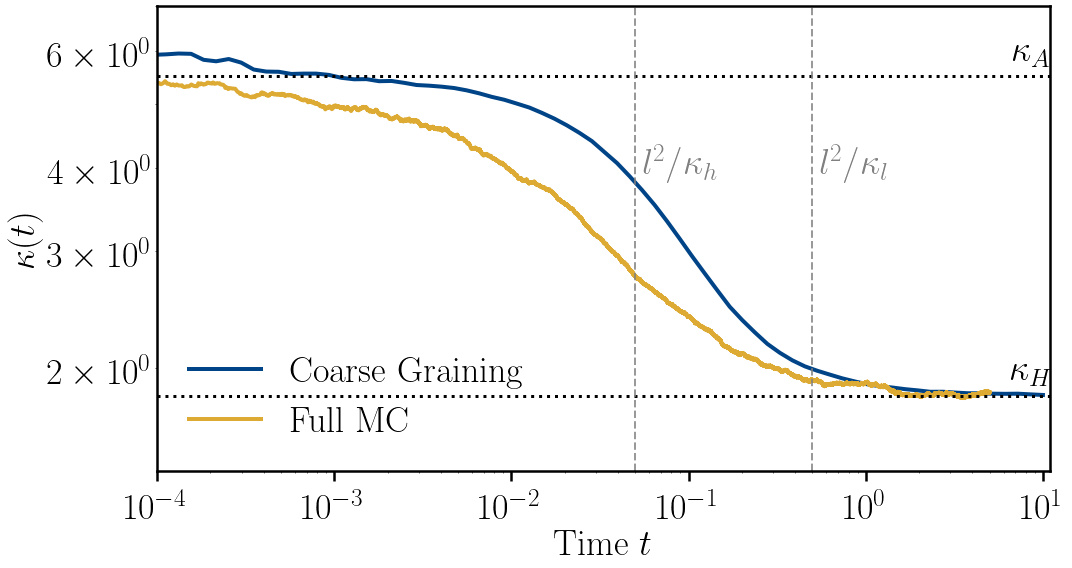}
    \caption{Comparison of the running diffusion coefficient $\kappa(t)$ obtained from coarse-grained accelerated MC and full MC simulations, with constant patch sizes and $\kappa_h/\kappa_l=10$. $\kappa(t)$ starts from the arithmetic mean $\kappa_A$ at early times and gradually decreases to the harmonic mean $\kappa_H$ at long times (dotted lines). Vertical gray dashed lines mark the characteristic diffusion timescales $l^2/\kappa_h$ and $l^2/\kappa_l$.}
    \label{fig:compare}
\end{figure}
With the analytical two-cell waiting time distributions $U_{L/R}(s)$ and splitting probabilities derived, we can construct a macroscopic, event-driven Accelerated Monte Carlo (AMC) scheme. Instead of integrating the SDE over time steps $\Delta t$, the AMC scheme evolves particles by jumping from interface to interface. At each cell boundary, the particle draws its next transition direction (left or right) based on the macroscopic splitting probabilities (Eq. \ref{eq:alphas_appendix}) and advances its internal clock \red{by a waiting time drawn from the corresponding first-passage distribution, either from its Prony approximation or from the exact distribution obtained by numerical inversion (Appendix~\ref{app:numerics}).}

To validate this coarse-grained approach, we compare the transport statistics of the AMC against a traditional full Monte Carlo simulation that resolves local scattering within cells. Figure \ref{fig:compare} tracks the running diffusion coefficient, $\kappa(t) = \langle x^2(t) \rangle / 2t$, for a periodic two-phase medium with a constant patch size and a diffusivity contrast of $\kappa_h/\kappa_l=10$. 

At early times ($t \ll l^2/\kappa_h$), particles diffuse within their initial cells before encountering interfaces, so the running diffusion coefficient remains near the arithmetic mean, $\kappa_A=\langle\kappa\rangle$. As particles begin to cross interfaces and encounter slow patches, $\kappa(t)$ drops toward the harmonic-mean limit. In the basic AMC scheme this crossover is slightly distorted, because particles are evolved interface-to-interface rather than through every internal scattering event.

\red{We correct this by changing the initialization. As in the fine-grained Monte Carlo, particles begin uniformly throughout the medium rather than at interfaces. Each particle is assigned to a cell with probability $p_i=L_i/\sum_j L_j$ placed at a random position within that cell, and evolved with resolved diffusion until it first reaches a cell boundary. The first exit is therefore treated exactly. Afterward, the particle follows the usual interface-to-interface jump scheme. This restores the correct short time first-passage statistics while adding only one resolved cell crossing per particle. At late times, $t\gg l^2/\kappa_l$, particles sample many cells and approach the ergodic limit, with $\kappa(t)\rightarrow\kappa_H$. The AMC reproduces this limit accurately, showing that coarse-graining the internal cell dynamics preserves the large-scale relaxation.}

The power of the AMC framework lies in its computational efficiency. In a standard full MC simulation, numerical stability (CFL-like condition for diffusion) determines the global timestep $\Delta t$. It must be strictly smaller than the crossing time of the smallest, fastest patch in the domain: $\Delta t \ll l_{\rm{min}}^2 / \kappa_h$. Simultaneously, to reach the steady-state macroscopic regime, the simulation must run for several multiples of the longest characteristic timescale, $t_{\rm{max}} \sim l_{\rm{max}}^2 / \kappa_l$. This creates a severe scale separation in time. Especially in a highly heterogeneous medium with a large spatial dynamic range and a strong diffusivity contrast ($\kappa_h/\kappa_l$), the number of required integration steps scales as $\mathcal{O}([l_{\rm max}/l_{\rm min}]^2 \cdot \kappa_h/\kappa_l)$, making simulations expensive. Because the AMC coarse-grain the internal scatterings within cells, it entirely bypasses this microscopic timestep constraint.

\begin{table}
\centering
\smallskip
\renewcommand{\arraystretch}{1.3}
\begin{tabular}{r l c c c}
\hline
$\kappa_h/\kappa_l$ & Geometry
  & CPU (Full MC) & CPU (Accel MC) & Speedup \\
\hline
\multicolumn{5}{c}{\textit{Equal-size cells} ($\ell_\mathrm{min}=\ell_\mathrm{max}=1$)} \\[2pt]
1       & $L=1$ & 1\,s                & 0.02\,s  & $47\times$ \\
10      & $L=1$ & 12\,s$^\dagger$     & 0.05\,s  & $214\times$ \\
100     & $L=1$ & 2\,min$^\dagger$    & 0.4\,s   & $311\times$ \\
10\,000 & $L=1$ & 3.2\,hr$^\dagger$   & 32\,s    & $\mathbf{368\times}$ \\[4pt]
\hline
\multicolumn{5}{c}{\textit{Variable cells} ($\ell_\mathrm{min}=0.1,\;\ell_\mathrm{max}=1$)} \\[2pt]
100     & $\alpha=2$   & 3.2\,hr$^\dagger$   & 9\,s$^\ddagger$     & $\mathbf{1\,240\times}$ \\
100     & $\alpha=3.5$ & 3.2\,hr$^\dagger$   & 19\,s$^\ddagger$    & $620\times$ \\
10\,000 & $\alpha=2$   & 13.5\,days$^\dagger$ & 13\,min$^\ddagger$ & $\mathbf{1\,470\times}$ \\
10\,000 & $\alpha=3.5$ & 13.5\,days$^\dagger$ & 26\,min$^\ddagger$ & $740\times$ \\
\hline
\end{tabular}
\caption{CPU time comparison between full and accelerated Monte Carlo to run up to $10 l_{\rm{max}}^2/\kappa_l$. (The speedup is independent of $T$ since both methods scale linearly with simulation duration.) The Full MC timestep obeys $\Delta t = 10^{-3}l_{\rm{min}}^2 / (\kappa_\text{max}) = 10^{-3}\tau_{\rm{min}}$. All runs use $10^4$ particles. The entries marked $^\dagger$ are extrapolated from cost per step, and entries marked $^\ddagger$ use the empirically calibrated jump-count ratio.}
\label{tab:benchmark}
\end{table}

Table \ref{tab:benchmark} quantifies the CPU time speedup of the AMC scheme against full MC. For equal-sized cells, the speedup grows sub-linearly with the diffusivity contrast, reaching a factor of 400 at $\kappa_h/\kappa_l = 10^4$. The gains become even more dramatic when applied to realistic, heterogeneous media with variable cell sizes drawn from a power-law distribution $P(l) \propto l^{-\alpha}$. In environments with both a high diffusivity contrast and a dynamic range of spatial scales ($\ell_\mathrm{min}=0.1, \ell_\mathrm{max}=1$), the AMC achieves speedups by three orders of magnitude ($\sim 1500\times$). 

Furthermore, flatter spatial distributions yield larger relative speedups because they contain a higher proportion of large, slow traps. In a full MC, particles spend many steps scattering within these deep traps, whereas the AMC resolves the entire trapping event in a single random draw. This makes it feasible to simulate global CR transport through intermittent, multiphase Galactic environments, while preserving the underlying fractional transport and memory dynamics.

\subsubsection{\red{Propagator and Large Scale Kernel}}
\label{subsec:recovering}
\begin{figure}
    \centering
    \includegraphics[width=0.99\linewidth]{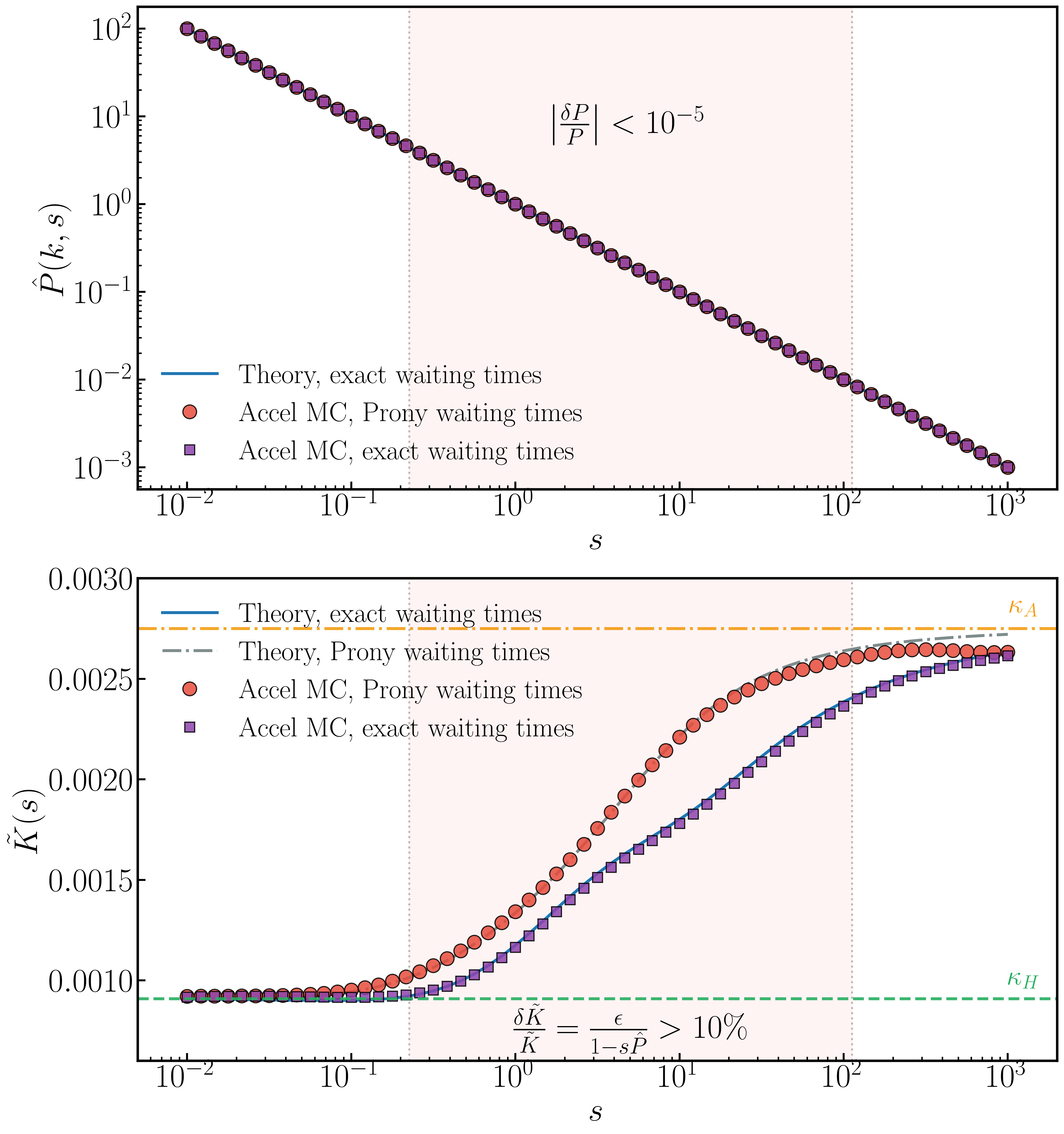}
    \caption{\red{Monte Carlo against theory in the Laplace domain for the constant cell medium, with theory curve constructed both with the exact waiting time distribution and the Prony approximation and MC runs using each of the two waiting time samplers. \emph{Top:} the propagator $P(k=0.1,s)$ agrees with the analytic prediction across all frequencies. \emph{Bottom:} the large scale memory kernel $\tilde{K}(s)$ obtained with the Prony approximation matches the exact theory curves at small $s$ near $\kappa_H$ and deviates at large $s$. The breakdown is the expected amplification of the systematic short time waiting-time approximation error amplified when extracting $\tilde{K}(s)$ from  $P(k,s)$.}}
    \label{fig:const_all}
\end{figure}

\red{From the propagator $P(k,s)$ we recover the large scale memory kernel $\tilde{K}(s)$, a scalar defined directly from $P(k,s)$ through Eq.~\ref{eq:Ks_from_P}.} 
\red{Since $P(k,s)$ is scalar, the hidden state information present in matrix MW has already been coarse-grained out.} 
\red{The waiting time transform, by contrast, is known analytically and we do not obtain it by inverting $P(k,s)$. We evaluate $\tilde{K}(s)$ at $k=0.1$, which Appendix~\ref{subsec:multiple_k} shows lies in the large-scale regime where Eq.~\ref{eq:Ks_from_P} becomes independent of $k$.}

\red{For the constant-cell case, $P(k,s)$ is given analytically by the two-state construction of Appendix~\ref{sec:analytic-MW}. To match the accelerated Monte Carlo, the first step is treated exactly in both approaches. In the theoretical calculation, the first-step terms in Eq.~(\ref{eq:P_general_firststep}) are averaged over a uniform birth position within each cell. The pre-exit contribution is the absorbing-cell propagator $\tilde S_{i,0}(k,s)$, while the exit kernels $V^{\pm}_{i,0}(k,s)$ carry the particle from its birth position to either cell boundary, where the interface-to-interface description begins (see \S\ref{sec:two-cell}). In the MC, each particle is followed explicitly from its birth position until it first reaches a cell boundary. The only remaining difference is therefore how the interface-to-interface waiting times are sampled.}

\red{
We extract $P(k,s)$ from our Monte Carlo trajectories and recover $\tilde{K}(s)$. Figure \ref{fig:const_all} compares it with the analytic prediction. 
The accelerated Monte Carlo draws interface waiting times from a Prony fit to $U_{L,R}(s)$. This fit is accurate at moderate and long times but overestimates the very short-time contribution. 
The effect of this approximation on the propagator is negligible. At $kL_i=0.01$, replacing the exact $U_{L,R}(s)$ with the fitted
form changes the propagator by less than $10^{-5}$, which is not visible in the top panel of Fig.~\ref{fig:const_all}. The extracted kernel is much more sensitive. The same replacement changes $\tilde K(s)$ by more than $10\%$ over $0.2\lesssim s\lesssim100$. Figure~\ref{fig:const_all} shows the theory built with each choice, and each Monte Carlo run follows the curve built from the waiting times it was given, to $3\%$. 
\red{Both calculations describe the same coarse-grained interface process.
The ``exact'' curve uses the exact first-passage kernels
$U_{L,R}(s)$, whereas the Prony curve makes the additional numerical
approximation of replacing those kernels by finite sums of exponentials.
Thus the difference between the two curves in Fig.~11 is due to a systematic short time error in the Prony approximation.}
We quantify this bias in Appendix~\ref{app:numerics}.
At large $s$, even when $P(k,s)$ has negligible discrepancy, the inversion amplifies small errors,
\begin{equation}
\frac{\delta \tilde{K}}{\tilde{K}}
=
-\frac{\delta P/P}{1-sP}
=
-\frac{\epsilon}{1-sP},
\label{eq:delta_K}
\end{equation}
which diverges as $sP\rightarrow1$ (i.e.\ $s\gg k^2\tilde{K}$, where $P\rightarrow1/s$). In practice, the successful recovery of $\tilde{K}(s)$ at small $s$ confirms that the propagator preserves the large scale transport statistics.
}

\red{These results show that the macroscopic propagator is useful for data analysis as well as theory. In a medium with hidden interface states, although a full description requires matrix MW, the scalar kernal $\tilde{K}(s)$ can still be obtained from the propagator $P(k,s)$.}

\subsection{\red{Variable Cell Size}}
\label{subsec:variable}

The multiphase framework can readily accommodate variation in both cell size and diffusivity by assigning each phase or cell its own pair $(L_i,\kappa_i)$, which then determines the corresponding coarse-grained jump and waiting-time statistics. For simplicity, and without losing the main physics of space--time coupling, we allow only the cell sizes to vary continuously while keeping the diffusivities fixed within each phase. We thus now generalize from equal-sized cells to a heterogeneous multiphase medium in which the high and low diffusivity patch lengths are drawn from truncated power laws, $P(\ell) \propto \ell^{-\alpha}$ over $\ell \in [\ell_{\min},\ell_{\max}]$. Unless otherwise stated, both phases share the same $\alpha$ and dynamic range. Once the cell lengths vary, the coarse-grained walk is no longer described by a single jump length and a single waiting-time law: the distance advanced at an interface and the time required for that advance are now coupled, and set by cells adjacent to the interface. 
\red{The process remains semi-Markov on the interface states. What changes here is that the pooled step kernel no longer factorizes into a spatial and a temporal part.}

\subsubsection{Approach to the Harmonic Mean: Initial Conditions, Filling Fraction}

\begin{figure}
    \centering
    \includegraphics[width=0.99\linewidth]{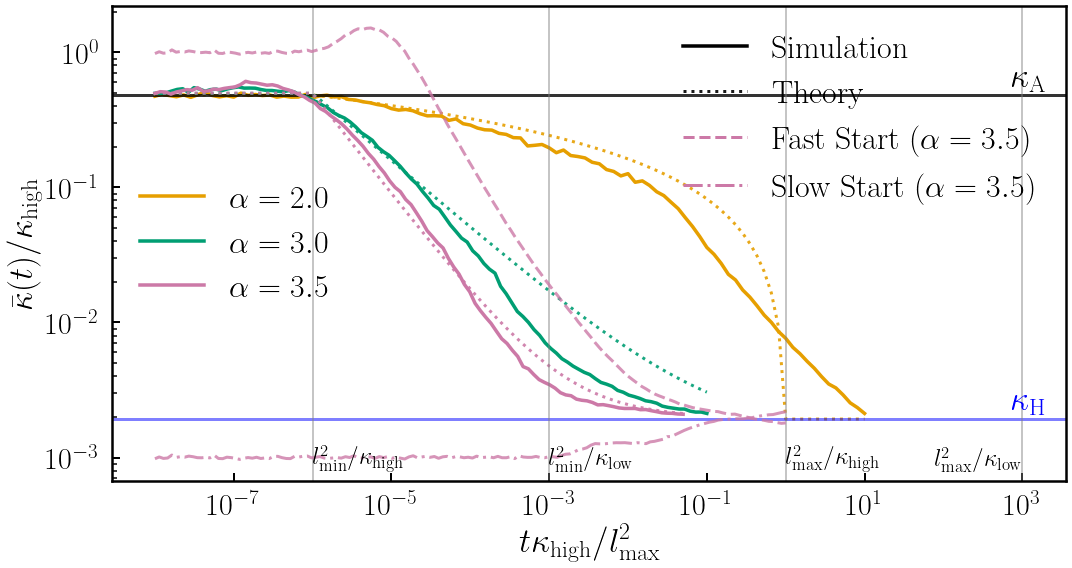}
    \caption{Normalized running diffusion coefficient for variable-cell media with patch lengths drawn from $P(\ell)\propto \ell^{-\alpha}$. Solid curves show Monte Carlo results and dotted curves show the scaling estimate above. For a uniform initial distribution, transport interpolates from the short-time arithmetic mean $\kappa_A$ to the long-time harmonic mean $\kappa_H$. Initial conditions restricted to fast (dashed) or slow (dash-dotted) cells instead probe the escape dynamics of the birth environment before converging to the same late-time limit. Vertical lines mark the characteristic minimum crossing times.}
    \label{fig:variable_cell}
\end{figure}

For a uniform spatial initialization, the normalized running diffusion coefficient in Figure~\ref{fig:variable_cell} evolves from the short-time arithmetic mean $\kappa_A$ to the long-time harmonic mean $\kappa_H$. At early times, particles mostly diffuse within their birth cells; the crossover begins once the diffusion length $\sqrt{\kappa_h t}$ becomes comparable to the sizes of the fast patches. Following \citet{ewart25}, this behavior is captured by the estimate
\begin{equation}
    \kappa_{\mathrm{theory}}(t)
    \sim
    \kappa_h
    \frac{\int_{\sqrt{\kappa_h t}}^{\ell_{\max}} P_h(\ell)\,\ell\,d\ell}
         {\int_{\ell_{\min}}^{\ell_{\max}} [P_h(\ell)+P_l(\ell)]\,\ell\,d\ell},
\end{equation}
which measures the fraction of particles that still reside in fast patches larger than the instantaneous diffusion length. The agreement between the Monte Carlo curves and this scaling estimate shows that the AMC scheme reproduces not only the asymptotic limits, but also the intermediate crossover.

Figure~\ref{fig:variable_cell} also shows that initial conditions matter. Particles born in fast cells approach $\kappa_H$ from above, whereas particles born in slow cells approach it from below, because the early evolution is controlled by escape from the birth environment rather than by transport through the full medium\footnote{Particles which are initially uniformly distributed are dominated by the diffusion of those born in the fast cells, hence the approach to the harmonic mean from above.}. This is potentially important since CR injection in the ISM is not uniform but concentrated in regions such as supernova remnants and pulsar wind nebulae, which are observationally known to have lower diffusion coefficients. We discuss this further in \S~\ref{subsec:local_global}. For $\kappa_l \ll \kappa_h$, the late-time diffusivity is approximately $\kappa_H \simeq [(1-f_l)/\kappa_h + f_l/\kappa_l]^{-1} \simeq \kappa_l/f_l$ \citep{ewart25}, so even if CRs sources are embedded in a very slow phase, over time the running diffusion coefficient can still converge to a much larger value if the slow phase has a small filling fraction. The effect appears mild in Fig. \ref{fig:variable_cell} since $\kappa_H/\kappa_l \sim f_l^{-1}$ and we adopt $f_l \sim 0.5$. In reality, we expect $f_l \ll 1$ and so much larger separation between $\kappa_l$ and $\kappa_{\rm H}$. 

\begin{figure}
    \centering
    \includegraphics[width=0.99\linewidth]{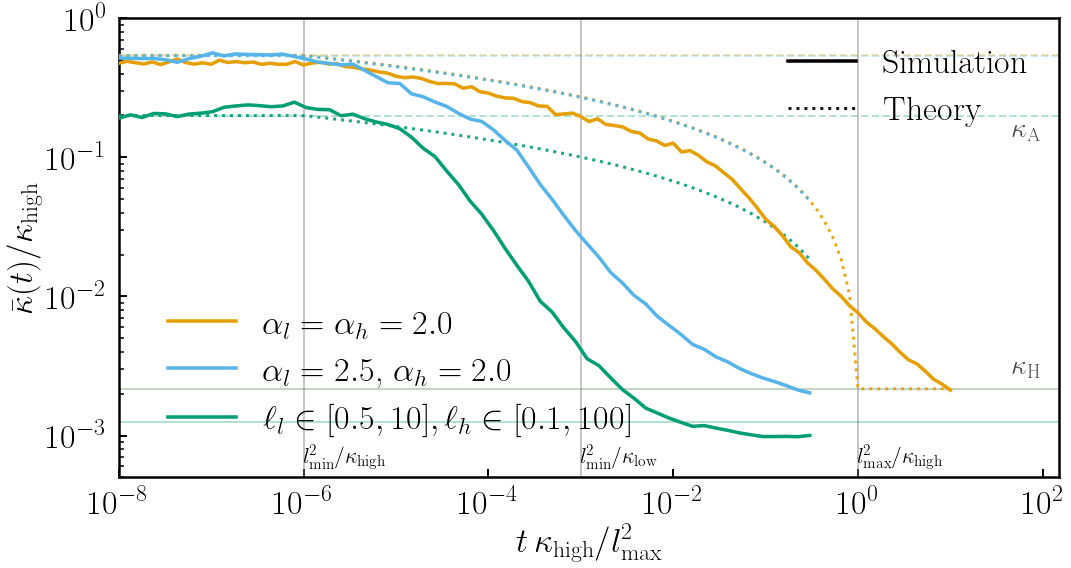}
    \caption{Normalized running diffusion coefficient for $\kappa_l=0.1$ and $\kappa_h=100$, with power-law patch lengths $P(\ell)\propto \ell^{-\alpha}$, comparing three slow-patch distributions: a symmetric baseline ($\alpha_l=\alpha_h=2$, $\ell\in[0.1,100]$), a steeper slow-phase exponent ($\alpha_l=2.5$), and a narrower slow-phase range ($\ell_l\in[0.5,10]$). Changing only the slow-phase geometry changes the filling fraction $f_l$ and therefore shifts both the asymptotic levels and the time at which the crossover toward $\kappa_H$ begins.}
    \label{fig:asym_dist}
\end{figure}

The crossover is also sensitive to how the slow phase is distributed geometrically. In Figure~\ref{fig:asym_dist}, we vary only the slow-patch distribution while holding $\kappa_h$ and $\kappa_l$ fixed. The main effect is through the slow-phase filling fraction $f_l$: narrowing the slow-patch distribution increases $f_{l}$ ($0.7$ vs.\ $0.5$). Making slow patches more common (larger $f_l$) reduces $\kappa_H/\kappa_l$ and shifts the crossover to earlier times, because particles encounter slow material sooner and more often. Reducing $f_l$ would have the opposite effect, delaying the crossover while increasing the contrast between the local slow diffusivity and the eventual harmonic mean. 

\subsubsection{Quenched versus annealed disorder}

The results above assume quenched disorder: particles move through a fixed spatial arrangement of fast and slow cells. This is the more realistic idealization for CR transport through a static multiphase medium, because the bottlenecks are tied to actual structures in space. We compare this with an annealed model in which the environment is redrawn after each coarse-grained transition. The annealed case is therefore best viewed as a reference problem---and a useful computational simplification---rather than as a literal model of a static ISM. It becomes relevant when the environment is effectively renewed between steps, for example in the rapid-reset limit discussed in \S~\ref{subsec:renewal}.

\begin{figure}
    \centering
    \includegraphics[width=0.99\linewidth]{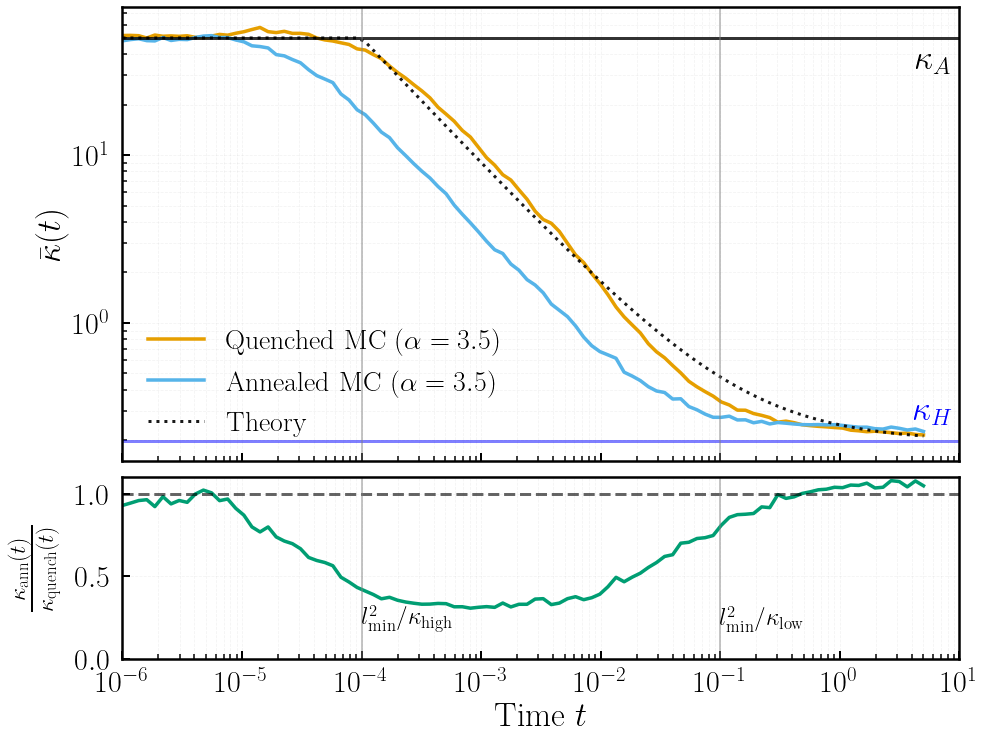}
    \caption{Running diffusion coefficient for quenched and annealed environments in the variable cell case. \emph{Top:} both cases interpolate from the short-time arithmetic mean $\kappa_A$ to the long-time harmonic mean $\kappa_H$. In the parameter range shown, the annealed curve departs only modestly from the quenched one, entering the crossover slightly earlier. \emph{Bottom:} the ratio $\kappa_{\mathrm{ann}}(t)/\kappa_{\mathrm{quench}}(t)$ isolates this difference.}
    \label{fig:ann_quenched}
\end{figure}

In the parameter range of Figure~\ref{fig:ann_quenched}, the difference is modest. Both cases share the same short-time limit $\kappa_A$ and the same late-time limit $\kappa_H$, and the annealed curve merely enters the crossover slightly earlier. The bottom panel shows that the largest discrepancy occurs between $\ell_{\min}^2/\kappa_h$ and $\ell_{\min}^2/\kappa_l$, where re-encounters with the same slow structures matter most. Annealing removes those fixed-geometry correlations, so it weakly accelerates convergence toward $\kappa_H$. We expect a larger separation between quenched and annealed transport when spatial correlations are stronger---for example for broader cell-size distributions, rarer but deeper traps, or other geometries that promote repeated encounters with the same bottlenecks. The close agreement here also explains why annealed sampling, which is computationally the most straightforward, can be a useful surrogate when one only needs bulk crossover behavior.

\subsubsection{\red{Recovering the Large Scale Kernel}}
\begin{figure}
    \centering
    \includegraphics[width=0.99\linewidth]{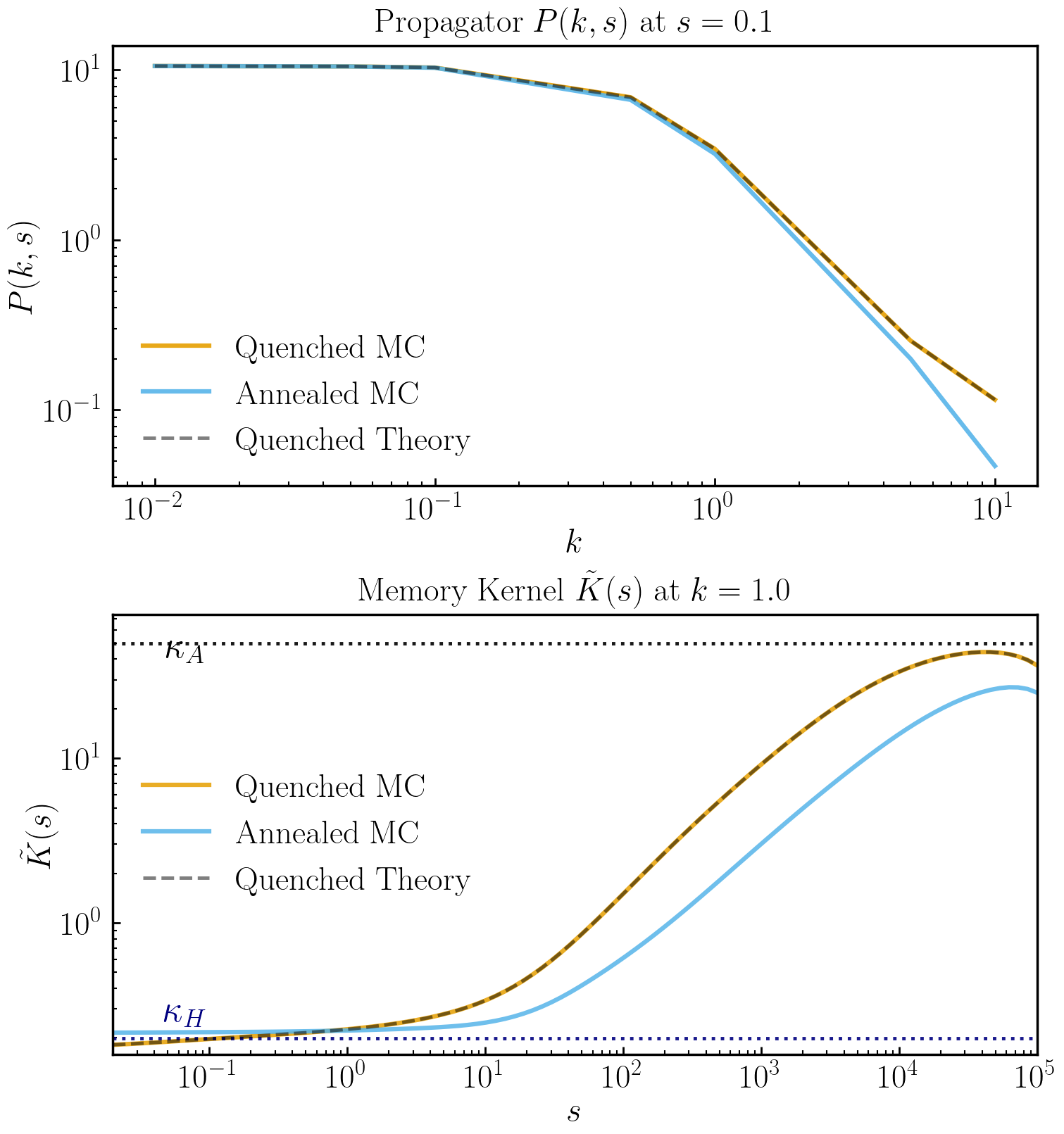}
    \caption{
    \red{\emph{Top:} the propagator $P(k,s)$ at fixed $s=0.1$ in the variable cell case. The quenched Monte Carlo agrees with the quenched matrix theory, while the annealed propagator remains close at small $k$ and deviates only mildly at larger $k$. \emph{Bottom:} the large scale memory kernel $\tilde{K}(s)$ at fixed $k=1.0$ for quenched and annealed environments, crossing over from $\kappa_H$ at low $s$ to $\kappa_A$ at high $s$.}}
    \label{fig:ann_quenched_Ks}
\end{figure}

\red{Figure~\ref{fig:ann_quenched_Ks} shows that the variable cell propagator can still be recovered from particle trajectories, and that the large scale memory kernel can be extracted from it.}
In the top panel, we first compare the quenched propagator $P(k,s)$ to expectations from analytic theory (Appendix \ref{subsec:matrix_MW}), and recover a very good match. This is gratifying, given that the analytic quenched propagator--which is the most realistic case for the ISM -- is non-trivial to calculate. The annealed propagator remains close over the low-$k$ range most relevant to the macroscopic limit and deviates only at larger $k$.

\red{
Because $P(k,s)$ is a scalar, it averages over the hidden interface states retained in the matrix kernel $\Phi_{ij}(k,s)$. That state resolved information is not recoverable from $P(k,s)$ alone. What is recoverable is the large scale memory kernel $\tilde{K}(s)$, which is defined directly from $P(k,s)$.
The bottom panel of Fig.~\ref{fig:ann_quenched_Ks} shows $\tilde{K}(s)$ recovered from $P(k,s)$ compared with the analytic matrix theory.
}

The bottom panel shows the Laplace space kernel recovered from $P(k,s)$, using equation \ref{eq:Ks_from_P}. In Appendix \ref{subsec:multiple_k}, we directly verify that the chosen wavenumber $k=1.0$ lies in the long wavelength limit where equation \ref{eq:Ks_from_P} is independent of $k$. We recover the frequency-dependent crossover in $\tilde K(s)$, bridging $\kappa_H$ at low $s$ and $\kappa_A$ at high $s$. Again, there is excellent agreement between analytic theory and the AMC results, since both the theory and the MC coarse-grain the intra-cell scattering in the exact same manner. 


\subsubsection{Separability and space-time coupling}

The main reason the \red{separable} MW description fails for variable cells is that space and time are no longer independent. In the constant-cell case, every renewal step spans the same distance and is governed by the same waiting law, so the \red{pooled} joint distribution factorizes. With variable cells, the same local geometry sets both the displacement and the escape time. Larger cells tend to produce both longer jumps and longer waiting times. This induces intrinsic space--time coupling.

A key assumption of the constant cell MW framework is the separability of the joint step/waiting distribution,
\begin{equation}
    W(\ell,\tau)=\lambda(\ell)\,\psi(\tau),
\end{equation}
which implies $\Phi(k,s)=\hat\lambda(k)\,\tilde\psi(s)$. 
If $\Phi(k_i,s_j)$ is sampled on a grid of $k$ and $s$, exact separability means that the resulting matrix is rank one. Singular value decomposition (SVD) provides a simple diagnostic: if only the leading singular value is appreciable, then $\sigma_2/\sigma_1 \ll 1$, whereas appreciable singular values for higher modes indicate departures from separability.

\begin{figure}
    \centering
    \includegraphics[width=1\linewidth]{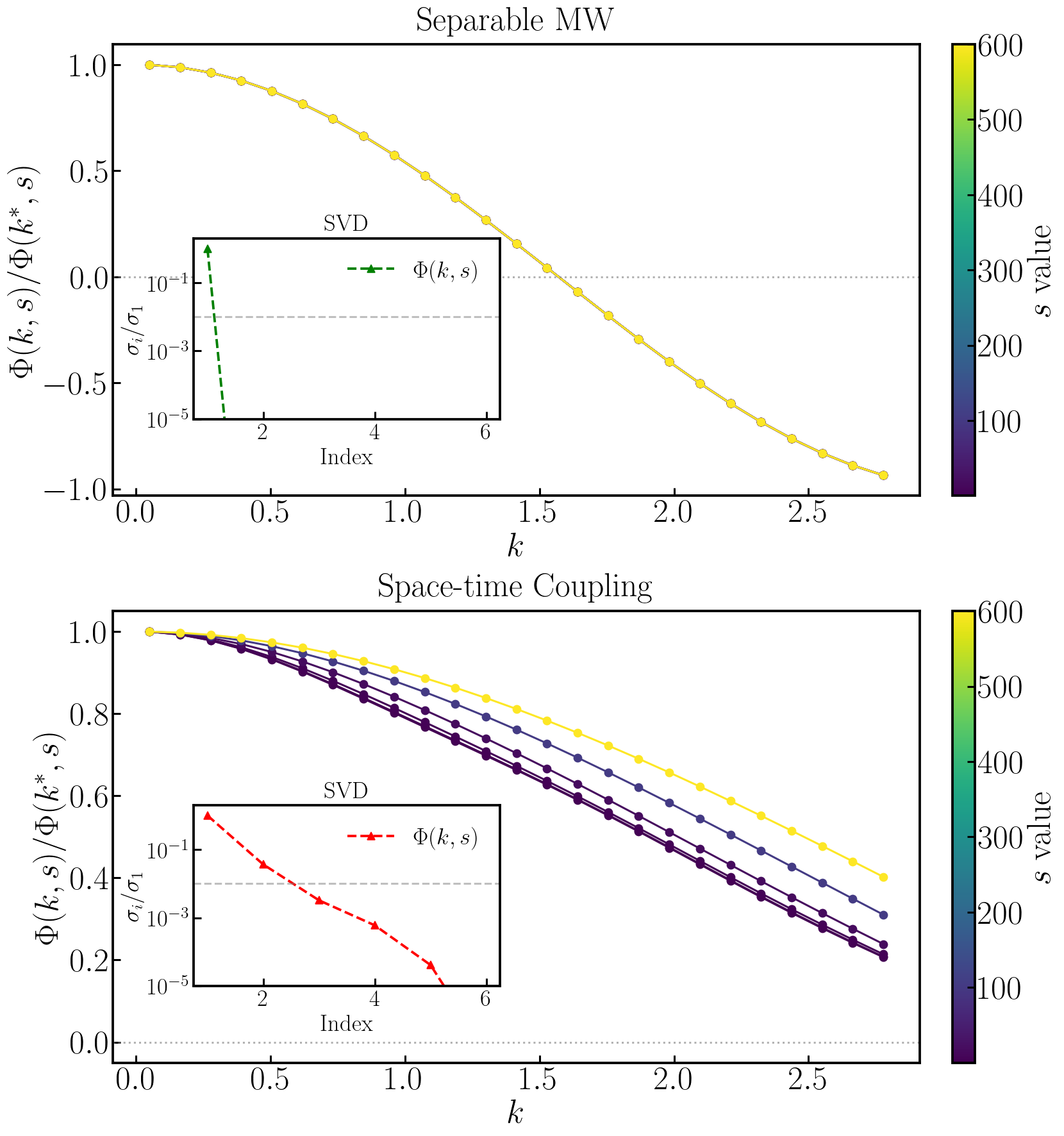}
    \caption{\red{Separability of the joint step transform $\Phi(k,s)=\langle e^{ik\Delta x - s\tau}\rangle$, measured from the $(\Delta x,\tau)$ pairs of the accelerated MC and normalized at $k^*=0.05$. \emph{Top:} constant cells. All $s$ collapse onto $\cos(kL)$, consistent with the separable form $\Phi(k,s)=\hat\lambda(k)\tilde\psi(s)$. The inset shows that the matrix $\Phi(k_i,s_j)$ is rank one. \emph{Bottom:} variable cells. The curves fan out with $s$ and higher singular values remain finite since larger cells produce both longer jumps and longer waits, showing space--time coupling within a coarse-grained step.}}
    \label{fig:separability}
\end{figure}

\red{
Unlike the equal-cell case, where the coarse-grained jump length is fixed and the only non-trivial renewal statistic is the waiting-time transform $\psi(s)$, the variable-cell problem requires the measured joint step kernel
\begin{equation}
    \Phi(k,s) \equiv \left\langle e^{ik\Delta x - s\tau} \right\rangle,
\end{equation}
since both the jump length $\Delta x$ and the waiting time $\tau$ vary from step to step, and are coupled in general.}

Figure~\ref{fig:separability} applies this diagnostic to $\Phi(k,s)$. 
\red{$\Phi(k,s)$ is pooled over the two interface states. The test therefore probes space--time separability within a coarse-grained step.} 
In the constant-cell case, the normalized curves collapse and the singular spectra are essentially rank one. In the variable-cell case, the normalized curves fan apart with $s$, and the higher singular values remain non-zero. The transport is therefore not captured by a single $\hat\lambda(k)$ and a single $\tilde\psi(s)$. \red{Notably, it is not a test of whether a scalar or matrix description is required, since the process here is matrix in either case.}

The joint kernel $\Phi(k,s)$ directly characterizes a single coarse-grained step, so non-separability in $\Phi$ is the clearest signature that jump lengths and waiting times are correlated. 
 
At the same time, we emphasize that changes in the parameters $(q_R,q_\tau,\alpha,\Delta)$ can have significant impact on the Laplace space kernel $\tilde{K}(s)$, particularly its width (the frequency range over which it transitions from the arithmetic to the harmonic mean) and recoverability (the frequency range over which the small $k$ approximation given by equation \ref{eq:Ks_from_P}) holds. We document this in Appendix \S\ref{app:parameters}. 

\subsection{Resistance, Residence Time, and Bottlenecks}
\label{subsec:interpret}
\begin{figure}
    \centering
    \includegraphics[width=1\linewidth]{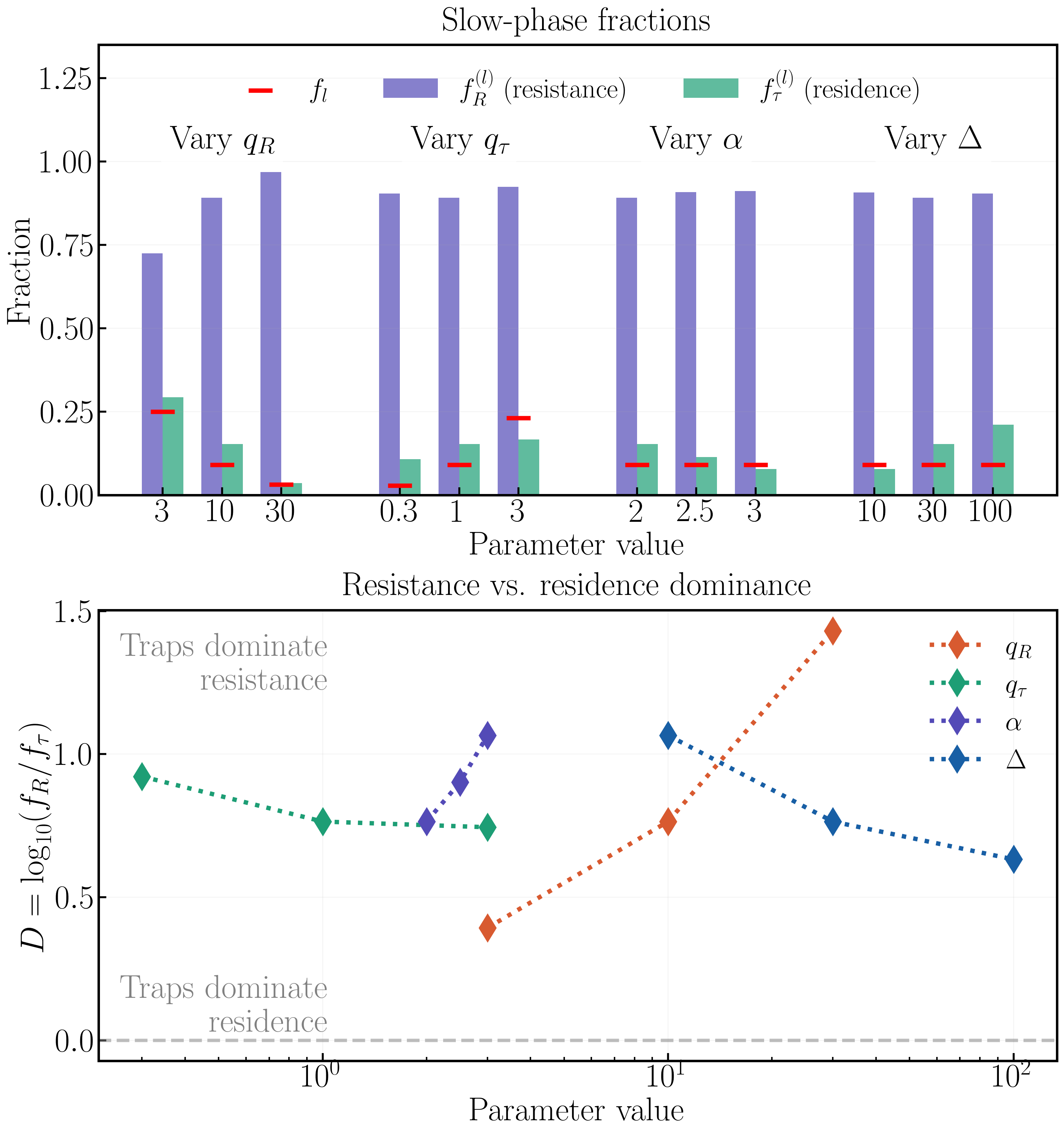}
    \caption{The slow fractions and resistance dominance across different parameters. \emph{Top:} Grouped bar chart shows the residence fraction ($f_\tau^{(s)}$), resistance fraction ($f_R^{(s)}$), and total slow-phase volume fraction ($f_{\rm slow}$) across sweeps of the parameters $q_R$, $q_\tau$, $\alpha$, and $\Delta$. \emph{Bottom:} The dominance $D = \log_{10}(f_R / f_\tau)$ as a function of the swept parameter values. All tests lead to positive values ($D > 0$), indicating that traps dominate resistance but not residence.}
    \label{fig:residence}
\end{figure}

The results above can be understood in terms of a simple diffusive resistance picture. For steady transport through a patch \(i\) of length \(L_i\) and diffusivity \(\kappa_i\), Fick's law gives
\begin{equation}
    J \approx -\kappa_i \frac{\Delta n_i}{L_i},
\end{equation}
\red{so the density drop across the patch is
\begin{equation}
    \Delta n_i \approx -J\mathcal{R}_i,
    \qquad
    \mathcal{R}_i\equiv\frac{L_i}{\kappa_i}=2R_i.
\end{equation}
Here $\mathcal{R}_i$ is the full-patch resistance, while $R_i=L_i/(2\kappa_i)$ denotes the half-patch resistance used in \S\ref{subsec:two_cells}. Thus $\mathcal{R}_i$ plays the role of a diffusive resistance:}
transport across a patch is impeded either because the patch is thick or because diffusion through it is slow. In a one-dimensional chain, these resistances add in series. Defining the effective diffusivity by \(\kappa_{\rm eff}\equiv J L_{\rm tot}/\Delta n_{\rm tot}\), one obtains
\red{\begin{equation}
    \frac{1}{\kappa_{\rm eff}}
    =
    \frac{1}{L_{\rm tot}}
    \sum_i \mathcal{R}_i
    =
    \frac{1}{L_{\rm tot}}
    \sum_i\frac{L_i}{\kappa_i}.
\end{equation}}
so the large-scale diffusivity is the length-weighted harmonic mean of the local diffusivities. A small number of high-resistance patches can therefore control the total flux even if they occupy only a small fraction of the volume.

The important point is that dominating the resistance is much easier than dominating the residence time. A low-diffusivity patch can regulate the global transport because every particle must eventually cross it, so its resistance enters the series sum above. But this does not imply that particles spend most of their time there. In our multiphase medium, slow regions act primarily as bottlenecks: particles can bounce off them many times before transmission, while still spending most of their time in the fast, volume-filling phase. High resistance therefore does not necessarily imply long residence time.

To quantify this distinction, we compare two measures extracted from the Monte Carlo trajectories: the fraction of the total diffusive resistance contributed by the low-diffusivity phase,
\begin{equation}
    f_R^{(l)}
    =
    \frac{\sum_{\rm slow} L_i/\kappa_i}{\sum_{\rm all} L_i/\kappa_i},
\end{equation}
and the corresponding residence-time fraction,
\begin{equation}
    f_\tau^{(l)}
    =
    \frac{\tau_{\rm slow}}{\tau_{\rm all}}.
\end{equation}
We then define the dominance ratio
\begin{equation}
    D \equiv \log_{10}\!\left(\frac{f_R^{(l)}}{f_\tau^{(l)}}\right).
\end{equation}
When \(D>0\), the slow patches dominate the macroscopic resistance more strongly than they dominate the residence-time budget; when \(D<0\), the converse is true.

Figure~\ref{fig:residence} summarizes this behavior across the parameter scan. The exact definitions of the control parameters \((q_R, q_\tau, \alpha, \Delta)\) are given in Appendix~\ref{app:parameters}, but their roles are simple. The parameter \(q_R\) controls how strongly the slow phase contributes to the total resistance, while \(q_\tau\) controls how strongly it contributes to the residence-time budget. The geometric parameters \(\alpha\) and \(\Delta\) control the breadth of the patch-size distribution: decreasing \(\alpha\) or increasing \(\Delta\) increases the importance of rare large structures. Across all of the cases shown, we find \(D>0\): the slow phase dominates the resistance budget more strongly than the residence-time budget.

This behavior follows from the different scalings of resistance and residence time. Roughly, the resistance contribution of a trap scales linearly with its size, \(R \sim L/\kappa\), whereas the time required to diffuse across it scales quadratically, \(t \sim L^2/\kappa\). As a result, shrinking the physical size of the slow patches suppresses their contribution to the residence-time budget much more strongly than their contribution to the total resistance. Small, highly scattering patches can therefore act as effective bottlenecks without monopolizing the total CR lifetime. We explain how this arises in terms of local vs. downstream resistance in Appendix \ref{app:residence}.

This distinction is physically important. If CRs spent most of their time inside slow, likely overdense traps, they would accumulate a disproportionate amount of grammage there. Instead, our results show that a realistic regime exists in which bottlenecks control the macroscopic transport while the particles still spend most of their time in the fast phase. In this sense, the slow regions regulate the throughput of the system without dominating its residence-time budget.

That CRs spend relatively little time in the slow regions that regulate their transport might seem counterintuitive. One might assume that slow regions impede transport because CRs are physically stuck in traps for a long time, which is true only when local resistance dominates. However, when downstream resistance dominates, the primary effect of slow regions is to act as bottlenecks with a low probability of transmission $T$. A CR must repeatedly bounce off the slow region $\sim 1/T$ times before successful transmission. This is directly analogous to how small, high-optical-depth patches control radiative transfer by repeatedly scattering photons, even though the photons spend most of their time in the volume-filling, low-opacity surrounding medium.

\subsection{Renewal Reset and the Breakdown of Harmonic Convergence}
\label{subsec:renewal}

Thus far, we have assumed that the multiphase environment is static, allowing particles to fully explore and traverse the slow regions. Under this assumption, the macroscopic transport inevitably asymptotes to the harmonic mean. However, in a realistic astrophysical plasma, this static approximation often breaks down. Magnetic structures and turbulent traps are continuously created and destroyed by the dynamic medium. Even if the magnetic structures are long-lived, CRs can undergo perpendicular, cross-field diffusion, jumping to an adjacent field line and effectively sampling a completely statistically independent environment. 

These processes act as ``renewal resets.'' When a reset occurs, the particle's history is wiped, and it begins a new walk from a randomly sampled phase. The macroscopic transport is then governed by a competition between two timescales: the characteristic trapping time of the slow phase $\tau_{\rm slow} \sim l^2/\kappa_l$ and the environmental reset time $\tau_{\rm renew}$. If $\tau_{\rm renew} \gg \tau_{\rm slow}$, particles have ample time to traverse the bottlenecks, and the transport safely converges to $\kappa_H$. However, if the environment resets rapidly ($\tau_{\rm renew} \lesssim \tau_{\rm slow}$), particles are liberated from the traps. This alters the long-time, large-scale behavior of the system, causing a breakdown of the harmonic convergence.

\begin{figure}
    \centering
    \includegraphics[width=0.99\linewidth]{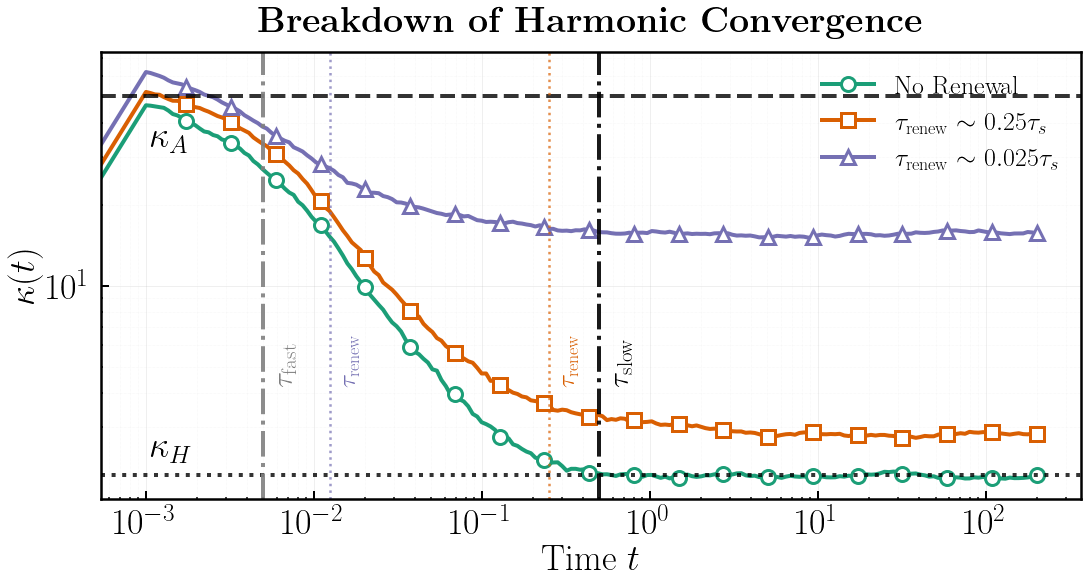}
    \caption{Running diffusion coefficient $\kappa(t)$ for a particle in a 1D alternating diffusivity landscape ($k_S=0.1, k_F=10.0$) with constant cell size. Without resetting (circles), $\kappa(t)$ converges to the harmonic mean (dotted). With fast resetting ($\tau_{\rm{renew}}\sim 0.02 \tau_{s}$, triangles), the environment renews rapidly, leading to breakdown of the convergence. Slow resetting ($\tau_{\rm{renew}}\sim 0.5 \tau_{s}$, squares) shows a crossover behavior.}
    \label{fig:renew}
\end{figure}

We demonstrate this quantitatively in Figure \ref{fig:renew}, which introduces a Poisson reset process into our 1D alternating diffusivity, constant cell size case. The running diffusion coefficient $\kappa(t)$ for the purely static case naturally decays from the arithmetic mean ($\kappa_A$) down to the harmonic mean ($\kappa_H$). When a slow reset $\tau_{\rm fast} < \tau_{\rm reset} < \tau_{\rm slow}$ is introduced (where $\tau_{\rm fast} \sim l^2/\kappa_{\rm h}$), the decay is interrupted, particles rarely experience the full delay of a trap, and the system reaches a steady-state effective diffusivity above $\kappa_H$. In the limit of fast resets, $\tau_{\rm reset} < \tau_{\rm fast}$, $\kappa(t)$ remains near the arithmetic limit. 
\red{Thus, within this one-dimensional Poisson reset model, the late time effective diffusivity remains above the static harmonic mean. More generally, dynamical environment or cross-field transport may prevent particles from fully sampling static bottlenecks and thereby keep the effective diffusivity above the corresponding static limit.}

\section{Discussion}
\label{sec:discussion}
An important theme that emerges from our work is that late-time diffusion is often the end state, not the whole story. If jump lengths are truncated and trapping times are finite, anomalous transport ultimately converges to standard diffusion \citep{liang25}. But physically interesting questions often concern the approach to that limit: how particles escape their birth environment, how bottlenecks shape the residence time distribution, and how different observables probe different parts of the transport. Those are the aspects encoded in $P(k,s)$ and in the memory kernel, not in the final asymptotic diffusion coefficient alone.

\subsection{Local Escape versus Global Transport}
\label{subsec:local_global}

A key advantage of the propagator-based picture is that it separates questions that are blurred together in the usual diffusion coefficient language. In a homogeneous Markovian medium, the same scalar $\kappa$ controls local escape, large-scale spreading, and the eventual residence-time budget. In a multi-phase, intermittent medium, these need not be the same question. Different observations can probe different stages of the same transport process, and can therefore infer different effective diffusion coefficients without requiring different underlying transport laws.

This point is especially relevant for the tension between very slow transport inferred near sources such as pulsar halos, supernova remnants, and molecular-cloud interfaces \citep[e.g.][]{abeysekara17,johannesson+19,yang23} and the much faster diffusion coefficient inferred on Galactic scales. In the framework developed here, a source embedded in a slow patch initially measures a local escape problem: particles must first leak through nearby bottlenecks before they sample the broader medium. The resulting near-source transport can therefore look much slower than the eventual large-scale transport, even when the late-time limit is ordinary diffusion. The important distinction is not simply that $\kappa$ varies from place to place, but that local and global measurements need not probe the same part of the propagator.

The bottleneck picture sharpens this further. In \S\ref{subsec:interpret}, slow regions can regulate escape much more easily than they dominate the total residence-time budget. In a static realization, the overall escape rate is set by the rate at which particles leak through bottlenecks, so even a modest amount of slow material can exert strong control over the transport. But particles need not actually reside mostly in the bottlenecks: they can spend much of their time in the fast, volume-filling phase while repeatedly failing to cross a small number of high-resistance barriers. Thus strong control of escape does not automatically imply that CRs spend most of their lifetime in dense or slow regions.

The static harmonic-mean limit should also be viewed with some caution. It is the natural late-time limit of a static medium that particles fully sample, but \S\ref{subsec:renewal} shows that it is not universal. If the trapping structures evolve on timescales comparable to particle transport, or if the particles undergo cross-field transport and hop onto a new field line, the effective large-scale transport can remain above the static limit. In that sense, the harmonic mean is best viewed as a limiting case for static bottlenecks, not as a generic prediction for time-dependent astrophysical turbulence.

\subsection{What Can Actually Test the Framework?}
\label{subsec:obs_tests}

Another payoff of the propagator is that it clarifies how different observables weight different parts of the transport. Secondary-to-primary ratios such as B/C are sensitive to the grammage, i.e. to the gas-density-weighted time integral of the CR distribution. In that sense they probe an integral of the propagator over both space and time, rather than a single diffusion coefficient. Radioactive clocks such as $^{10}$Be/$^{9}$Be weight the same residence-time distribution more strongly at long times through decay. Near-source systems probe the early-time local escape problem. Large-scale $\gamma$-ray emissivity and CR anisotropy are potentially sensitive to the spatial structure and geometry of transport. These are therefore not all measurements of the same thing, and one should not expect a single scalar $\kappa(E)$ to summarize them equally well.

This perspective suggests several concrete possibilities. Near-source slow diffusion may reflect local bottlenecks rather than a globally small effective diffusion coefficient. Radial variations in the mixture of fast and slow environments could help reshape the Galactic $\gamma$-ray emissivity profile relative to homogeneous-diffusion models. The energy dependence of the inferred transport law need not come only from a power-law $\kappa(E)$: it may also reflect a change in transport regime, for example if particles become progressively less trapped once their gyroradii exceed the scales of the structures that previously confined them. Likewise, the dipole amplitude and phase may depend not only on the effective large-scale diffusion coefficient, but also on the geometry and timing of escape and on which transport channels dominate at a given rigidity. We do not claim to resolve these problems here, but the propagator provides a more natural language in which to pose them.

The same logic also points to cleaner tests outside the Galactic steady-state problem. Heliophysics data, particularly in the Parker Solar Probe and Solar Orbiter era \citep{malandraki+23,whitman+23}, probe shorter timescales and more localized transport physics, and therefore offer a promising laboratory for non-diffusive effects. SEP dropouts, delayed cross-field escape, and trapping by coherent near-Sun structures \citep{mazur+00,giacalone+00,laitinen+16,engelbrecht+22} such as helical flux ropes \citep{pecora+21}, may provide a more direct window into broad waiting-time statistics and pre-asymptotic transport than integrated Galactic observables. Transient sources such as pulsar wind nebulae or UHECR echoes \citep{taylor23} provide another promising test, because arrival-time distributions and early-time CR profiles probe the propagator more directly than any late-time effective diffusivity.

\subsection{From Propagators to Predictive Models}
\label{subsec:program_outlook}

The value of the Montroll--Weiss framework is not only interpretive, but also practical. The Prony representation developed in \S\ref{subsec:approx_K} turns a broad memory kernel into a small number of local relaxation modes. This provides a tractable route for embedding history-dependent transport into two-fluid CR-MHD without storing the full past gradient at every point. In that sense, the framework does more than diagnose non-diffusive transport: it provides a route to using it.

In practice, this framework connects three levels of the problem. Propagators can be measured directly from particle trajectories in frozen or live MHD turbulence, and in kinetic particle-in-cell (PIC) simulations, so that trapping, mirroring, field-line wandering, and temporal decorrelation are absorbed into empirically calibrated kernels rather than guessed diffusion coefficients. Those kernels can then be confronted with observations, including grammage constraints, radioactive clocks, near-source transport measurements, and potentially anisotropy and gamma-ray structure. The same kernels can also be deployed in CR-MHD through Prony closures, allowing controlled comparisons between models with identical late-time diffusivities but different early-time transport histories.

That last point is especially important for CR feedback. In many dynamical problems, the crucial question is not the eventual asymptotic diffusivity, but where and when CR pressure is deposited before the transport has relaxed to that limit. Two models with the same late-time $\kappa_{\rm eff}$ can therefore produce different wind launching, mass loading, or non-thermal emission if their early-time transport differs. The propagator and memory kernel provide a compact way to carry that information into macroscopic models.

While our focus here is CR spatial transport, our approach could prove fruitful in other contexts, including hydrogen Ly$\alpha$ radiative transport \citep{dijkstra19}, radio-wave scattering through turbulent plasma \citep{boldyrev05} and CR momentum diffusion during acceleration \citep{lemoine20}.

\section{Conclusions}
\label{sec:conclusion}
\red{The main message of this paper is that a diffusion coefficient is not
the full transport law. A diffusion coefficient captures only the growth of the variance. Distinct transport processes can have the same
mean-squared displacement, or even the same asymptotic diffusion
coefficient, while differing substantially in their spatial distributions
and transient evolution. This additional information is contained in the
propagator $P(k,s)$, which describes transport simultaneously as a
function of spatial scale and timescale. It also reveals whether the
large-scale transport has \emph{memory}: whether the CR flux at a given
time depends only on the instantaneous density gradient, as in ordinary
diffusion, or also depends on gradients experienced at earlier
times. In the latter case, this history dependence is described by a
frequency-dependent transport kernel $\tilde K(s)$, or equivalently
a time-domain memory kernel $K(t)$. A frequency-dependent quantity
measured at finite $k$ represents genuine large-scale memory only if it
converges to a non-trivial, $k$-independent $\tilde K(s)$ as
$k\rightarrow0$.}

\red{We developed this framework using the Montroll--Weiss formalism, which coarse-grains unresolved microscopic motion into a sequence of transport events, each described by a step size and a waiting time between steps. We
applied it to a simple model of CR transport through a one-dimensional
multiphase medium composed of alternating regions with high and low
diffusivity, based on \citet{ewart25}. On scales much larger than an individual region, resolving
diffusion within each patch is both unnecessary and
computationally expensive. We therefore coarse-grain the motion into
first-passage events between neighboring cell interfaces. This provides a controlled example in which the microscopic
diffusion, the reduced event statistics, the propagator, and the resulting
macroscopic transport law can all be calculated and compared directly.}

\red{Our main conclusions are:}
\begin{itemize}
\item
\textbf{The propagator contains substantially more information than a
diffusion coefficient.}
For instance, particles can each undergo ordinary diffusion but with
different diffusivities. The ensemble then has an exactly linear
mean-squared displacement,
$\langle x^2(t)\rangle=2\langle D\rangle t$, while its propagator is
non-Gaussian and contains finite-scale structure that cannot be described
by a single diffusion coefficient. The propagator can likewise capture
transient subdiffusion or superdiffusion before eventual convergence
toward ordinary Gaussian diffusion \citep{liang25}. Thus processes that
look identical according to their mean-squared displacement are
readily distinguishable in $P(k,s)$.

\item
\textbf{Memory arises when dynamically relevant transport states are
hidden.}
The usual diffusion description is itself a coarse-grained, asymptotic
description: complicated microscopic transport is replaced by a diffusion
coefficient relating the instantaneous CR flux to the density gradient.
In doing so, information about the local transport state and its history
is discarded---for example, whether particles have recently sampled fast
or slow regions, magnetic traps, or other coherent structures. If these
unresolved states persist long enough to influence subsequent motion,
their elimination produces memory: the present flux can depend on the
past history of the density gradient even when the underlying microscopic
dynamics is Markovian. Our multiphase model provides a controlled example
of this general phenomenon. Coarse-graining the local diffusion into
interface-to-interface first-passage events makes the hidden-state
physics explicit through non-exponential waiting statistics and a
non-trivial memory kernel.

\item
\textbf{Hidden transport states require a multistate description, and
cannot in general be reconstructed from the scalar propagator $P(k,s)$ alone.}
$P(k,s)$ describes the observable spatial transport, but not any hidden internal states. When such states matter, their transition kernels cannot be recovered from $P(k,s)$ alone, and only the scalar large-scale kernel $\tilde{K}(s)$ can be extracted. When the statistics of the next
transport event depend on the current internal state, the forward
problem requires a multistate, matrix Montroll--Weiss description. We
demonstrated this in the multiphase medium, where the relevant states
are the local phase/interface configuration. In addition, when cell
sizes vary, the distance traveled in a coarse-grained step and the time
required for that step become coupled, so that the jump and waiting-time
statistics are no longer separable.

\item
\textbf{Slow regions can regulate transport without dominating the
residence-time budget.}
In a static multiphase medium, relatively small regions in which
transport is slow can act as bottlenecks that control the overall rate
of CR escape. However, this does not imply that particles spend most of
their time in those regions. Slow structures can therefore strongly
regulate the large-scale transport even when most of the CR residence
time is spent elsewhere.
 
\item
\textbf{The static late time limit need not be reached in a
time-dependent environment.}
If trapping structures evolve, or particles change field lines, before
they have fully sampled the static bottlenecks, the transport can be
continually reset before reaching the static asymptotic regime. In the
one-dimensional reset models explored here, this leaves the effective
diffusivity above the corresponding static harmonic-mean value. The
macroscopic transport therefore depends both on the spatial distribution
of scattering and trapping environments (e.g.\ magnetic mirrors or
field-line structures) and on their lifetimes relative to the particle
transport time.

\end{itemize}
\red{The accelerated scheme makes these calculation practical by replacing diffusion within each cell with a single first passage step, avoiding the small time step and speeding up the calculation by two to three orders of magnitude (Table~\ref{tab:benchmark}).} Taken together, these results provide a practical bridge between particle trajectories, full propagators, history-dependent transport, and reduced transport models. They also motivate a broader program. In future work, we will use this framework to infer transport kernels directly from particle tracing in realistic turbulent fields, confront those kernels with Galactic and heliospheric observations, and build CR-MHD closures that retain the physically important pre-asymptotic transport history rather than collapsing everything into a single effective diffusion coefficient.

\section*{Acknowledgements}

We thank Patrick Reichherzer, Robert Ewart, and Ka Wai Ho for stimulating conversations. \red{We also thank the anonymous referee for a particularly thoughtful and perceptive report which greatly improved the paper.} We acknowledge NSF grant AST240752 for support. 

\section*{Data Availability Statement}
The data supporting this study are available from the corresponding author upon reasonable request.


\begin{thebibliography}{}
\makeatletter
\relax
\def\mn@urlcharsother{\let\do\@makeother \do\$\do\&\do\#\do\^\do\_\do\%\do\~}
\def\mn@doi{\begingroup\mn@urlcharsother \@ifnextchar [ {\mn@doi@} {\mn@doi@[]}}
\def\mn@doi@[#1]#2{\def\@tempa{#1}\ifx\@tempa\@empty \href {http://dx.doi.org/#2} {doi:#2}\else \href {http://dx.doi.org/#2} {#1}\fi \endgroup}
\def\mn@eprint#1#2{\mn@eprint@#1:#2::\@nil}
\def\mn@eprint@arXiv#1{\href {http://arxiv.org/abs/#1} {{\tt arXiv:#1}}}
\def\mn@eprint@dblp#1{\href {http://dblp.uni-trier.de/rec/bibtex/#1.xml} {dblp:#1}}
\def\mn@eprint@#1:#2:#3:#4\@nil{\def\@tempa {#1}\def\@tempb {#2}\def\@tempc {#3}\ifx \@tempc \@empty \let \@tempc \@tempb \let \@tempb \@tempa \fi \ifx \@tempb \@empty \def\@tempb {arXiv}\fi \@ifundefined {mn@eprint@\@tempb}{\@tempb:\@tempc}{\expandafter \expandafter \csname mn@eprint@\@tempb\endcsname \expandafter{\@tempc}}}

\bibitem[\protect\citeauthoryear{{Abeysekara} et~al.,}{{Abeysekara} et~al.}{2017}]{abeysekara17}
{Abeysekara} A.~U.,  et~al., 2017, \mn@doi [Science] {10.1126/science.aan4880}, \href {https://ui.adsabs.harvard.edu/abs/2017Sci...358..911A} {358, 911}

\bibitem[\protect\citeauthoryear{{Ackermann} et~al.,}{{Ackermann} et~al.}{2011}]{ackermann11}
{Ackermann} M.,  et~al., 2011, \mn@doi [\apj] {10.1088/0004-637X/726/2/81}, \href {http://adsabs.harvard.edu/abs/2011ApJ...726...81A} {726, 81}

\bibitem[\protect\citeauthoryear{{Aharonian}, {Yang}  \& {de O{\~n}a Wilhelmi}}{{Aharonian} et~al.}{2019}]{aharonian+19}
{Aharonian} F.,  {Yang} R.,   {de O{\~n}a Wilhelmi} E.,  2019, \mn@doi [Nature Astronomy] {10.1038/s41550-019-0724-0}, \href {https://ui.adsabs.harvard.edu/abs/2019NatAs...3..561A} {3, 561}

\bibitem[\protect\citeauthoryear{{Almada Monter} \& {Gronke}}{{Almada Monter} \& {Gronke}}{2024}]{monter24}
{Almada Monter} S.,  {Gronke} M.,  2024, \mn@doi [\mnras] {10.1093/mnrasl/slae074}, \href {https://ui.adsabs.harvard.edu/abs/2024MNRAS.534L...7A} {534, L7}

\bibitem[\protect\citeauthoryear{Beylkin \& Monz{\'o}n}{Beylkin \& Monz{\'o}n}{2010}]{beylkin10}
Beylkin G.,  Monz{\'o}n L.,  2010, Applied and Computational Harmonic Analysis, 28, 131

\bibitem[\protect\citeauthoryear{{Bloemen}, {Dogiel}, {Dorman}  \& {Ptuskin}}{{Bloemen} et~al.}{1993}]{bloemen93}
{Bloemen} J.~B.~G.~M.,  {Dogiel} V.~A.,  {Dorman} V.~L.,   {Ptuskin} V.~S.,  1993, \aap, \href {https://ui.adsabs.harvard.edu/abs/1993A&A...267..372B} {267, 372}

\bibitem[\protect\citeauthoryear{{Boldyrev} \& {Gwinn}}{{Boldyrev} \& {Gwinn}}{2005}]{boldyrev05}
{Boldyrev} S.,  {Gwinn} C.~R.,  2005, \mn@doi [\apj] {10.1086/428919}, \href {https://ui.adsabs.harvard.edu/abs/2005ApJ...624..213B} {624, 213}

\bibitem[\protect\citeauthoryear{Bouchaud \& Georges}{Bouchaud \& Georges}{1990}]{bouchaud90}
Bouchaud J.-P.,  Georges A.,  1990, \mn@doi [Physics Reports] {https://doi.org/10.1016/0370-1573(90)90099-N}, 195, 127

\bibitem[\protect\citeauthoryear{{Butsky}, {Hopkins}, {Kempski}, {Ponnada}, {Quataert}  \& {Squire}}{{Butsky} et~al.}{2024}]{butsky24-CRscatter}
{Butsky} I.~S.,  {Hopkins} P.~F.,  {Kempski} P.,  {Ponnada} S.~B.,  {Quataert} E.,   {Squire} J.,  2024, \mn@doi [\mnras] {10.1093/mnras/stae276}, \href {https://ui.adsabs.harvard.edu/abs/2024MNRAS.528.4245B} {528, 4245}

\bibitem[\protect\citeauthoryear{Chechkin, Seno, Metzler  \& Sokolov}{Chechkin et~al.}{2017}]{chechkin+17}
Chechkin A.~V.,  Seno F.,  Metzler R.,   Sokolov I.~M.,  2017, \mn@doi [Physical Review X] {10.1103/physrevx.7.021002}, 7

\bibitem[\protect\citeauthoryear{Cressoni, Viswanathan, Ferreira  \& da Silva}{Cressoni et~al.}{2012}]{cressoni+12}
Cressoni J.~C.,  Viswanathan G.~M.,  Ferreira A.~S.,   da Silva M. A.~A.,  2012, \mn@doi [Phys. Rev. E] {10.1103/PhysRevE.86.022103}, 86, 022103

\bibitem[\protect\citeauthoryear{{Dijkstra}}{{Dijkstra}}{2019}]{dijkstra19}
{Dijkstra} M.,  2019, \mn@doi [Saas-Fee Advanced Course] {10.1007/978-3-662-59623-4_1}, \href {https://ui.adsabs.harvard.edu/abs/2019SAAS...46....1D} {46, 1}

\bibitem[\protect\citeauthoryear{Effenberger et~al.,}{Effenberger et~al.}{2025}]{effenberger+25}
Effenberger F.,  et~al., 2025, \mn@doi [Space Science Reviews] {10.1007/s11214-025-01203-4}, 221, 75

\bibitem[\protect\citeauthoryear{{Engelbrecht} et~al.,}{{Engelbrecht} et~al.}{2022}]{engelbrecht+22}
{Engelbrecht} N.~E.,  et~al., 2022, \mn@doi [\ssr] {10.1007/s11214-022-00896-1}, \href {https://ui.adsabs.harvard.edu/abs/2022SSRv..218...33E} {218, 33}

\bibitem[\protect\citeauthoryear{Evoli \& Dupletsa}{Evoli \& Dupletsa}{2023}]{evoli2023}
Evoli C.,  Dupletsa U.,  2023, Phenomenological models of {Cosmic} {Ray} transport in {Galaxies}, \mn@doi{10.48550/arXiv.2309.00298}, \url {https://ui.adsabs.harvard.edu/abs/2023arXiv230900298E}

\bibitem[\protect\citeauthoryear{{Ewart} et~al.,}{{Ewart} et~al.}{2025}]{ewart25}
{Ewart} R.~J.,  et~al., 2025, \mn@doi [arXiv e-prints] {10.48550/arXiv.2507.19044}, \href {https://ui.adsabs.harvard.edu/abs/2025arXiv250719044E} {p. arXiv:2507.19044}

\bibitem[\protect\citeauthoryear{{Gabici}, {Evoli}, {Gaggero}, {Lipari}, {Mertsch}, {Orlando}, {Strong}  \& {Vittino}}{{Gabici} et~al.}{2019}]{gabici19}
{Gabici} S.,  {Evoli} C.,  {Gaggero} D.,  {Lipari} P.,  {Mertsch} P.,  {Orlando} E.,  {Strong} A.,   {Vittino} A.,  2019, \mn@doi [International Journal of Modern Physics D] {10.1142/S0218271819300222}, \href {https://ui.adsabs.harvard.edu/abs/2019IJMPD..2830022G} {28, 1930022}

\bibitem[\protect\citeauthoryear{{Giacalone}, {Jokipii}  \& {Mazur}}{{Giacalone} et~al.}{2000}]{giacalone+00}
{Giacalone} J.,  {Jokipii} J.~R.,   {Mazur} J.~E.,  2000, \mn@doi [\apjl] {10.1086/312564}, \href {https://ui.adsabs.harvard.edu/abs/2000ApJ...532L..75G} {532, L75}

\bibitem[\protect\citeauthoryear{{Hopkins}}{{Hopkins}}{2025}]{hopkins25-review}
{Hopkins} P.~F.,  2025, \mn@doi [arXiv e-prints] {10.48550/arXiv.2509.07104}, \href {https://ui.adsabs.harvard.edu/abs/2025arXiv250907104H} {p. arXiv:2509.07104}

\bibitem[\protect\citeauthoryear{{Hopkins}, {Squire}, {Butsky}  \& {Ji}}{{Hopkins} et~al.}{2022}]{hopkins22}
{Hopkins} P.~F.,  {Squire} J.,  {Butsky} I.~S.,   {Ji} S.,  2022, \mn@doi [\mnras] {10.1093/mnras/stac2909}, \href {https://ui.adsabs.harvard.edu/abs/2022MNRAS.517.5413H} {517, 5413}

\bibitem[\protect\citeauthoryear{{J{\'o}hannesson}, {Porter}  \& {Moskalenko}}{{J{\'o}hannesson} et~al.}{2019}]{johannesson+19}
{J{\'o}hannesson} G.,  {Porter} T.~A.,   {Moskalenko} I.~V.,  2019, \mn@doi [\apj] {10.3847/1538-4357/ab258e}, \href {https://ui.adsabs.harvard.edu/abs/2019ApJ...879...91J} {879, 91}

\bibitem[\protect\citeauthoryear{{Kempski} \& {Quataert}}{{Kempski} \& {Quataert}}{2022}]{kempski22}
{Kempski} P.,  {Quataert} E.,  2022, \mn@doi [\mnras] {10.1093/mnras/stac1240}, \href {https://ui.adsabs.harvard.edu/abs/2022MNRAS.514..657K} {514, 657}

\bibitem[\protect\citeauthoryear{{Kempski}, {Fielding}, {Quataert}, {Galishnikova}, {Kunz}, {Philippov}  \& {Ripperda}}{{Kempski} et~al.}{2023}]{kempski23}
{Kempski} P.,  {Fielding} D.~B.,  {Quataert} E.,  {Galishnikova} A.~K.,  {Kunz} M.~W.,  {Philippov} A.~A.,   {Ripperda} B.,  2023, \mn@doi [\mnras] {10.1093/mnras/stad2609}, \href {https://ui.adsabs.harvard.edu/abs/2023MNRAS.525.4985K} {525, 4985}

\bibitem[\protect\citeauthoryear{{Kempski}, {Fielding}, {Quataert}, {Ewart}, {Grete}, {Kunz}, {Philippov}  \& {Stone}}{{Kempski} et~al.}{2025}]{kempski25}
{Kempski} P.,  {Fielding} D.~B.,  {Quataert} E.,  {Ewart} R.~J.,  {Grete} P.,  {Kunz} M.~W.,  {Philippov} A.~A.,   {Stone} J.,  2025, \mn@doi [arXiv e-prints] {10.48550/arXiv.2507.10651}, \href {https://ui.adsabs.harvard.edu/abs/2025arXiv250710651K} {p. arXiv:2507.10651}

\bibitem[\protect\citeauthoryear{{Laitinen}, {Kopp}, {Effenberger}, {Dalla}  \& {Marsh}}{{Laitinen} et~al.}{2016}]{laitinen+16}
{Laitinen} T.,  {Kopp} A.,  {Effenberger} F.,  {Dalla} S.,   {Marsh} M.~S.,  2016, \mn@doi [\aap] {10.1051/0004-6361/201527801}, \href {https://ui.adsabs.harvard.edu/abs/2016A&A...591A..18L} {591, A18}

\bibitem[\protect\citeauthoryear{{Lazarian} \& {Vishniac}}{{Lazarian} \& {Vishniac}}{1999}]{lazarian99}
{Lazarian} A.,  {Vishniac} E.~T.,  1999, \mn@doi [\apj] {10.1086/307233}, \href {https://ui.adsabs.harvard.edu/abs/1999ApJ...517..700L} {517, 700}

\bibitem[\protect\citeauthoryear{{Lazarian} \& {Xu}}{{Lazarian} \& {Xu}}{2021}]{lazarian21}
{Lazarian} A.,  {Xu} S.,  2021, \mn@doi [\apj] {10.3847/1538-4357/ac2de9}, \href {https://ui.adsabs.harvard.edu/abs/2021ApJ...923...53L} {923, 53}

\bibitem[\protect\citeauthoryear{{Lemoine}}{{Lemoine}}{2023}]{lemoine23}
{Lemoine} M.,  2023, \mn@doi [Journal of Plasma Physics] {10.1017/S0022377823000946}, \href {https://ui.adsabs.harvard.edu/abs/2023JPlPh..89e1701L} {89, 175890501}

\bibitem[\protect\citeauthoryear{{Lemoine} \& {Malkov}}{{Lemoine} \& {Malkov}}{2020}]{lemoine20}
{Lemoine} M.,  {Malkov} M.~A.,  2020, \mn@doi [\mnras] {10.1093/mnras/staa3131}, \href {https://ui.adsabs.harvard.edu/abs/2020MNRAS.499.4972L} {499, 4972}

\bibitem[\protect\citeauthoryear{{Liang} \& {Oh}}{{Liang} \& {Oh}}{2025}]{liang25}
{Liang} N.,  {Oh} S.~P.,  2025, \mn@doi [\mnras] {10.1093/mnras/staf1474}, \href {https://ui.adsabs.harvard.edu/abs/2025MNRAS.543.1911L} {543, 1911}

\bibitem[\protect\citeauthoryear{{L{\"u}bke}, {Reichherzer}, {Aerdker}, {Effenberger}, {Wilbert}, {Fichtner}  \& {Grauer}}{{L{\"u}bke} et~al.}{2025a}]{lubke25_mag}
{L{\"u}bke} J.,  {Reichherzer} P.,  {Aerdker} S.,  {Effenberger} F.,  {Wilbert} M.,  {Fichtner} H.,   {Grauer} R.,  2025a, \mn@doi [arXiv e-prints] {10.48550/arXiv.2505.18155}, \href {https://ui.adsabs.harvard.edu/abs/2025arXiv250518155L} {p. arXiv:2505.18155}

\bibitem[\protect\citeauthoryear{{L{\"u}bke}, {Effenberger}, {Wilbert}, {Fichtner}  \& {Grauer}}{{L{\"u}bke} et~al.}{2025b}]{lubke25b}
{L{\"u}bke} J.,  {Effenberger} F.,  {Wilbert} M.,  {Fichtner} H.,   {Grauer} R.,  2025b, \mn@doi [arXiv e-prints] {10.48550/arXiv.2509.15320}, \href {https://ui.adsabs.harvard.edu/abs/2025arXiv250915320L} {p. arXiv:2509.15320}

\bibitem[\protect\citeauthoryear{{Malandraki} et~al.,}{{Malandraki} et~al.}{2023}]{malandraki+23}
{Malandraki} O.~E.,  et~al., 2023, \mn@doi [Physics of Plasmas] {10.1063/5.0147683}, \href {https://ui.adsabs.harvard.edu/abs/2023PhPl...30e0501M} {30, 050501}

\bibitem[\protect\citeauthoryear{{Mazur}, {Mason}, {Dwyer}, {Giacalone}, {Jokipii}  \& {Stone}}{{Mazur} et~al.}{2000}]{mazur+00}
{Mazur} J.~E.,  {Mason} G.~M.,  {Dwyer} J.~R.,  {Giacalone} J.,  {Jokipii} J.~R.,   {Stone} E.~C.,  2000, \mn@doi [\apjl] {10.1086/312561}, \href {https://ui.adsabs.harvard.edu/abs/2000ApJ...532L..79M} {532, L79}

\bibitem[\protect\citeauthoryear{Metzler \& Klafter}{Metzler \& Klafter}{2000}]{metzler00}
Metzler R.,  Klafter J.,  2000, Physics reports, 339, 1

\bibitem[\protect\citeauthoryear{Montroll \& Weiss}{Montroll \& Weiss}{1965}]{montroll65}
Montroll E.~W.,  Weiss G.~H.,  1965, Journal of Mathematical Physics, 6, 167

\bibitem[\protect\citeauthoryear{Mori}{Mori}{1965}]{Mori1965}
Mori H.,  1965, \mn@doi [Progress of Theoretical Physics] {10.1143/PTP.33.423}, 33, 423

\bibitem[\protect\citeauthoryear{{Pecora} et~al.,}{{Pecora} et~al.}{2021}]{pecora+21}
{Pecora} F.,  et~al., 2021, \mn@doi [\mnras] {10.1093/mnras/stab2659}, \href {https://ui.adsabs.harvard.edu/abs/2021MNRAS.508.2114P} {508, 2114}

\bibitem[\protect\citeauthoryear{{Ragot} \& {Kirk}}{{Ragot} \& {Kirk}}{1997}]{ragot+97}
{Ragot} B.~R.,  {Kirk} J.~G.,  1997, \mn@doi [\aap] {10.48550/arXiv.astro-ph/9708041}, \href {https://ui.adsabs.harvard.edu/abs/1997A&A...327..432R} {327, 432}

\bibitem[\protect\citeauthoryear{{Ruszkowski} \& {Pfrommer}}{{Ruszkowski} \& {Pfrommer}}{2023}]{ruszkowski23}
{Ruszkowski} M.,  {Pfrommer} C.,  2023, \mn@doi [\aapr] {10.1007/s00159-023-00149-2}, \href {https://ui.adsabs.harvard.edu/abs/2023A&ARv..31....4R} {31, 4}

\bibitem[\protect\citeauthoryear{{Sampson}, {Beattie}, {Krumholz}, {Crocker}, {Federrath}  \& {Seta}}{{Sampson} et~al.}{2022}]{sampson22}
{Sampson} M.~L.,  {Beattie} J.~R.,  {Krumholz} M.~R.,  {Crocker} R.~M.,  {Federrath} C.,   {Seta} A.,  2022, arXiv e-prints, \href {https://ui.adsabs.harvard.edu/abs/2022arXiv220508174S} {p. arXiv:2205.08174}

\bibitem[\protect\citeauthoryear{Shin, Kim, Talkner  \& Lee}{Shin et~al.}{2010}]{shin+10}
Shin H.~K.,  Kim C.,  Talkner P.,   Lee E.~K.,  2010, \mn@doi [Chemical Physics] {10.1016/j.chemphys.2010.05.019}, 375, 316–326

\bibitem[\protect\citeauthoryear{{Taylor}, {Matthews}  \& {Bell}}{{Taylor} et~al.}{2023}]{taylor23}
{Taylor} A.~M.,  {Matthews} J.~H.,   {Bell} A.~R.,  2023, \mn@doi [\mnras] {10.1093/mnras/stad1716}, \href {https://ui.adsabs.harvard.edu/abs/2023MNRAS.524..631T} {524, 631}

\bibitem[\protect\citeauthoryear{{Whitman} et~al.,}{{Whitman} et~al.}{2023}]{whitman+23}
{Whitman} K.,  et~al., 2023, \mn@doi [Advances in Space Research] {10.1016/j.asr.2022.08.006}, \href {https://ui.adsabs.harvard.edu/abs/2023AdSpR..72.5161W} {72, 5161}

\bibitem[\protect\citeauthoryear{{Xu} \& {Yan}}{{Xu} \& {Yan}}{2013}]{xu13}
{Xu} S.,  {Yan} H.,  2013, \mn@doi [\apj] {10.1088/0004-637X/779/2/140}, \href {https://ui.adsabs.harvard.edu/abs/2013ApJ...779..140X} {779, 140}

\bibitem[\protect\citeauthoryear{{Yan} \& {Lazarian}}{{Yan} \& {Lazarian}}{2008}]{yan08}
{Yan} H.,  {Lazarian} A.,  2008, \mn@doi [\apj] {10.1086/524771}, \href {http://adsabs.harvard.edu/abs/2008ApJ...673..942Y} {673, 942}

\bibitem[\protect\citeauthoryear{Yan, Yan, Pavaskar, Hou  \& Liu}{Yan et~al.}{2026}]{yan+26}
Yan K.,  Yan H.,  Pavaskar P.,  Hou C.,   Liu R.-Y.,  2026, Non-Markovian Cosmic-Ray Pitch-Angle Transport from Mirror Interactions (\mn@eprint {arXiv} {2603.19037}), \url {https://arxiv.org/abs/2603.19037}

\bibitem[\protect\citeauthoryear{{Yang}, {Li}, {Wilhelmi}, {Cui}, {Liu}  \& {Aharonian}}{{Yang} et~al.}{2023}]{yang23}
{Yang} R.-z.,  {Li} G.-X.,  {Wilhelmi} E. d.~O.,  {Cui} Y.-D.,  {Liu} B.,   {Aharonian} F.,  2023, \mn@doi [Nature Astronomy] {10.1038/s41550-022-01868-9}, \href {https://ui.adsabs.harvard.edu/abs/2023NatAs...7..351Y} {7, 351}

\bibitem[\protect\citeauthoryear{{Zhang} \& {Xu}}{{Zhang} \& {Xu}}{2023}]{zhang+23}
{Zhang} C.,  {Xu} S.,  2023, \mn@doi [\apjl] {10.3847/2041-8213/ad0fe5}, \href {https://ui.adsabs.harvard.edu/abs/2023ApJ...959L...8Z} {959, L8}

\bibitem[\protect\citeauthoryear{{Zimbardo} \& {Perri}}{{Zimbardo} \& {Perri}}{2013}]{zimbardo+13}
{Zimbardo} G.,  {Perri} S.,  2013, \mn@doi [\apj] {10.1088/0004-637X/778/1/35}, \href {https://ui.adsabs.harvard.edu/abs/2013ApJ...778...35Z} {778, 35}

\bibitem[\protect\citeauthoryear{Zweibel}{Zweibel}{2017}]{zweibel17}
Zweibel E.~G.,  2017, \mn@doi [Physics of Plasmas] {10.1063/1.4984017}, 24, 055402

\makeatother
\end{thebibliography}
\input{revised_ref.bbl}

\appendix
\section{Two-Cell First-Passage Problem}
\label{sec:two-cell}
This appendix derives the analytic first-passage results used in \S\ref{subsec:two_cells}. Consider two adjacent cells with diffusivities $\kappa_l$ and $\kappa_h$ and widths $L_l$ and $L_h$, spanning $x=-L_l$ to $x=L_h$, with the interface at $x=0$ and absorbing boundaries at the two outer edges. In the accelerated Monte Carlo, a coarse-grained step is determined by two ingredients: the probability of exiting through either boundary, and the distribution of exit times.

As noted in the main text, the familiar L\'evy--Smirnov scaling applies to one-sided first passage on a semi-infinite domain, where arbitrarily long excursions generate a $t^{-3/2}$ tail. Here the particle is confined between two absorbing boundaries, so those long excursions are cut off: the short-time behavior is similar, but the late-time tail is exponentially truncated.

\subsection{Splitting probabilities}

Let $u(x)$ be the probability that a particle starting at $x$ exits through the right boundary before the left. Because this is a function of the starting position, it satisfies
\begin{equation}
    \frac{d}{dx}\!\left[\kappa(x)\frac{du}{dx}\right]=0,
\end{equation}
with
\begin{equation}
    \kappa(x)=
    \begin{cases}
        \kappa_l, & -L_l < x < 0,\\[3pt]
        \kappa_h, & 0 < x < L_h.
    \end{cases}
\end{equation}
The boundary conditions are
\begin{equation}
    u(-L_l)=0, \qquad u(L_h)=1,
\end{equation}
together with continuity at the interface,
\begin{equation}
    u(0^-)=u(0^+),
\end{equation}
and continuity of diffusive flux,
\begin{equation}
    \kappa_l\,u'(0^-)=\kappa_h\,u'(0^+).
\end{equation}
The solution is linear on each side, and evaluating it at the injection point $x=0$ gives the branching ratios
\begin{equation}
    \alpha_h \equiv u(0)=\frac{R_l}{R_l+R_h},\qquad
    \alpha_l \equiv 1-u(0)=\frac{R_h}{R_l+R_h},
    \label{eq:alphas_appendix}
\end{equation}
where
\begin{equation}
    R_i \equiv \frac{L_i}{2\kappa_i}
\end{equation}
is the diffusive resistance of half of cell $i$. Thus escape is biased toward the lower-resistance side. For equal cell sizes, $L_h=L_l$, this reduces to
\begin{equation}
    \alpha_h=\frac{\kappa_h}{\kappa_h+\kappa_l},
    \qquad
    \alpha_l=\frac{\kappa_l}{\kappa_h+\kappa_l}.
\end{equation}

\subsection{First-passage kernels}

Let $U_h(x,t)$ denote the probability density for first exit through the right boundary at time $t$, given a starting point $x$, and define its Laplace transform by
\begin{equation}
    U_h(x,s)=\int_0^\infty e^{-st}U_h(x,t)\,dt.
\end{equation}
As a function of the starting point $x$, it obeys
\begin{equation}
    s\,U_h(x,s)=\kappa(x)\frac{d^2U_h(x,s)}{dx^2},
\end{equation}
with boundary conditions
\begin{equation}
    U_h(-L_l,s)=0, \qquad U_h(L_h,s)=1,
\end{equation}
together with interface matching,
\begin{equation}
    U_h(0^-,s)=U_h(0^+,s),
\end{equation}
and
\begin{equation}
    \kappa_l\,\partial_x U_h(0^-,s)=\kappa_h\,\partial_x U_h(0^+,s).
\end{equation}

The solutions are elementary combinations of hyperbolic functions on each side. Writing
\begin{equation}
    \lambda_i=\sqrt{\frac{s}{\kappa_i}},\qquad
    a_i=\lambda_i L_i,\qquad
    A_i=\kappa_i\lambda_i=\sqrt{\kappa_i s},
\end{equation}
the interface-started kernel $U_h(s)\equiv U_h(0,s)$ is
\begin{equation}
    U_h(s)=
    \frac{A_h\,\mathrm{csch}(a_h)}
         {A_l\,\coth(a_l)+A_h\,\coth(a_h)}.
    \label{eq:Uh_appendix}
\end{equation}
By symmetry, the corresponding kernel for first exit through the left boundary is
\begin{equation}
    U_l(s)=
    \frac{A_l\,\mathrm{csch}(a_l)}
         {A_l\,\coth(a_l)+A_h\,\coth(a_h)}.
    \label{eq:Ul_appendix}
\end{equation}

These kernels reduce to the splitting probabilities in the $s\to0$ limit,
\begin{equation}
    U_h(0)=\alpha_h,\qquad U_l(0)=\alpha_l,
\end{equation}
and provide the basic building blocks for the coarse-grained Montroll--Weiss description used throughout \S\ref{sec:multiphase}.

\section{Analytic Expressions for the MW Propagator in a Multiphase Medium}
\label{sec:analytic-MW}

\subsection{General Interface-State Formalism: Quenched Propagator}
\label{subsec:matrix_MW}

For a general quenched one-dimensional multiphase medium, the cell lengths $L_i$ and diffusivities $\kappa_i$ vary from cell to cell, so a scalar renewal description is no longer sufficient. Instead, we treat the system as a Markov chain on the set of cell interfaces. Each interface defines an internal state, and each coarse-grained event consists of diffusing away from that interface until first passage to one of its two neighboring interfaces.

\red{We follow the convention introduced in \S\ref{subsec:generalization}: $\Phi_{ij}(k,s)$ denotes the Fourier--Laplace transform of the joint probability for a particle starting in state $i$ to wait a time $\tau$, make a jump $\Delta x$, and arrive in destination state $j$. If $g_i(k,s)$ denotes the contribution to the propagator conditioned on the the walk starting in interface state $i$, then the matrix MW equation is}
\begin{equation}
    \mathbf{g}(k,s)
    =
    \left[\mathbf{I}-\mathbf{\Phi}(k,s)\right]^{-1}
    \tilde{\mathbf{S}}(s),
    \label{eq:matrix_mw_general_app}
\end{equation}
with
\begin{equation}
    \tilde S_i(s)=\frac{1-\tilde\psi_i(s)}{s},
    \qquad
    \tilde\psi_i(s)=U_i^+(s)+U_i^-(s).
    \label{eq:Si_general_app}
\end{equation}

Because the transport is one-dimensional, a particle at interface $i$ can only jump to its immediate neighbors, $i+1$ and $i-1$. The only non-zero matrix elements are therefore
\red{
\begin{equation}
\Phi_{i,i+1}(k,s)=e^{ikL_i}\,U_i^{+}(s),\qquad
\Phi_{i,i-1}(k,s)=e^{-ikL_{i-1}}\,U_i^{-}(s),
\end{equation}
}
where $L_i$ is the length of the cell to the right of interface $i$, and $L_{i-1}$ is the length of the cell to the left. Here $U_i^\pm(s)$ are the Laplace-transformed first-passage time distributions for escape to the right $(+)$ or left $(-)$ from interface $i$, obtained from the local two-cell solution in Appendix~\ref{sec:two-cell}. In a periodic medium, indices are understood modulo the number of interfaces. Equation~(\ref{eq:matrix_mw_general_app}) is thus a tri-diagonal linear system and can be solved in $\mathcal{O}(N)$ time with standard sparse solvers.

Equation~(\ref{eq:matrix_mw_general_app}) describes the propagator once the dynamics has been coarse-grained to the interface states. \red{The resulting propagator is
therefore a coarse-grained interface propagator, intended to describe
transport on scales larger than an individual cell rather than the exact
particle-position propagator on sub-cell scales.} For a uniform initial particle distribution in the physical domain, however, the process does not begin at an interface. The first step is therefore different from subsequent renewal steps: a particle must first diffuse from a random point inside its initial cell to one of the two bounding interfaces. This is the theory counterpart of the modified-start accelerated Monte Carlo used in our simulations: the first waiting time is drawn from the first-exit distribution of a uniformly sampled point inside the initial cell, after which the standard interface-to-interface renewal dynamics resumes.

Consider a particle initially placed uniformly inside cell $i$, which lies between interfaces $i$ and $i+1$. Let $x\in[0,L_i]$ denote the particle position measured from the left interface of that cell, and define
\begin{equation}
    \lambda_i=\sqrt{\frac{s}{\kappa_i}},
    \qquad
    a_i=\lambda_i L_i.
\end{equation}
The Laplace transforms of the first-passage densities to the left and right interfaces are
\begin{equation}
    V_i^-(x,s)=\frac{\sinh[\lambda_i(L_i-x)]}{\sinh(a_i)},
    \qquad
    V_i^+(x,s)=\frac{\sinh(\lambda_i x)}{\sinh(a_i)}.
    \label{eq:V_pm_x}
\end{equation}
Averaging over a uniform starting point inside the cell gives the first-step kernels
\begin{equation}
\begin{aligned}
    V_{i,0}^-(k,s)
    &=
    \frac{1}{L_i}\int_0^{L_i} dx\, e^{-ikx}\,V_i^-(x,s) \\
    &=
    \frac{
        \lambda_i \coth(a_i)
        - \lambda_i e^{-ikL_i}\,\mathrm{csch}(a_i)
        - ik
    }{
        L_i(\lambda_i^2+k^2)
    },
\end{aligned}
\label{eq:V_i0_minus}
\end{equation}
\begin{equation}
\begin{aligned}
    V_{i,0}^+(k,s)
    &=
    \frac{1}{L_i}\int_0^{L_i} dx\, e^{ik(L_i-x)}\,V_i^+(x,s) \\
    &=
    \frac{
        \lambda_i \coth(a_i)
        - \lambda_i e^{ikL_i}\,\mathrm{csch}(a_i)
        + ik
    }{
        L_i(\lambda_i^2+k^2)
    }.
\end{aligned}
\label{eq:V_i0_plus}
\end{equation}
At $k=0$, these reduce to
\begin{equation}
    V_{i,0}^\pm(0,s)
    =
    \frac{\cosh(a_i)-1}{a_i \sinh(a_i)},
    \label{eq:V_i0_k0}
\end{equation}
so the Laplace transform of the first waiting-time distribution is
\begin{equation}
    \tilde\psi_{i,0}(s)
    =
    V_{i,0}^-(0,s)+V_{i,0}^+(0,s)
    =
    \frac{2\left[\cosh(a_i)-1\right]}{a_i \sinh(a_i)}.
    \label{eq:psi_i0}
\end{equation}

The corresponding contribution to the propagator before the first interface encounter is
\begin{equation}
    \tilde S_{i,0}(k,s)
    =
    \frac{1}{L_i}\int_0^{L_i} dx
    \int_0^{L_i} dy\, e^{ik(y-x)}\,\tilde G_i^{\rm abs}(y,x;s),
    \label{eq:S0_general}
\end{equation}
where
\begin{equation}
    \tilde G_i^{\rm abs}(y,x;s)
    =
    \frac{
        \sinh\!\bigl[\lambda_i \min(x,y)\bigr]\,
        \sinh\!\bigl[\lambda_i(L_i-\max(x,y))\bigr]
    }{
        \kappa_i \lambda_i \sinh(a_i)
    }
\end{equation}
is the Laplace-space absorbing propagator inside cell $i$ with absorbing boundaries at $y=0$ and $y=L_i$. At $k=0$,
\begin{equation}
    \tilde S_{i,0}(0,s)
    =
    \frac{1-\tilde\psi_{i,0}(s)}{s}.
    \label{eq:S_i0}
\end{equation}

For a uniform spatial injection, the probability of starting in cell $i$ is proportional to its length,
\begin{equation}
    f_i=\frac{L_i}{\sum_j L_j}.
    \label{eq:fi_general}
\end{equation}
The full propagator is then
\begin{equation}
    P(k,s)
    =
    \sum_i f_i
    \Bigl[
        \tilde S_{i,0}(k,s)
        + V_{i,0}^-(k,s)\,g_i(k,s)
        + V_{i,0}^+(k,s)\,g_{i+1}(k,s)
    \Bigr].
    \label{eq:P_general_firststep}
\end{equation}
Equation~(\ref{eq:P_general_firststep}) is the correctly initialized propagator for the same coarse-grained process used in the modified-start accelerated Monte Carlo. \red{The explicit birth-cell term in Eq.~(B14) resolves the initial excursion
before the first interface encounter; subsequent excursions remain
coarse-grained at the interface level.}

If the distinction between the initial escape from the interior of the starting cell and subsequent renewal steps is neglected, Eq.~(\ref{eq:P_general_firststep}) reduces to the interface-weighted approximation
\begin{equation}
    P(k,s)\approx \sum_i p_i^{(0)} g_i(k,s),
    \qquad
    p_i^{(0)}=\frac{L_{i-1}+L_i}{2\sum_j L_j}.
    \label{eq:interface_weight_approx}
\end{equation}
This approximation is asymptotically correct at late times, but it does not capture the short-time first-step correction associated with a uniform spatial initialization.

\subsection{Constant Cell Sizes as an $N=2$ Special Case}
\label{subsec:scalar_MW}

For an alternating medium with constant high- and low-diffusivity cells, the general interface-state formalism reduces to a two-state problem. Let $A$ denote an interface with a low-diffusivity cell on the left and a high-diffusivity cell on the right, and let $B$ denote the opposite interface. After the first interface encounter, the process alternates deterministically between these two interface states. Equations~(\ref{eq:matrix_mw_general_app})--(\ref{eq:P_general_firststep}) then reduce to the expressions below.

The interface-to-interface kernels are
\begin{eqnarray}
    \Phi_A(k,s)&=&e^{ikL_h}U_h(s)+e^{-ikL_l}U_l(s),
    \label{eq:Phi_A}
    \\
    \Phi_B(k,s)&=&e^{ikL_l}U_l(s)+e^{-ikL_h}U_h(s),
    \label{eq:Phi_B}
\end{eqnarray}
with
\begin{equation}
    \tilde S_{A/B}(s)=\frac{1-\Phi_{A/B}(0,s)}{s}.
    \label{eq:SA_SB}
\end{equation}
The propagators conditioned on starting at interface $A$ or $B$ are
\begin{equation}
\begin{aligned}
    P_A(k,s)
    &=
    \frac{\tilde S_A(s)+\Phi_A(k,s)\tilde S_B(s)}
         {1-\Phi_A(k,s)\Phi_B(k,s)},\\
    P_B(k,s)
    &=
    \frac{\tilde S_B(s)+\Phi_B(k,s)\tilde S_A(s)}
         {1-\Phi_A(k,s)\Phi_B(k,s)}.
\end{aligned}
\label{eq:PAB_corrected}
\end{equation}

The first-step kernels are given by Eqs.~(\ref{eq:V_i0_minus})--(\ref{eq:V_i0_plus}) with $i\rightarrow h,l$, where
\begin{equation}
    \lambda_\nu=\sqrt{\frac{s}{\kappa_\nu}},
    \qquad
    a_\nu=\lambda_\nu L_\nu,
    \qquad \nu\in\{h,l\}.
\end{equation}
The corresponding first waiting-time transforms and survival terms are
\begin{equation}
    \tilde\psi_{\nu,0}(s)=V_{\nu,0}^-(0,s)+V_{\nu,0}^+(0,s),
    \quad
    \tilde S_{\nu,0}(0,s)=\frac{1-\tilde\psi_{\nu,0}(s)}{s},
    \label{eq:Shl0}
\end{equation}
where $\nu\in\{h,l\}$. For finite $k$, the quantities $\tilde S_{h,0}(k,s)$ and $\tilde S_{l,0}(k,s)$ are defined by Eq.~(\ref{eq:S0_general}) using the corresponding absorbing propagators for the high- and low-diffusivity cells.

If a particle starts in a high-diffusivity cell, its first exit to the left reaches interface $A$, while its first exit to the right reaches interface $B$. Thus
\begin{equation}
    Q_h(k,s)
    =
    \tilde S_{h,0}(k,s)
    + V_{h,0}^-(k,s)\,P_A(k,s)
    + V_{h,0}^+(k,s)\,P_B(k,s).
    \label{eq:Qh}
\end{equation}
Similarly, for a particle starting in a low-diffusivity cell,
\begin{equation}
    Q_l(k,s)
    =
    \tilde S_{l,0}(k,s)
    + V_{l,0}^-(k,s)\,P_B(k,s)
    + V_{l,0}^+(k,s)\,P_A(k,s).
    \label{eq:Ql}
\end{equation}
The correctly initialized propagator is then
\begin{equation}
    P(k,s)=f_h\,Q_h(k,s)+f_l\,Q_l(k,s),
    \quad
    f_i=\frac{L_i}{L_h+L_l},
    \label{eq:P_const_firststep}
\end{equation}
and $i\in\{h,l\}$. For equal cell sizes, $f_h=f_l=1/2$.

If the first-step correction is neglected, Eq.~(\ref{eq:P_const_firststep}) reduces to the interface-averaged expression used previously. The difference appears primarily at large $s$, where short-time behavior is sensitive to the fact that particles are initialized in cell interiors rather than at renewal interfaces. This correction leaves the long-time behavior unchanged, but it can be important for derived quantities such as $\tilde K(s)$, where small high-$s$ differences in $P(k,s)$ are amplified by the inversion.

\subsection{Variable Cell Size: Annealed Propagator}

The variable cell calculation in Appendix \ref{subsec:matrix_MW} is for the quenched case, i.e., a particle diffusing in a fixed realization of the cell sequence, rather than the annealed case, where the new cell properties are randomly chosen each time the particle crosses an interface\footnote{In the constant cell size case, there is no distinction between the quenched and annealed propagator, since cell properties are periodic rather than random.}. The quenched case is more realistic, but at least for the region of parameter space we have explored, the difference between the quenched and annealed propagators is relatively modest (see Fig \ref{fig:ann_quenched_Ks}). The annealed case is analytically much simpler, because the two-state interface formalism of \S\ref{subsec:scalar_MW} closes after averaging the interface kernels over the cell-size distributions.

Specifically, one replaces the kernels in Eqs.~(\ref{eq:Phi_A})--(\ref{eq:Phi_B}) by their average over the cell-size distribution,
\begin{equation}
    \langle \Phi_{A/B}(k,s) \rangle
    \equiv
    \int dL_l\,p_l(L_l)\int dL_h\,p_h(L_h)\,
    \Phi_{A/B}(k,s;L_l,L_h),
    \label{eq:Phi_ann}
\end{equation}
where $\Phi_{A/B}(k,s;L_l,L_h)$ are the fixed-length expressions from Eqs.~(B16)--(B17), evaluated for a given local pair $(L_l,L_h)$. The corresponding survival terms and interface-conditioned propagators are then obtained directly from Eqs.~(B18)--(B19) with the replacement
\begin{equation}
    \Phi_{A/B}(k,s)\;\rightarrow\;\langle \Phi_{A/B}(k,s)\rangle.
\end{equation}
For a uniform spatial initialization, the first-step correction is treated in the same way: one averages Eqs.~(B22)--(B23) over the length-biased starting-cell distributions, and then uses Eq.~(B24) with
\begin{equation}
    f_i=\frac{\langle L_i\rangle}{\langle L_h\rangle+\langle L_l\rangle},
    \qquad i\in\{h,l\}.
\end{equation}

\section{\red{Finite k Convergence of Transport Kernel}}
\label{subsec:multiple_k}
\begin{figure*}
    \centering
    \includegraphics[width=0.8\linewidth]{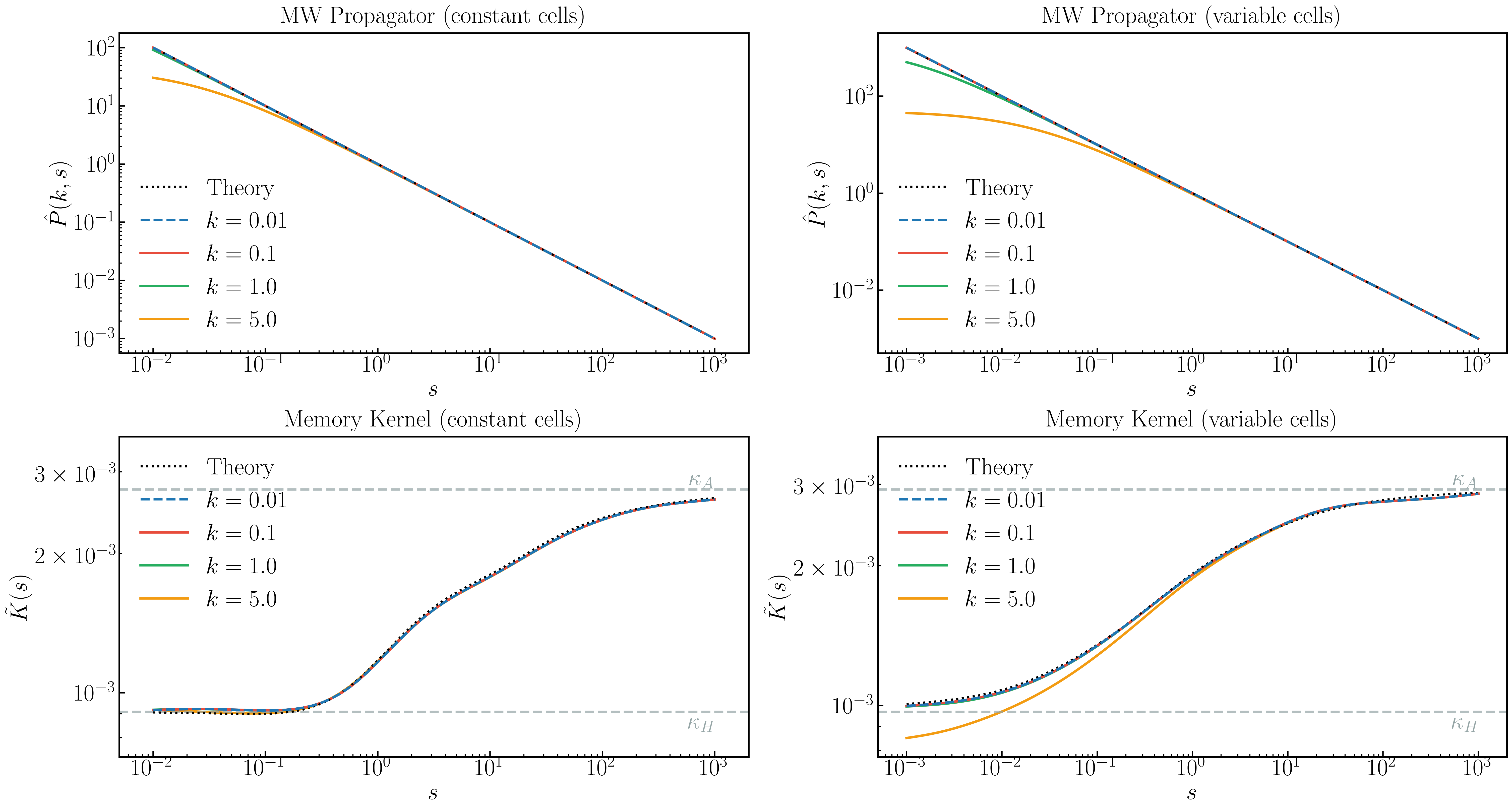}
    \caption{\red{Finite-$k$ convergence of the transport kernel for constant cells (left) and variable cells (right). Colored curves show the propagator $\hat P(k,s)$ (top) and extracted kernel $\tilde K(s)$ (bottom) for $k=0.01$, $0.1$, $1$, and $5$. Dotted curves show the analytic prediction at $k=0.1$. The kernels converge once $k$ is small compared with the inverse cell size. Relative to $k=0.01$, the $k=0.1$ and $k=1$ results agree to within $0.5\%$, while $k=5$ shows finite-$k$ deviations of $2\%$ for constant cells and $15\%$ for variable cells.}}
    \label{fig:k_test}
\end{figure*}

\red{The MW memory kernel $\tilde{K}(s)$ is formally defined in the large-scale, coarse-grained limit ($k \to 0$). However, when extracting these properties empirically from MC trajectories, one must compute the Fourier-Laplace transforms at finite extraction wavenumbers $k$. }

To reliably extract memory effects, one must ensure the extraction is performed in the asymptotic low-$k$ regime where the results become scale-independent. If the chosen $k$ is too large, the corresponding spatial scale $\lambda = 2\pi/k$ becomes comparable to the microscopic correlation lengths or individual patch sizes of the medium. 

\red{More quantitatively, the long-wavelength expansion may be written as
\begin{equation}
P^{-1}(k,s)=s+k^2\tilde K(s)+k^4\tilde K_4(s)+O(k^6),
\end{equation}
so that
$\tilde K_{\rm eff}(k,s)=\tilde K(s)
+k^2\tilde K_4(s)+O(k^4)$.
The upper end of the useful $k$ range is therefore set by requiring
$k^2|\tilde K_4/\tilde K|\ll1$, where $\tilde K(s), \tilde K_4(s)$ are obtained from direct fits to the data. There is also a practical
lower limit: as $k\rightarrow0$, $P\rightarrow1/s$ and statistical
errors in $P$ are strongly amplified when recovering $\tilde K$ as $1-sP\simeq k^2\tilde K/s$ becomes small
(see Eq.~\ref{eq:delta_K}). For a target fractional kernel error $\eta_{\rm K}$ and
fractional propagator uncertainty $\epsilon_P$, one approximately
requires
\begin{equation}
\frac{\epsilon_P s}{k^2\tilde K(s)} \lesssim \eta_{\rm K}.
\end{equation}
Thus $\tilde K(s)$ is best measured in an intermediate
``hydrodynamic window'' where $k$ is small enough for the $O(k^4)$
corrections to be negligible, but large enough that the $O(k^2)$
transport signal remains measurable. The curve collapse shown below
empirically identifies this window.}

\red{To systematically verify the robustness of our reconstruction and determine this valid regime, we compute the constant cell MW propagator $P(k,s)$ from accelerated MC trajectories across multiple wavenumbers $k \in \{0.01,0.1, 1.0, 5.0\}$, for the constant cell medium of Fig.~\ref{fig:const_all} and for a variable cell medium.} As shown in Figure \ref{fig:k_test}, the measurements are sensitive to $k$ at small spatial scales. \red{At $k=5$ the kernel departs from the small $k$ limit by $2\%$ for constant cells and by $15\%$ for variable cells, whose largest patches are ten times bigger. As $k$ decreases to $1.0$ and below, the curves cleanly collapse onto a single macroscopic master curve that matches the analytic prediction at $k=0.1$.}

\section{Parameter Scan for $\tilde{K}(\lowercase{s})$ properties: width and recoverability }
\label{app:parameters}

This appendix collects the precise definitions of the control parameters used in the variable-cell parameter scan, and summarizes how they affect the width of the Laplace-space kernel $\tilde{K}(s)$ and the range over which that kernel can be reliably recovered from $P(k,s)$.

We consider a two-phase medium with high- and low-diffusivity patches, $\kappa_h$ and $\kappa_l$, whose lengths are drawn from truncated power-law distributions. The four control parameters are:
\begin{itemize}
    \item \textbf{Spatial distribution exponent ($\alpha$):} the power-law exponent characterizing the distribution of patch lengths,
    \begin{equation}
        P(L) \propto L^{-\alpha}, \qquad L\in[L_{\min},L_{\max}] .
    \end{equation}

    \item \textbf{Spatial dynamic range ($\Delta$):} the ratio of the maximum to minimum patch length,
    \begin{equation}
        \Delta \equiv \frac{L_{\max}}{L_{\min}} .
    \end{equation}
    Both $\alpha$ and $\Delta$ are taken to be the same in the high- and low-diffusivity phases.

    \item \textbf{Resistance scaling parameter ($q_R$):} the ratio of the characteristic resistances of the slow and fast phases,
    \begin{equation}
        q_R \equiv \frac{R_l}{R_h}
        = \frac{\langle L_l\rangle}{\langle L_h\rangle}\,\frac{\kappa_h}{\kappa_l} .
    \end{equation}

    \item \textbf{Residence-time scaling parameter ($q_\tau$):} the ratio of the characteristic diffusive crossing times in the slow and fast phases,
    \begin{equation}
        q_\tau \equiv \frac{t_l}{t_h}
        = \frac{\langle L_l^2\rangle}{\langle L_h^2\rangle}\,\frac{\kappa_h}{\kappa_l} .
    \end{equation}
\end{itemize}

To quantify how these parameters affect the Laplace-space kernel $\tilde K(s)$, obtained from equation~\ref{eq:ks-limit}, we use two summary statistics: the memory width, $W_K$, and the valid recovery window, $\Delta\log_{10}s_{\rm valid}$.

The memory width measures the span of the crossover between the asymptotic diffusion coefficients $\kappa_A$ and $\kappa_H$ in Laplace space,
\begin{equation}
    W_K \equiv \log_{10}\!\left(\frac{s_{90}}{s_{10}}\right) .
\end{equation}
Here $s_{10}$ and $s_{90}$ are defined from the normalized logarithmic kernel
\begin{equation}
    m(s) \equiv \frac{\log \tilde K(s)-\log \kappa_H}
    {\log \kappa_A-\log \kappa_H} ,
\end{equation}
such that $m(s_{10})=0.1$ and $m(s_{90})=0.9$.

The recovery window $\Delta\log_{10}s_{\rm valid}$ measures the frequency range over which the kernel can be reliably recovered from the propagator in the small-$k$ limit. Using $\tilde K(s;k)$ extracted from equation~\ref{eq:Ks_from_P}, we define
\begin{equation}
    \delta_K(s)\equiv
    \frac{\mathrm{std}_{k}\!\left[\tilde K(s;k)\right]}
         {\overline{\tilde K}(s)},
    \qquad
    \overline{\tilde K}(s)\equiv \langle \tilde K(s;k)\rangle_{k},
\end{equation}
where the standard deviation and mean are taken over several small extraction wavenumbers $k$. We then define $\Delta\log_{10}s_{\rm valid}$ as the span in $\log_{10}s$ over which
\begin{equation}
    \delta_K(s) < \delta_{\rm crit},
\end{equation}
with $\delta_{\rm crit}\approx 0.2$.

\begin{figure}
    \centering
    \includegraphics[width=\linewidth]{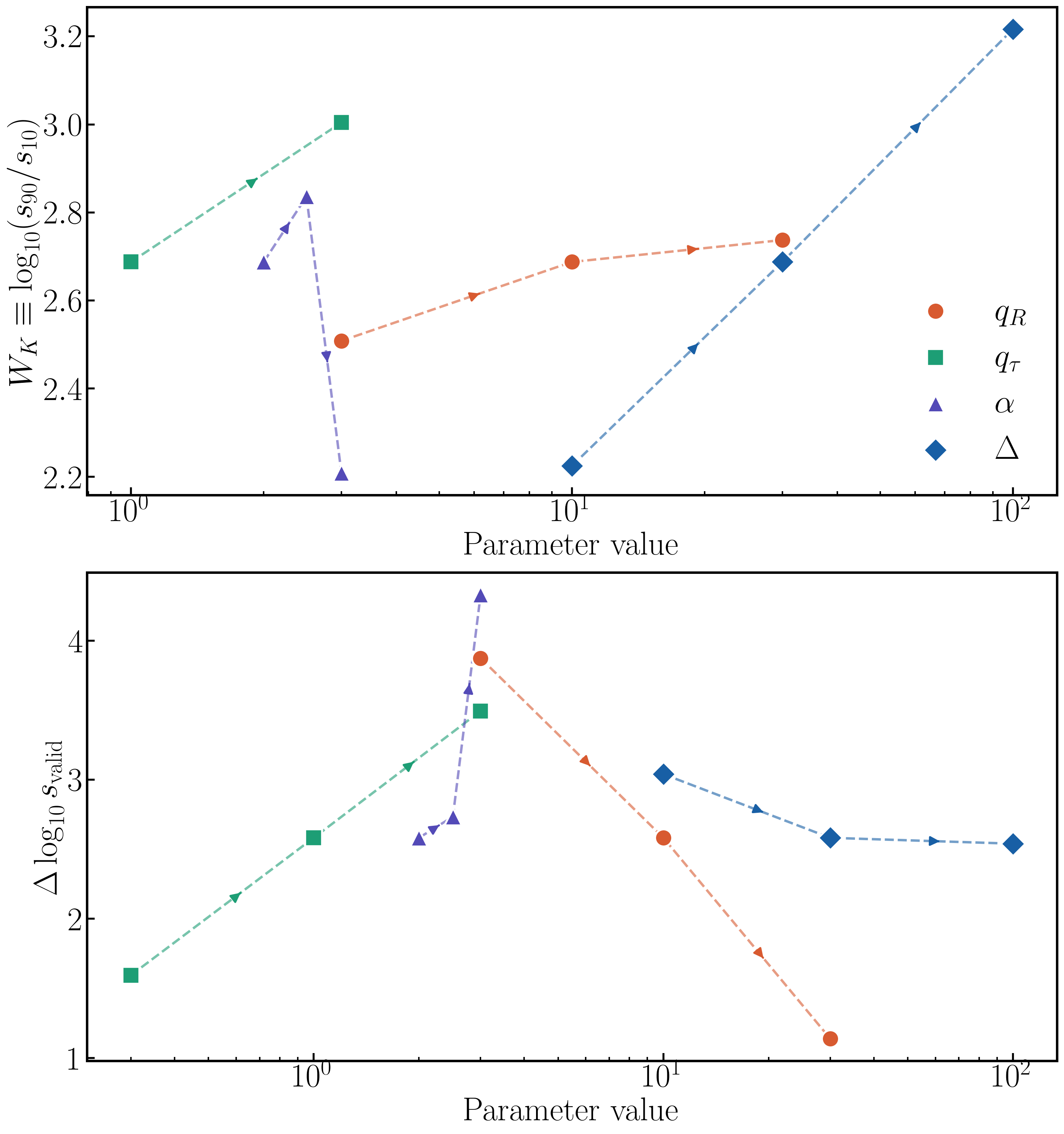}
    \caption{
    \emph{Top:} dependence of the memory width $W_K$ on the control parameters $q_R$, $q_\tau$, $\alpha$, and $\Delta$. Arrows indicate the direction of increasing parameter value along each sweep. \emph{Bottom:} recoverability of the Laplace-space kernel $\tilde K(s)$ from the MW propagator $P(k,s)$, quantified by the span $\Delta\log_{10}s_{\rm valid}$ over which the extracted kernel is $k$-independent, defined by $\delta_K<0.2$ across several small $k$. Note that both axes are logarithmic.
    }
    \label{fig:memory_scan}
\end{figure}

The top panel of Figure~\ref{fig:memory_scan} shows how the physical parameters of the multiphase medium change the width of the memory kernel. Because the characteristic diffusion time across a patch scales as $\tau\sim L^2/\kappa$, increasing the dynamic range $\Delta=L_{\max}/L_{\min}$ broadens the hierarchy of relaxation timescales and therefore increases $W_K$. Conversely, increasing the spatial exponent $\alpha$ shifts the medium toward smaller patches, compressing the range of trap lifetimes and narrowing the memory kernel. By contrast, varying $q_R$ and $q_\tau$ changes $W_K$ less strongly, since these parameters primarily shift the asymptotic endpoints $(\kappa_A,\kappa_H)$ rather than the breadth of the crossover itself.

The bottom panel shows that recoverability is tied to the shape of the kernel, especially its overall contrast and width. If the contrast between $\kappa_A$ and $\kappa_H$ is large, or if the kernel spans many decades in frequency, then the asymptotic long-wavelength limit required for the approximation
\begin{equation}
    P(k,s)\approx [s+k^2\tilde K(s)]^{-1}
\end{equation}
is pushed to smaller $k$ and smaller $s$. This delays the onset of scale-independent transport and shrinks the valid extraction window $\Delta\log_{10}s_{\rm valid}$.

Because the axes in Figure~\ref{fig:memory_scan} are logarithmic, changes in these diagnostics can be fairly substantial. Larger $q_R$ and larger $\Delta$ reduce recoverability because they increase the kernel contrast and broaden the hierarchy of trap lifetimes, respectively. By contrast, larger $q_\tau$ and steeper $\alpha$ improve recoverability: increasing $q_\tau$ reduces the disparity between the asymptotic diffusivities, while increasing $\alpha$ suppresses rare large traps and narrows the spectrum of relaxation times.

\section{CR Residence Times}
\label{app:residence}

This appendix makes explicit why a phase can dominate the total diffusive resistance without dominating the residence-time budget, as discussed in \S\ref{subsec:interpret}. Consider a static one-dimensional chain of patches, with injection on the left and escape at $x=L$. In steady state, the diffusive flux $J$ is constant, and within patch $i$ of length $L_i$ and diffusivity $\kappa_i$, Fick's law gives
\begin{equation}
    J = -\kappa_i \frac{dn}{dx}.
\end{equation}
Writing the patch as $x_i < x < x_{i+1}$ with $x_{i+1}-x_i=L_i$, the density profile is therefore
\begin{equation}
    n_i(x) = n_{i+1} + \frac{J}{\kappa_i}(x_{i+1}-x),
\end{equation}
where $n_{i+1}\equiv n(x_{i+1})$ is the density just to the right of the patch.

The column density stored in patch $i$ is
\begin{equation}
    N_i \equiv \int_{x_i}^{x_{i+1}} n_i(x)\,dx
    = \frac{J L_i^2}{2\kappa_i} + n_{i+1}L_i.
\end{equation}
It is convenient to define the downstream resistance
\begin{equation}
    R_i^{\rm d} \equiv \frac{n_{i+1}}{J}
    = \int_{x_{i+1}}^{L}\frac{dx}{\kappa(x)},
\end{equation}
together with the local half-patch resistance
\begin{equation}
    R_i^{\rm loc} \equiv \frac{L_i}{2\kappa_i}.
\end{equation}
Then
\begin{equation}
    N_i = J\,L_i\left(R_i^{\rm loc}+R_i^{\rm d}\right).
    \label{eq:Ni_appendix}
\end{equation}

In steady state, the column density stored in a patch is simply the throughput times the mean dwell time there: a region contains a lot of material either because the flux through it is large or because particles spend a long time inside it. Since the diffusive flux $J$ is the same through every patch in a static one-dimensional chain, the fraction of the total residence time spent in patch $i$ is equal to the fraction of the total column density stored there,
\begin{equation}
    \frac{t_i}{t_{\rm tot}}=\frac{N_i}{N_{\rm tot}}.
\end{equation}
This steady-state identity is known as Little's law. Equation~(\ref{eq:Ni_appendix}) therefore gives the residence-time weighting directly. In particular, when the downstream resistance dominates the local one,
\begin{equation}
    R_i^{\rm d} \gg R_i^{\rm loc},
\end{equation}
one finds
\begin{equation}
    N_i \approx J\,L_i R_i^{\rm d},
\end{equation}
so the residence time in that patch depends mainly on its size and on the resistance of the material downstream, and only weakly on its own diffusivity. This regime is typical for interior patches in a long chain; only very near the escape boundary does the local term become comparable.

Summing over all low-diffusivity patches gives the slow-phase residence fraction
\begin{equation}
    f_\tau^{(l)}
    = \frac{\sum_{i\in l} N_i}{\sum_{\rm all} N_i}
    = \frac{\sum_{i\in l} L_i\left(R_i^{\rm loc}+R_i^{\rm d}\right)}
           {\sum_{\rm all} L_i\left(R_i^{\rm loc}+R_i^{\rm d}\right)}.
\end{equation}
This makes clear why the slow phase can contribute strongly to the total series resistance while contributing only modestly to the total residence-time budget, consistent with the Monte Carlo diagnostics discussed in \S\ref{subsec:interpret}. These arguments apply to a static quenched medium; renewal resets (\S\ref{subsec:renewal}) break the steady-state series-resistance picture.

\red{
\section{Numerical Methods and Convergence}
\label{app:numerics}
We collect here the numerical details needed to assess the central results: the
continuous-time random walk (CTRW) simulations, the Laplace transform and kernel extraction, the Prony
fit, and the convergence with respect to particle number, run length, wavenumber, and mode count. The
memory kernel of \S3 and the multiphase transport of \S4 share the same estimator (Eq.~\ref{eq:Ks_from_P}).}

\subsection{CTRW simulations}
\label{app:ctrw}
Each walker takes Gaussian jumps of variance $\sigma^2$ separated by waiting times drawn from a
Pareto law $\psi(\tau)\propto\tau^{-1-\alpha}$ ($0<\alpha<1$) with lower cutoff $\tau_0$. We initialize a
single Fourier mode, $n(x,0)\propto 1+\epsilon\cos(kx)$ with $k=2\pi/L$, and evolve $N$ independent
walkers by an event-driven scheme, recording the propagator
$P(k,t)=n_k(t)/n_k(0)$ on a logarithmically spaced time grid on $[0,T_{\max}]$, where $n_k(t)$ is given by equation \ref{eq:nk_measure}. Log-time
sampling is adequate because $P(k,t)$ relaxes smoothly. It reduces the transform cost by two orders of
magnitude relative to a uniform grid at no loss of accuracy. The per-$\alpha$ parameters are listed in
Table~\ref{tab:runs}. Heavier tails (smaller $\alpha$) require larger $T_{\max}$ to reach the same lowest
frequency (\S\ref{app:conv}).

\begin{table}
\centering
\caption{CTRW run parameters for the memory-kernel measurement (common frequency window
$[s_{\min},s_{\max}]=[10^{-6},10^{-3}]$).}
\label{tab:runs}
\begin{tabular}{cccc}
\hline
$\alpha$ & $T_{\max}$ & $N$ & $\tau_0$ \\
\hline
0.3 & $4\times10^{7}$ & $1.0\times10^{5}$ & 1 \\
0.5 & $4\times10^{7}$ & $1.2\times10^{5}$ & 1 \\
0.7 & $1\times10^{7}$ & $3.0\times10^{4}$ & 1 \\
\hline
\end{tabular}
\end{table}

\subsection{Laplace transform and kernel extraction}
\label{app:extract}
The Laplace transform $\tilde P(k,s)=\int_0^{T_{\max}}P(k,t)\,e^{-st}\,dt$ is computed by the
trapezoidal rule on the log-time grid, and the frequency-dependent kernel is obtained from
Eq.~\ref{eq:Ks_from_P},
The estimate is reliable only over a finite window: below $s\sim$ a few$/T_{\max}$ the finite run time
truncates the transform, and above $s_{\max}$ the measured $\tilde K(s)$ departs from $s^{1-\alpha}$
as the walk leaves the asymptotic regime. We fit only within $[s_{\min},s_{\max}]$; for the kernels of
\S3 this common window is $[10^{-6},10^{-3}]$, over which the measured Laplace-space slopes of
$\tilde K(s)$ are $0.68$, $0.50$, $0.31$ (Fig.~\ref{fig:Ks}).

\subsection{Prony fitting}
\label{app:prony}
We represent the positive primitive $\tilde K(s)/s$ by a non-negative sum,
$\tilde K(s)/s=\sum_{i=1}^{N}a_i/(s+1/\tau_i)$, with the relaxation times $\tau_i$ fixed on a
logarithmic grid spanning $[1/(s_{\max}\,p),\,p/s_{\min}]$ ($p=10$) and the amplitudes $a_i\geq0$
obtained by non-negative least squares on fractional residuals. Fixing the grid and solving only for the
amplitudes makes the fit convex and reproducible. We use $N=10$ modes. Table~\ref{tab:modeconv} shows
that the relative $L^2$ error falls from $\sim10^{-2}$ at $N=4$ to $\sim10^{-5}-10^{-4}$ by $N=10$ and then
saturates, so $N=10$ is comfortably converged. The resulting coefficients are given in
Table~\ref{tab:prony_coefficients}; over $[10^{-6},10^{-3}]$ the maximum fractional residual is
$\lesssim2\times10^{-4}$ for all $\alpha$.

\begin{table}
\centering
\caption{Convergence of the Prony fit (relative $L^2$ error on $\tilde K(s)/s$) with the number of
modes $N$.}
\label{tab:modeconv}
\begin{tabular}{cccc}
\hline
$N$ & $\alpha=0.3$ & $\alpha=0.5$ & $\alpha=0.7$ \\
\hline
4  & $1.3\times10^{-2}$ & $1.5\times10^{-2}$ & $1.0\times10^{-2}$ \\
6  & $2.5\times10^{-4}$ & $3.5\times10^{-4}$ & $3.6\times10^{-4}$ \\
8  & $3.9\times10^{-5}$ & $6.8\times10^{-6}$ & $2.4\times10^{-4}$ \\
10 & $1.7\times10^{-5}$ & $1.5\times10^{-5}$ & $1.4\times10^{-4}$ \\
16 & $2.8\times10^{-6}$ & $1.1\times10^{-6}$ & $4.5\times10^{-5}$ \\
\hline
\end{tabular}
\end{table}

\begin{table}
\centering
\caption{Prony coefficients for $\tilde{K}(s)/s=\sum_i a_i/(s+1/\tau_i)$, from non-negative least
squares over $s\in[10^{-6},10^{-3}]$ with $N=10$ modes. The $\tau_i$ lie on a fixed logarithmic grid
common to all $\alpha$; $a_i\geq0$. Entries of $0$ denote modes not selected by the fit.}
\label{tab:prony_coefficients}
\begin{tabular}{cc ccc}
\hline
 & & \multicolumn{3}{c}{$a_i$}\\
\cline{3-5}
$i$ & $\tau_i$ & $\alpha=0.3$ & $\alpha=0.5$ & $\alpha=0.7$ \\
\hline
1  & $1.0\times10^{7}$ & $2.2\times10^{-6}$ & $9.9\times10^{-5}$ & $1.8\times10^{-3}$ \\
2  & $2.8\times10^{6}$ & $6.0\times10^{-6}$ & $4.4\times10^{-5}$ & $0$ \\
3  & $7.7\times10^{5}$ & $6.0\times10^{-6}$ & $1.3\times10^{-4}$ & $1.4\times10^{-3}$ \\
4  & $2.2\times10^{5}$ & $3.0\times10^{-5}$ & $2.6\times10^{-4}$ & $9.1\times10^{-4}$ \\
5  & $6.0\times10^{4}$ & $4.3\times10^{-5}$ & $4.3\times10^{-4}$ & $2.8\times10^{-3}$ \\
6  & $1.7\times10^{4}$ & $1.6\times10^{-4}$ & $9.7\times10^{-4}$ & $2.2\times10^{-3}$ \\
7  & $4.6\times10^{3}$ & $3.1\times10^{-4}$ & $1.6\times10^{-3}$ & $5.3\times10^{-3}$ \\
8  & $1.3\times10^{3}$ & $9.7\times10^{-4}$ & $3.0\times10^{-3}$ & $8.7\times10^{-3}$ \\
9  & $3.6\times10^{2}$ & $5.9\times10^{-4}$ & $5.4\times10^{-3}$ & $0$ \\
10 & $1.0\times10^{2}$ & $1.5\times10^{-2}$ & $2.4\times10^{-2}$ & $5.3\times10^{-2}$ \\
\hline
\end{tabular}
\end{table}

\begin{figure}
\centering
\includegraphics[width=0.99\linewidth]{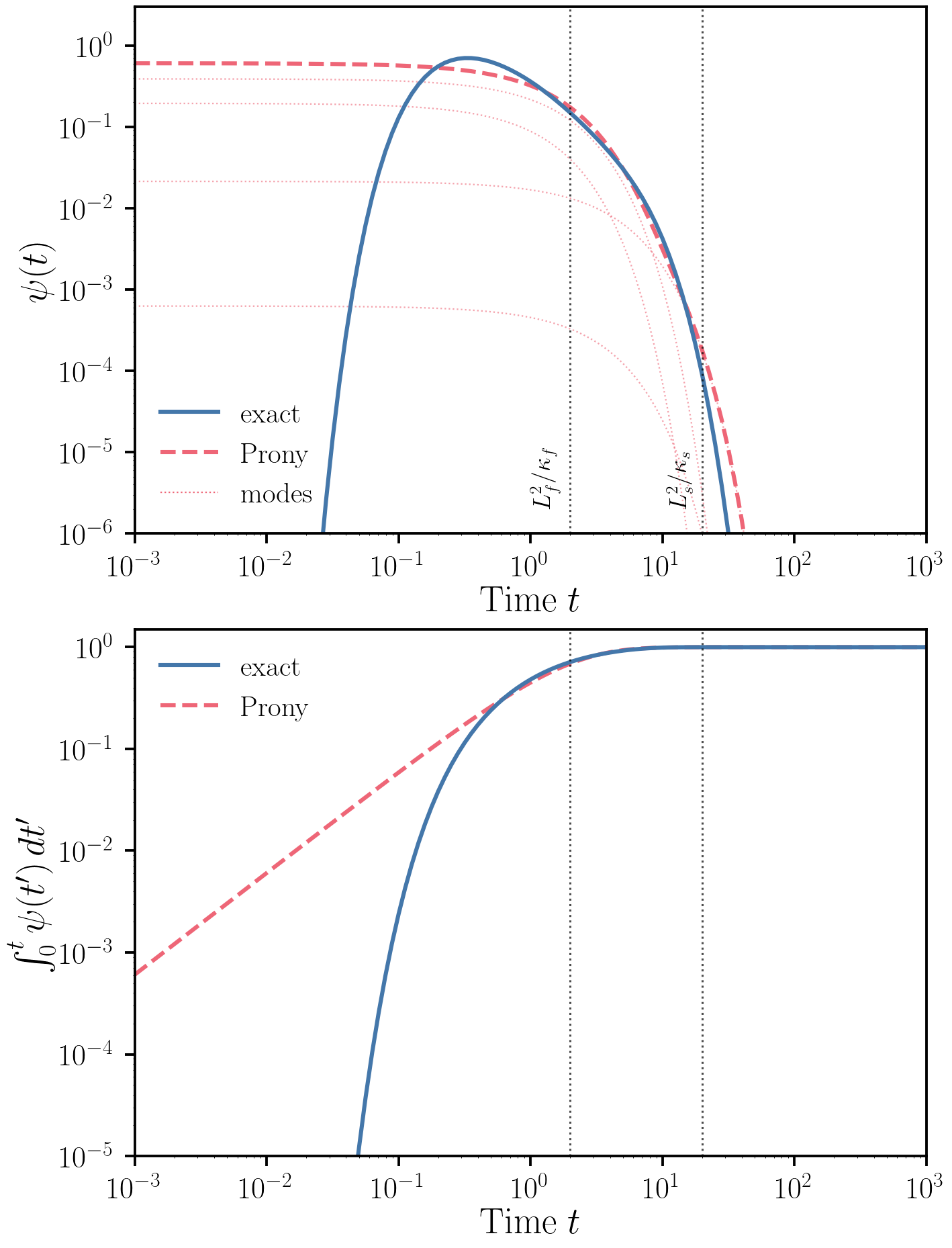}
\caption{\red{Waiting time density $\psi(t)=\mathcal{L}^{-1}[U_L+U_R]$ for the constant-cell medium. \emph{Top:} Exact density (solid), the non-negative exponential sum used in the accelerated Monte Carlo (dashed), and its individual modes (dotted). Vertical lines mark the crossing times $L_f^2/\kappa_f$ and $L_s^2/\kappa_s$. \emph{Bottom:} The corresponding cumulative distributions. The exponential sum matches the exact distribution over the relevant crossing times, but has an artificial short time excess because it remains finite as $t\to0$. The exact density is exponentially suppressed.}}
\label{fig:prony_fpt}
\end{figure}

\red{The accelerated Monte Carlo discussed in this paper uses the same representation. The two-cell first-passage transforms $U_{L,R}(s)$ are fitted by non-negative sums of $\lambda_m/(s+\lambda_m)$, and waiting times are drawn from the corresponding mixture of exponentials. Figure~\ref{fig:prony_fpt} compares this mixture with the exact waiting time density for the constant cell medium of Fig.~\ref{fig:const_all}. The fit is accurate over the crossing times that control the transport, from $L_f^2/\kappa_f$ to $L_s^2/\kappa_s$ and longer, but it overestimates the weight at short times. This is because any finite exponential sum has a finite density at $t=0$, while the exact first-passage density is exponentially suppressed,
\begin{equation}
\psi(t)\propto e^{-L^2/4\kappa t},
\end{equation}
because a particle released at one interface must diffuse a distance $L$ before reaching the next. In the fit, the cumulative probability below $t=0.05\,L_f^2/\kappa_f$ is $6\times10^{-2}$, compared with $2\times10^{-3}$ in the exact distribution. This is the same artificial short-time response noted below Eq.~(\ref{eq:K_N_t}). Adding more exponential modes does not remove it, because no finite exponential sum can reproduce the essential singularity at the origin. It can be reduced by including the known short-time asymptotic in the basis, or avoided entirely by sampling from a pretabulated exact cumulative distribution, as in the exact-waiting-time runs of Fig.~\ref{fig:const_all}. 
In practice, we numerically invert $U_{L,R}(s)/s$ once for each interface type and tabulate the cumulative distributions $F_{L,R}(t)$ on a logarithmic grid from $10^{-4}\tau_{\rm fast}$ to $60\tau_{\rm slow}$. Each hop then draws the exit side from $U_L(0)/[U_L(0)+U_R(0)]$ and the waiting time by inverse interpolation of the corresponding cumulative distribution in $\log t$. No inversion is performed during the simulation. The constant-cell medium requires two tables, while the variable-cell ring uses one per interface.
This short-time excess has little effect on $P(k,s)$, but it is amplified in the recovered kernel, as discussed in \S\ref{subsec:recovering}.}

\subsection{Convergence and finite-$k$ checks}
\label{app:conv}
\emph{Particle number.} At fixed $T_{\max}$ the scatter of $\tilde K(s)$ about $s^{1-\alpha}$ falls as
$N^{-1/2}$ (e.g.\ from $0.046$ to $0.011$ dex as $N$ increases from $2\times10^4$ to $3.2\times10^5$).
The compact time-domain reconstruction is less sensitive: its deviation from the reference power law
plateaus at $\sim0.1$ dex once $N\gtrsim10^5$ and does not improve further, because the residual
undulation is set by the finite mode count, not by particle noise.

\emph{Run length.} $T_{\max}$ sets $s_{\min}$; extending it widens the reliable window toward low $s$
(large $t$). Heavier tails converge more slowly, so smaller $\alpha$ requires larger $T_{\max}$ to reach a
given $s_{\min}$ (Table~\ref{tab:runs}).

\emph{Finite $k$.} The kernel is extracted at small but finite $k$. For a linear-MSD (random-diffusivity)
medium the finite-$k$ correction is $\tilde K(s)=\langle D\rangle-k^2\,\mathrm{Var}(D)/s+O(k^4)$, so
the apparent frequency dependence there is an $O(k^2)$ effect. For the subdiffusive kernels
of \S3 and the multiphase transport of \S4 we verified that the extracted $\tilde K(s)$ is
$k$-independent over the wavenumbers used, consistent with the finite-$k$ tests in the former sections \S C and \S D. 

\emph{Mode count.} Convergence in $N$ is quantified in Table~\ref{tab:modeconv}.

\subsection{Spatially varying diffusivity, smoothed interfaces, and the interface problem}
\label{app:multiphase_numerics}
The multiphase medium of \S4 is a one-dimensional tiling of cells whose diffusivity alternates between
$\kappa_{\rm low}$ and $\kappa_{\rm high}$, with cell lengths drawn from a power law on $[l_{\min},l_{\max}]$
(filling fraction $f$). We treat this medium with two independent methods and cross-check them.
 
\emph{Direct (full) Monte Carlo.} We construct $\kappa(x)$ on a fine spatial grid similar as in \citep{ewart25}, replacing each sharp interface at a cell edge $x_0$ by a hyperbolic tangent transition of width $\sim1/b$,
\begin{equation}
\kappa(x)=\kappa_0+\sum_{\text{edges}}\tfrac12\,\Delta\kappa\,[1+\tanh(b(x-x_0))]
\end{equation}
Thus we have $\kappa'(x)$ defined as
\begin{equation}
    \kappa'(x)=\sum_{\text{edges}}\tfrac12\,\Delta\kappa\,b\,\mathrm{sech}^2\!\big(b(x-x_0)\big).
\end{equation}
Walkers are evolved with the It\^o--Langevin update
\begin{equation}
x\rightarrow x+\kappa'(x)\,\Delta t+\sqrt{2\kappa(x)\,\Delta t}\;\xi,\qquad \xi\sim\mathcal{N}(0,1),
\end{equation}
whose Fokker--Planck equation is the conservative diffusion equation $\partial_t n=\partial_x[\kappa(x)\,\partial_x n]$, and the drift $\kappa'(x)$ is the It\^o correction required to reproduce the Fickian flux.
In the timing benchmarks, we use $\Delta t=10^{-3}\ell_{\rm min}^2/\kappa_{\max}$, as listed in Table~\ref{tab:benchmark}. The timestep must also resolve the smoothed interface, $\Delta t\lesssim(1/b)^2/\kappa_{\max}$, so larger $b$ requires a smaller step. We tested convergence by repeating the constant-cell run of Fig.~\ref{fig:compare} with $b=30$, $50$, and $100$, and found no significant change. The results shown use $b=30$.
 
\emph{Interface renewal problem and accelerated Monte Carlo.} In the sharp-interface limit, diffusion within
a cell reduces to a first-passage problem between neighboring interfaces. For a two-cell unit (lengths
$L_L,L_R$, diffusivities $k_L,k_R$) the joint (exit-side, first-passage-time) transforms $U_L(s),U_R(s)$ are
obtained analytically; the splitting probabilities are $U_{L,R}(0)$ and the mean transit time follows from the $s\to0$ slope of $U_L+U_R$. The accelerated Monte Carlo evolves walkers directly on the interface lattice using these kernels. Each transform is represented by a non-negative exponential sum
\begin{equation}
U_{L,R}(s)=\sum_m w^{L,R}_m\,\frac{\lambda_m}{s+\lambda_m},
\end{equation}
fit by non-negative least squares on a fixed rate grid $\{\lambda_m\}$, with the fit constrained to reproduce
the exit probabilities $U_{L,R}(0)$ and the mean transit time. Each interface-to-interface hop then draws a
mode with probability $\propto w_m$, an exponential waiting time $\Delta t=-\lambda_m^{-1}\ln\mathcal{U}$
($\mathcal{U}$ sampled from a uniform distribution), and a step $\pm L$ to the adjacent interface; the two interface parities (the cell ordering seen from either side) use separate fits. The two schemes describe the same transport: the accelerated Monte Carlo is the sharp-interface ($1/b\to0$) limit of the full simulation, with the analytic kernels $U_L,U_R$ replacing the explicitly resolved intra-cell diffusion, and the two propagators agree in Fig.~\ref{fig:compare}.
\end{document}